\documentclass[12pt,preprint]{aastex}
\slugcomment{draft version}
\RequirePackage{longtable}%

\shorttitle{Cross-correlation and time lag of Cyg X-2}
\shortauthors{Y.J. Lei et al.}

\begin{document}

\title{Evolution of cross-correlation and time lag of Cyg X-2 along the branches}

\author{Y.J. Lei\altaffilmark{1}, J.L. Qu\altaffilmark{1}, L.M. Song\altaffilmark{1},
  C.M. Zhang\altaffilmark{1,2}, S. Zhang \altaffilmark{1},  F. Zhang \altaffilmark{1}, J.M.
  Wang\altaffilmark{1}, Z.B. Li\altaffilmark{1}, G.B. Zhang\altaffilmark{3}}

\altaffiltext{1}{Particle Astrophysics Center, Institute of High Energy
Physics, Chinese Academy of Sciences, Beijing 100049, P.R. China;
leiyj@mail.ihep.ac.cn}
 \altaffiltext{2}{National
Astronomical Observatories, Chinese Academy of Sciences, Beijing
100012, P.R. China}
\altaffiltext{3}{Binhai College, Qingdao, Shandong, 266555, P.R. China}

\begin{abstract}

We report the detections of the anti-correlated soft and hard X-rays, and the  time lags  
of $\sim$  hecto-second from the neutron star low-mass X-ray binary  Cyg X-2, a well-known Z-type luminous source. 
Both the anti-correlation and the positive correlation were detected during the low-intensity 
states, while  only the latter showed up during high-intensity states. 
Comparing with the lower part of normal branch and flaring branch, more observations located on  the horizontal and the upper normal branches are accompanied with the anti-correlation, 
implying   the occurrence of the anti-correlation   under circumstance of a low mass accretion rate.
So far  the anti-correlated hard lag of thousand-second timescale are only reported 
 from the Galactic black hole candidates  in their hard states. 
 Here we provide the first evidence that a similar feature can also 
establish in a neutron-star system like Cyg X-2. Finally, 
the possible origins of the observed time lags are discussed 
under the current LMXB models.

\end{abstract}


\section{Introduction}

The compact X-ray binary system hosts  either a neutron star
(NS) or a black hole (BH) candidate. Both systems show the similarity in their dependences
on the X-ray luminosity and spectral shape ( van
der Klis 1994a, 1994b, 2000; Belloni et al. 2002). The spectra of
X-ray binaries are generally dominated by soft component, probably from the
accretion disk and/or the surface of neutron star, and hard component from thermal Comptonization of
soft photons in a hot plasma close to the compact object (Nowak et al. 1999b; Church 2001; Di
Salvo et al. 2006b).
Low-Mass X-ray Binary (LMXB)  contains a
 companion  with the mass
 $\lesssim$ 1 $M_{\odot}$.  NS LMXBs are divided into Z-type  and Atoll-type
according to the track shape in the color-color diagrams
(CCDs), where the  different position on the track correlates with  the different timing behavior 
(Hasinger \& van der Klis 1989; Hasinger
1990).

Cyg X-2 is a  persistent LMXB, where  the type-I X-ray
bursts are detected, suggesting its compact star is a NS with a low
magnetic field (e.g. Kahn \& Grindlay 1984), and its mass is
measured  as 1.78 $\pm$ 0.23 $M_{\odot}$ (Orosz \& Kuulkers 1999).
The binary system has an evolved, late-type companion V1341 Cyg, with an orbital period
of $\sim$ 9.8 days (Cowley et al. 1979; Casares et al. 1998).
Cyg X-2 is classified as a Z source according to its track on the
 CCD (Hasinger \& van der Klis 1989; Hasinger 1990; van der Klis 2000),
where the so-called horizontal branch (HB) is located at the top, and 
the normal branch (NB) and the flaring branch (FB) at the bottom.
Along with the evolution of the source, it  moves
continuously on the Z-track, with each position related to the different mass accretion rate. 
The accretion rate reaches a minimum at the left end of the HB and  a
maximum at the right end of the FB (Hasinger \& van der Klis 1989;
Hasinger 1990). In addition, the quasi-periodic oscillations (QPOs)
of Cyg X-2 are connected with the position on CCD as well:   the
frequency varies between $\sim$ 15 and $\sim$ 60 Hz on the HB and
the upper part of the NB,  and the low frequency  about 5-7 Hz
  on the lower part of the NB (so-called NBOs) (Piraino et
al. 2002; van der Klis 2000, 2006).

Although Z sources are the well-known  LMXBs, the origin and evolution of the observed 
hard tails ($\gtrsim$ 30 keV) are still remaining unclear.
The hard power-law tails have been detected in the Z sources of
e.g. Sco X-1 (Manchanda 2006), GX 5-1 (Asai et al. 1994),
GX 17+2 (Di Salvo et al. 2000),  GX 349+2 (Di Salvo et al. 2001),
Cyg X-2 (Frontera et al. 1998), GX 340+0 (Lavagetto et al. 2004),
and Cir X-1 (Iaria et al. 2001; Ding et al. 2003).
Except for Sco X-1 whose hard tail does not show any clear
correlation with the position of CCD as detected by 
the Rossi X-Ray Timing Explorer ({\it RXTE}) (D'Amico et al. 2001; Manchanda
2006), and could be correlated with the periods of radio flaring
(Strickman \& Barret 2000).
 for most of the others, the hard component appears to become
weaker at the higher accretion rates.
However, for Sco X-1, Di Salvo et al. (2006a) find that the
flux of the power-law component slightly decreases when the source
moves in the CCD from HB to NB, and becomes not significantly detected
in FB with {\it INTEGRAL} and {\it RXTE}.
%
%
The jets  are also observed in the low magnetic field
NS systems (Fender \& Kuulkers 2001).
For Cyg X-2, the hard tail is detected only in the HB spectra, which is
consistent  with the fact that the radio flux is observed to have
a maximum in HB (Hasinger et al. 1990; Di Salvo et al. 2002).
Therefore, both the hard tail and radio emission could be produced by a jet
(Penninx et al. 1988; Hasinger et al. 1990; Di Salvo et al. 2002).
Such an accretion-ejection phenomena
(Das \& Chakrabarti 1999; Merier 2001) for the low magnetic NS systems 
is well addressed in the model of advection-dominated accretion flow by Narayan \& Yi (1994).

Recently, an anti-correlated time lag ($\lesssim 2000 s$) of the
hard X-rays (20-50 keV) with respect to the soft X-rays (2-7 keV) is
found in the hard state of some black hole X-ray binaries (BHXBs)
(Choudhury \& Rao 2004; Choudhury et al. 2005; Sriram et al. 2007).
The timescale of the delay could be ascribed to the viscous
timescale of matter flow in the optically thick accretion disk
(Choudhury \& Rao 2004). In addition, the outflow as a jet appears
in the low-hard state of all BH sources and the truncated accretion
disk also appears in low-hard state, which implies that the
accretion disk emission is connected with the jet emission (Fender
2001; Choudhury et al. 2005).
The purpose of this paper is to
investigate whether the anti-correlation between the soft  (2-5 keV)  and hard
(16-30 keV) X-rays exists as well in the NS LMXB system of Cyg X-2.
The paper is  organized as the follows. In section 2,
the observations and data analysis of  {\it RXTE} are introduced.
In section 3, the results obtained on the  QPO
and  the cross-correlation between the soft and hard
X-rays are presented. Finally, in section 4 are  the discussions and  summary.

\section{Observation and Data Reduction}

The observations analyzed in this paper are from the Proportional Counter Array
(PCA) on board the {\it RXTE} satellite.
The PCA consists of 5 non-imaging, coaligned Xe multiwire
proportional counter units (PCUs). The lightcurves are extracted with 4 ms
resolution from the observations when PCUs were working and, 
accordingly, the  power density spectra (PDS) are produced for QPO 
analysis. The data of the mode is B\_2ms\_16A\_0\_35\_Q, 
where the  time resolution  is 
2 ms (channel: 0-35\footnote{For the different gain
epochs of RXTE, the energy channels 0-35 correspond to the different
energy bands, the energy channels 0-35 of the five gain epochs
correspond to 1.5-9.5 keV, 1.6-11 keV, 1.9-13 keV, 2-15 keV and 2-15
keV, respectively.}). The PDS is produced by using the  XRONOS tool of 
``{\it powspec}'' (see Fig. 1).

For the analyses other than  PDS, only  PCU2 data are adopted due 
to the longest observational duration. For observations that containing segments longer than 
 2000 s, the lightcurves are  extracted from the data of standard 2 mode,  
 with bin size of 16 s, for estimating the cross-correlation via 
the XRONOS tool ``{\it crosscor}'', 
between the soft X-rays (2-5 keV) and the hard X-rays (16-30 keV). 
 With the FFT algorithm, ``{\it crosscor}''
computes the coefficient normalized by the square root of
the number of good newbins in the lightcurves, and hence provides 
the cross covariance of the two lightcurves.
The cross-correlation results are grouped into  
the positive  (see Fig. 2) , the ambiguous
 (see Fig. 3) and the anti-correlated (see Fig. 4). To investigate 
the accompanying time lag,  we fit the cross-correlation curve 
with an inverted Gaussian function for the anti-correlated delay part of 
cross-correlations. 
As shown in Figure 4 is the anti-correlations 
between the soft and hard X-rays, accompanying with  
the pivoting spectra from the data with different hardness ratios,  
here the lightcurves of those with the count-rate ratio of 16-30 keV/2-5 keV 
$>$ 10\% than the average are regarded as the hard  regions, and among the remaining $<$ 10\% than 
of the average as the soft regions.

For CCDs analysis,  the
soft and the hard colors are defined as well: as the count-rate ratio 4-6 keV/2-4 keV and 9.5-16 keV/6-9.5 keV 
(also see O\'Brien et
al. 2004), respectively, and the intensity is referred to the count-rate 
in the energy band 2-16 keV. The lightcurves are extracted from the 4-6
keV, 2-4 keV, 9.5-16 keV and 6-9.5 keV, with the  background subtracted. The
voltage settings of the PCUs on board {\it RXTE} are changed on
three occasions, defining the five {\it RXTE} gain epochs. To
correct the changes in effective area between the different {\it
RXTE} gain epochs and the gain drifts within these epochs, we
calculate the colors and intensity of the Crab Nebula with PCU2, which are supposed
 to be  constant. 
We divide the obtained color points of each observation by the corresponding Crab values that are 
closest in time but within the same gain epoch (van Straaten et al. 2005).
Figures 5-7 show  the revised CCDs and
HIDs with the bin size of 1024 s for different intensity states.

\section{Results}

\subsection{Color-Color and Hardness-Intensity Diagrams}

The intensity  of Cyg X-2 is visible to vary on the time scales of 
days to months, and accordingly has the low-, medium- and high-intensity 
states. These states are
associated with systematic changes in position and shape of the Z
track in CCD and hardness-intensity diagrams (HID) (Kuukers et al.
1996; Wijnands et al. 1996). In high level the Z pattern is shifted
to higher overall intensity with respect to the pattern of a medium
level. In the low intensity of  Cyg X-2, we find several observations 
with peculiar CCDs and HIDs  that are away from the general Z-pattern (see 
panels A and B in Fig. 5). 
The observations in panel A are from 28/29 September, 1997  and 16/17 
December, 1999; in panel  B from 23/24/26 December, 1999. 
Once Kuulkers et al. (1999) ascribed part of data (28/29 September, 1997) 
in the panel A  to NB and FB. One sees in  the CCD (the Panel B) that the 
data from 23/24/26 December, 1999 form a  curved
track, which seemly tracing the
 NB in the lower part and the FB. But the corresponding 
evidences are not presented in the plot of HID.

Figure 6 shows the  CCD well constructed from the observations of the 
dominated  medium-intensity states and some low-intensity states. Unlikely shown in Figure 5, 
these observations of low-intensity state seem to be extensions towards lower 
intensity on the HB in HID and clearly show HB QPO. The
CCD and HID of high-intensity states are shown in Figure 7. During
the high-intensity level, the hard color of the HB is lower than that of 
 the medium-intensity level. In the HID, the FB connects
to the lower part of the NB, but this is not seen  in the CCD. This implies that
 the intensity drops while the color remains the same.

\subsection{Quasi-periodic Oscillation and Branch}
While Cyg X-2 traces out a Z-track in CCD on timescales of days as a
consequence of the change in the mass accretion rate, the morphology
and position of the Z-track can vary on timescales of weeks to
months (Kuulkers et al. 1996). Generally, the individual observation
does not form a complete Z-track, and hence the branch information can not be 
obtained from the CCD only. 
However, it is well-known that the
shape of the PDS and the centroid frequency of QPO evolve with the
position in  CCD for Z (Atoll) source (see e.g., M\'endez \& van der Klis 1999; M\'endez 2000; Piraino et al. 2002).
Therefore, we can estimate to which branch the lightcurve belongs,
according to the CCD morphology, the PDS, and the centroid frequency of QPO.

For Cyg X-2, in general, QPOs with centroid frequency between
$\sim15$ and $\sim60$ Hz are on the HB and upper part of the NB.  
Kuulkers et al. (1999) find a weak QPO signal at $\sim$40 Hz from the 
observations on  28 September, 1997, and a weak noise component peaking 
around  6-7 Hz from the observations 29 September,1997. Kuulkers
et al. thus ascribe these observations to the NB and the FB. 
We calculate the PDS and find no obvious QPO signal from the observations 
on 16/17 December, 1999, implying a classification to the NB and the FB. 
The PDS  shows only
power-law noise component for the observations on 23/24 December and part observation of 26 December, 1999, and 
a QPO signal at $\sim$ 6-7 Hz on other part of December 26.

For the observations in Figure 6 where the soft and hard X-rays are
anti-correlated, the shape of PDS is power law and no NBO is found
on the lower part of the NB and the FB.
The HBOs is generally thought as the frame-dragging induced
 nodal frequency of tilt disk (Stella \& Vietri 1998; Stella et al. 1999),
   the magnetoacoustic wave frequency (Titarchuk \& Wood 2002), or  
 the Alfv\'en wave oscillation (Zhang et al. 2007).
Since we use the  QPOs for probing the branch location, and details 
 on QPOs origin are beyond the content of this paper.


\subsection{Cross-correlation between the Soft and Hard X-rays}

We analyze the cross-correlation between the soft (2-5 keV) and the
hard (16-30 keV) X-rays from all the observations  available  for Cyg X-2. 
As mentioned previously, there are three classes of the 
relationships according to the sign and the value of the derived correlation
coefficient. In addition,  observations showing simultaneously both 
positive and anti-correlation with different values of the time lag, 
as also classified as the ambiguous. Example to this is shown in Figure 3.

As shown in Figure 5, there are some special observations of Cyg X-2 
during its  low-intensity state. The soft and hard X-rays are mostly 
anti-correlated in the panel A of Figure 5, ambiguous in a few cases but 
not positive correlated. However, in panel B of Figure 5, 
only positive and  ambiguous correlations show up.

The majority of the observations on Cyg X-2 are enclosed into the CCD plot 
(Fig. 6), which is divided into  four regions.
The data in region `I' are on the vertical HB and HB, in region `II' are on the HB and upper NB, in regions `III'/`IV' are on the low NB
and FB. The hard and the soft X-rays are 26\% anti-correlated and  
28\% positively correlated in region `I'; 14\%  anti-correlated and 
51\% positively correlated in region `II'; $\sim$7\%  anti-correlated and 
$\sim$70\% positively correlated in the regions `III'/`IV' (Tab. 2).
Obviously, there are more observations with
anti-correlation in regions `I' and `II' than that in the
regions `III' and `IV'.

Figure 7 shows that the CCD and HID of the observations during the
high-intensity state. From the right pattern of Figure 7, we can
see that the intensity of the hard vertex (the transition between
the HB and NB) is $> 2000$ $counts$ $s^{-1}$. The soft X-rays of all
these observations are positively correlated with the hard X-rays.

%
%

Figure 4 shows the lightcurves of 2-5 keV and 16-30 keV and their
cross-correlation functions for the observations owning  the
anti-correlations. Table 1 shows the ObsID, the time lag obtained with the FFT 
tool ``{\it crosscor}", and the FFT coefficient. For substantiating the
correlation results from the FFT algorithm, we also estimate the Pearson
coefficient using the Pearson's test after correcting for the delay
for each  observation. The anti-correlation coefficients 
as derived with  FFT and Pearson are consistent.
Finally we have in these ObsIDs where the anti-correlations exist totally  $\sim$ 20\% soft time lags, 
$\sim$ 60\% hard time lags and $\sim$ 20\%
 time lag not obvious. The timescale of soft lag is less than 200 s, and 
that of hard lag is less than 400 s.

In addition, for the observations having anti-correlation, we 
show the spectral evolution in Figure 4, where the two spectra
correspond to the soft and hard regions of the lightcurves,
respectively.
All the spectra present the pivoting in the energies 5-16 keV,
further substantiating the existence of the anti-correlation between
the soft and hard X-rays.
Moreover,  we also have tested all the observations of showing  the
positive correlation, and do not find pivoting in the  spectral
evolution, for instance  in the ObsID 90030-01-32-00 as  shown in
Figure 2.  For the ambiguous cases, the spectral pivoting of some
observations  exists,  but is not obvious.

We also investigate on the   cross-correlation on  each individual observational segment of continuous data for  
 those ObsIDs with  anti-correlations detected.
We find that, for each individual interval,
some observations own the persistent  anti-correlations, 
while some others  have both the positive and anti-correlations.
If we get rid of the observations that the cross-correlation
functions of each individual interval of continuous data are  not
same, the distribution of anti-correlation is not affected. In this
paper, only long timescale of cross-correlation function is
considered.

\section{Discussion}

We have analyzed all the  data available from {\it RXTE} observations on the NS Cyg X-2, 
and detected the anti-correlation between the soft (2-5 keV)
and hard X-rays (16-30 keV). We find that, during the low-intensity state of Cyg X-2, 
there exists different observations showing separately either 
the anti-correlation or the positive correlation. These  observations are located on 
the NB and FB in the CCD. But for the majority of the observations, 
the fraction of
the observations with anti-correlation in HB and upper part of NB
are larger than that in low part of NB and FB. 
During the high-intensity state only positive correlation is detected.
Specially, during the FB of high-intensity
state, the intensities decrease, but the colors remain the same.

The pivoting of the wide-band X-ray spectrum is also detected in all
the observation with anti-correlation.   The mass accretion rate of
individual Z source increases along HB, NB and FB (van der Klis
2000; 2006). We discuss only the result in Figure 6 where most
observations are included. The anti-correlation between the soft and
hard X-rays is mostly found on the HB and upper parts of NB, which
suggest that the anti-correlation could be associated with the low
mass accretion rate. In the low-hard state of BHXBs, an
anti-correlation of the soft (2-7 keV) and hard X-rays (20-50 keV)
is detected, which suggests the existence of a truncated accretion
disk (Choudhury et al. 2005).
At low accretion rates, the
disk is truncated far away from the compact object and the X-ray
spectrum is dominated by a thermal Compton spectrum.
With the increase of the accretion rate, the location of the disk
truncation moves in, resulting in the Comptonizing cloud to decrease
as well as the amount of seed thermal photons to increase, which
cools the Comptonizing cloud more efficiently, giving rise to the
anti-correlation of hard and soft X-rays and the pivoting of the
wide-band X-ray spectrum (Choudhury et al. 2005).

The timescale of this anti-correlated hard X-ray time lag is
about  thousand seconds in BHXBs, which is explained to be
attributed to the viscous timescale of the matter flow in the
radiation pressure dominated optically thick accretion disk
(Choudhury \& Rao 2004), i.e., the influence by the matter flow
propagates from the outer disk (soft photon region) to the innermost
disk (hard photon region). Our results show that, for NS LMXB Cyg
X-2, the anti-correlated hard X-ray time lag is about
several hundred seconds. It is considered that the hard X-ray
emission could come from the Comptonizing of the soft seed photons
for X-ray binary system. For BHXBs, the soft seed photons are
produced from the accretion disk. However, for NS systems, the soft
seed photons are produced from the surface of NS and/or the
accretion disk,
and the Comptonizing cloud might be a hot corona, a hot
flared inner disk, or even the boundary layer between the NS and
accretion disk if it is sufficiently thin and hot (also see Popham
\& Sunyaev 2001).
If the soft seed photon comes from the NS surface, the hard photons come
from the inverse Compton scattering of soft photons by hot electrons,
and a hard lag of this Compton process is expected with timescale of about  $<$ 1s (Hasinger 1987;
van der Klis et al. 1987; Nowak et al. 1999a).
If the soft X-rays come from the cool and dense blob,
with the blob spiraling inward, it emits the soft X-ray
radiation that is Comptonized in an hot and dense corona, which results
in Comptonized X-ray spectrum to be harder with time,
causing the hard time lags of timescale $\sim$ 1 s (B{\"o}ttcher \&
Liang 1999).
Therefore, the above-mentioned models should be ruled out as
the possible mechanisms for the several hundred-second time lags,
and our results suggest that the hecto-second  time lag should
be produced by other mechanisms.

The timescale of the anti-correlated hard X-ray time lag is comparative
 to  the viscous timescale of the compact systems, therefore,
Choudhury and Rao (2004) ascribe the time lag of BH system to the
viscous timescale of the matter flow in the radiation pressure
dominated optically thick accretion disk.
Assuming that the soft
seed photons of Cyg X-2 come from the accretion disk, the viscous
time $t_{vis}$ in the disk can be calculated by
$$t_{vis} = 30 \alpha^{-1} M^{-1/2} R^{7/2} \dot{M}^{-2} s$$ where $\alpha$ is the
viscosity parameter in units of 0.01, M is the mass of the compact
object in solar mass, R is the radial location in the accretion disk
in units of 10$^7$ cm, $\dot{M}$ is the mass accretion rate in units
of 10$^{18}$ g/s (see Fender \& Belloni 2004). For Cyg X-2, taking
$\alpha$ = 1, M = 1.78 and $\dot{M}$ = 1.3 ($\sim$
2$\times$$10^{-8}$ M$_\odot$ $yr^{-1}$, see Smale 1998), we get R
$\sim$ 1.8 for a viscous timescale of $\sim$ 100 s. Thus, the
observed delay will infer  the location of the truncation of $\sim$
18 NS radius. Choudhury \& Rao (2004) obtain the location of the
disk truncation of $\sim$ 25 Schwarzschild radius for the BHXB Cyg X-3.
Table 1 shows the locations of the truncation of every
ObsID with anti-correlated hard X-rays time lag, by assuming  $\alpha$
= 1 and $\dot{M}$ = 1.3. The results of Table 1 show the locations
of the truncation are not evolutionary  along the branches. However,
unlikely the BH binaries, for the Z sources, the X-ray intensity may
not be proportional to $\dot{M}$, or $\dot{M}$ can not be estimated
accurately. In addition, the viscosity parameter $\alpha$ could be
different for the different position of Z-track. Therefore, we can
not obtain the locations of the truncation accurately, maybe,
different  from the results of Table 1, the locations of the
truncations are evolutionary  along Z-track.

The above model can explain the anti-correlated hard X-ray lag,
however, it can not explain the anti-correlated soft X-ray lag. The
anti-correlated soft X-ray lag ($< 200 s$) is also detected in Cyg
X-2, whose timescale is comparative to  the viscous timescale. If a
fluctuation produced from the innermost accretion disk could
propagate to the outer accretion disk and modulates the soft X-ray
emission regions, the soft lag is about to be explained.
It is possible to produce   such a fluctuation, since the model by
Li et al. (2007) provides a probability for it, which  ascribes the
accretion disk to be thermally unstable and to exhibit the
limit-cycle behavior (see also e.g. Kato et al. 1998). In the cycle
process, the expansion wave in the innermost disk can be formed and
moves outward with time, which perturbs the outer material,
modulating the soft X-ray emission, and resulting in the soft X-ray
lag, and the timescale is several hundred seconds.

The anti-correlation in BH binary systems implies that the
geometric structure of the disk-jet could be connected with a
truncated disk (Choudhury \& Rao 2004; Choudhury et al. 2005).
The truncated accretion disk could occur in the low-hard state where
the outflow in the form of a jet is present. In NS LMXB Cyg X-2, a
hard power-low tail is significantly detected only in the HB spectra
(Di Salvo et al. 2002). The strong non-thermal radio flares are
observed in the HB and upper NB, and the source is radio quiet in
the lower NB and FB (Hasinger et al. 1990). Therefore, the hard
power-law component could be related with the radio flares, and the
hard power-law component is related to the high-velocity electrons
(probably from a jet) of producing the radio emission (Di Salvo et
al. 2002). Our results show that in Cyg X-2 the
anti-correlations mostly present in HB and upper NB, at where the
hard power-law component and radio flares also occur. 

\acknowledgements
The authors are grateful to the anonymous referee for the helpful comments. We
are thankful for S.N. Zhang for the useful discussions. This
research has made use of data obtained through the high-energy
Astrophysics Science Archive Research Center Online Service,
provided by the NASA/Goddard Space Flight Center. We acknowledge the
RXTE data teams at NASA/GSFC for their help. This work is subsidized
by the Special Funds for Major State Basic Research Projects and by
the  Natural Science Foundation of China for support via NSFC
10473010, 10273010, 10773017, 10373013, 10733010, 10521001 and
10325313, CAS key project via KJCX2-YWT03.

\clearpage

\begin{table*}
\footnotesize
\caption{\bf ~~Details of ObsIDs which the anti-correlation of the hard and soft X-rays are detected.}
\scriptsize{}
\label{table:1}
\newcommand{\m}{\hphantom{$-$}}
\newcommand{\cc}[1]{\multicolumn{1}{c}{#1}}
\renewcommand{\tabcolsep}{0.6pc} 
\renewcommand{\arraystretch}{1.2} 
\medskip
\begin{center}
\begin{tabular}{c c c c c c c}
\hline
     & Date &  & Delay(Error)&\multicolumn{2}{c}{statistical coefficient} & R \\
\cline{5-6}
     ObsID      & Year-Month-Day        & Location       & (s)       &FFT(Error)
      & Pearson (Null Probability) &($10^{7}cm$) \\
\hline
   20057-01-01-000 & 1997-09-28 &  A/Fig. 5             & -2($\pm$16)    &-0.53($\sim$ 0.01)    & -0.51($<10^{-5}$)   &\\
   20057-01-01-00  & 1997-09-28 &  A/Fig. 5             & 64($\pm$16)    &-0.44($\sim$ 0.02)    & -0.43($<10^{-5}$)   &1.6\\
   20057-01-01-010 & 1997-09-29 &  A/Fig. 5             & -157($\pm$28)    &-0.28($\sim$ 0.01)  & -0.29($<10^{-5}$)   &\\

   40019-04-03-00  & 1999-12-16 &  A/Fig. 5           & 279($\pm$17)  &-0.28($\sim$ 0.01)     & -0.18($<10^{-5}$)   & 2.4\\
   40019-04-04-000/00 & 1999-12-17 &  A/Fig. 5      & 74($\pm$13)  &-0.57($\sim$ 0.01)     & -0.73($<10^{-5}$)   & 1.6\\
   40019-04-04-01  & 1999-12-17 &  A/Fig. 5          & 229($\pm$45)  &-0.21($\sim$ 0.01)     & -0.15(4$\times$$10^{-2}$)  &2.3\\

  40017-02-16-02  & 1999-11-20 & I/Fig. 6             & -29($\pm$30)   &-0.22($\sim$ 0.02)    & -0.24(3$\times$$10^{-3}$)   &\\
  40019-04-02-000/00 & 1999-09-23 & I/Fig. 6           & 146($\pm$21)   &-0.18($\sim$ 0.01)     & -0.34($<10^{-5}$)   & 2.0\\
 90030-01-46-00  & 2004-10-08 & I/Fig. 6               & 348($\pm$18)  &-0.24($\sim$ 0.01)     & -0.16(6$\times$$10^{-3}$)   &2.4\\
30046-01-12-00  & 1998-09-25 & I/Fig. 6               & 6($\pm$15)   &-0.33($\sim$ 0.01)    & -0.38($<10^{-5}$)   &\\
 90030-01-16-00  & 2004-05-13 & I/Fig. 6               & -37($\pm$32)  &-0.37($\sim$ 0.01)      & -0.40($<10^{-5}$)   &\\
  10066-01-01-00  & 1996-03-27 & I/Fig. 6              & -3($\pm$8)  &-0.36($\sim$ 0.01)      & -0.39($<10^{-5}$)    &\\
  90030-01-91-00  & 2005-12-25 &  I/Fig. 6              & 121($\pm$35)  &-0.41($\sim$ 0.02)      & -0.39($<10^{-5}$)   &1.9\\
   90030-01-95-00  & 2006-01-14 &  I/Fig. 6              & 11($\pm$30)  &-0.16($\sim$ 0.01)      & -0.19($\sim$$10^{-4}$)  &\\

 20053-04-01-03 & 1997-07-02    & II/Fig. 6        & -54($\pm$9)    &-0.30($\sim$ 0.01)    & -0.32($<10^{-5}$)    &    \\
90030-01-88-00  & 2005-12-10 &  II/Fig. 6        & 233($\pm$29)  &-0.20($\sim$ 0.01)     & -0.20(4$\times$$10^{-4}$)   &2.3\\
40017-02-18-00  & 1999-12-27   &  II/Fig. 6              & 255($\pm$28)  &-0.49($\sim$ 0.01)     & -0.38($<$$10^{-5}$)   &2.3\\
 91009-01-32-00  & 2005-08-04 &  II/Fig. 6        &337($\pm$19)&-0.56($\sim$ 0.01)        & -0.55($<10^{-5}$)  &2.5\\
60417-01-02-00  & 2001-08-03  & II/Fig. 6         & -184($\pm$29)  &-0.33($\sim$ 0.03)    & -0.19(4$\times$$10^{-2}$)   &\\
90030-01-72-00  & 2005-02-13 &  II/Fig. 6        & 62($\pm$21)  &-0.29($\sim$ 0.01)      & -0.23($\sim$$10^{-3}$)   &1.6\\
40017-02-07-00  & 1999-05-05 &  II/Fig. 6        & 376($\pm$25)   &-0.39($\sim$ 0.02)    & -0.16($\sim$$10^{-5}$)   &2.6\\
90030-01-36-00  & 2004-08-20 & II/Fig. 6         & -35($\pm$25)  &-0.30($\sim$ 0.02)     & -0.34($<10^{-5}$)   &\\
90030-01-31-00  & 2004-07-26  & II/Fig. 6         & 277($\pm$70)  &-0.36($\sim$ 0.01)     & -0.33($<10^{-5}$)   &1.8\\
30046-01-01-00  & 1998-07-15   & II/Fig. 6        & 83($\pm$9)  &-0.52($\sim$ 0.01)       & -0.51($<10^{-5}$)   &1.7\\
40021-02-01-00 &  2000-01-14  & II/Fig. 6         & 178($\pm$15) & -0.38($\sim$ 0.01) &-0.26($<10^{-5}$)  &   2.1   \\
   30046-01-05-01  & 1998-08-10 & III/Fig. 6       & 46($\pm$14)  & -0.38($\sim$ 0.01)     &-0.41($<10^{-5}$)   &1.4\\
   00029-04-01-00  & 1996-01-27 & III/Fig. 6       & 17($\pm$29)  &-0.40($\sim$ 0.02)      & -0.39($<10^{-5}$)    &\\
  90022-08-02-00  & 2004-12-08  & III/Fig. 6       & 124($\pm$39)  &-0.40($\sim$ 0.06)     & -0.25(2$\times$$10^{-3}$)   &1.9\\
   90030-01-68-00  & 2005-01-24 & III/Fig. 6       & 227($\pm$19)  &-0.32($\sim$ 0.01)     & -0.22($\sim$$10^{-5}$)   &2.2\\

   90030-01-67-00  & 2005-01-19 & IV/Fig. 6  & -41($\pm$16)  &-0.50($\sim$ 0.02)     & -0.53($<10^{-5}$)   &\\

\hline
\end{tabular}
\end{center}
\footnotesize
\tablecomments{R is the locations of the truncation of the accretion disk, a viscous timescale
of 100s is corresponding to 18 NS radius.
40019-04-02-000/00 is the observations of 40019-04-02-000 and 40019-04-02-00,
and 40019-04-04-000/00 is the observations of 40019-04-04-000 and 40019-04-04-00.}
\end{table*}

\begin{table*}
\begin{center}
\caption{\bf ~~The percent of positive and anti-correlation for each
region of Figure 6.}
\medskip
\begin{tabular}{l c c c c}
\hline
       & I      & II & III &  IV \\
\hline
  positive-correlation &  28\%  & 51\% & 76\% &65\%\\
\hline
    anti-correlation   &  26\%  & 14\%  & 7\% &7\%\\

\hline
   ambiguous           & 46\%  &  35\% &17\% & 28\%\\

\hline
\end{tabular}
\end{center}
\end{table*}

\clearpage

\begin{figure*}
\includegraphics[width=3.1cm,angle=270,clip]{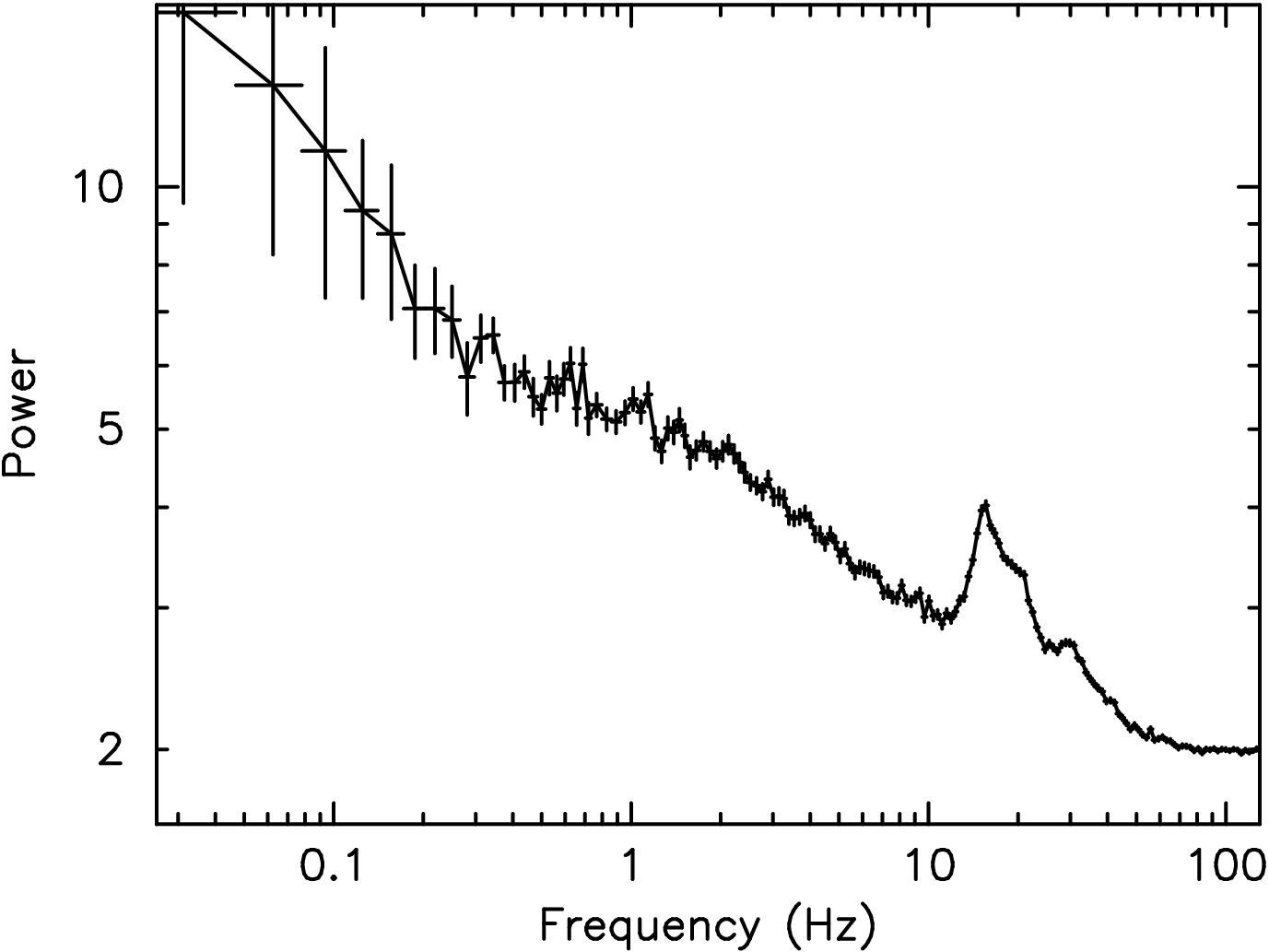}
\includegraphics[width=3.1cm,angle=270,clip]{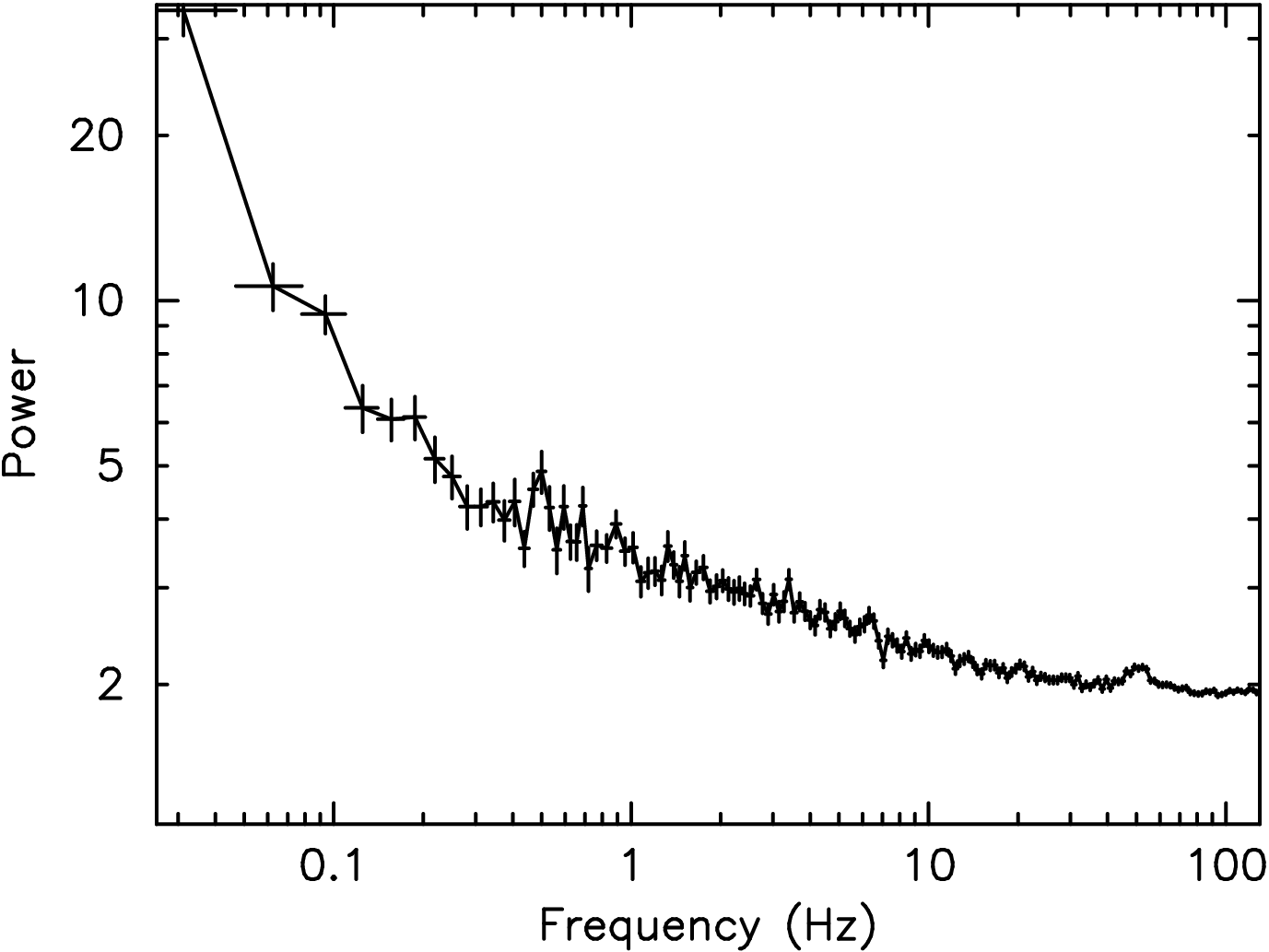}
\includegraphics[width=3.1cm,angle=270,clip]{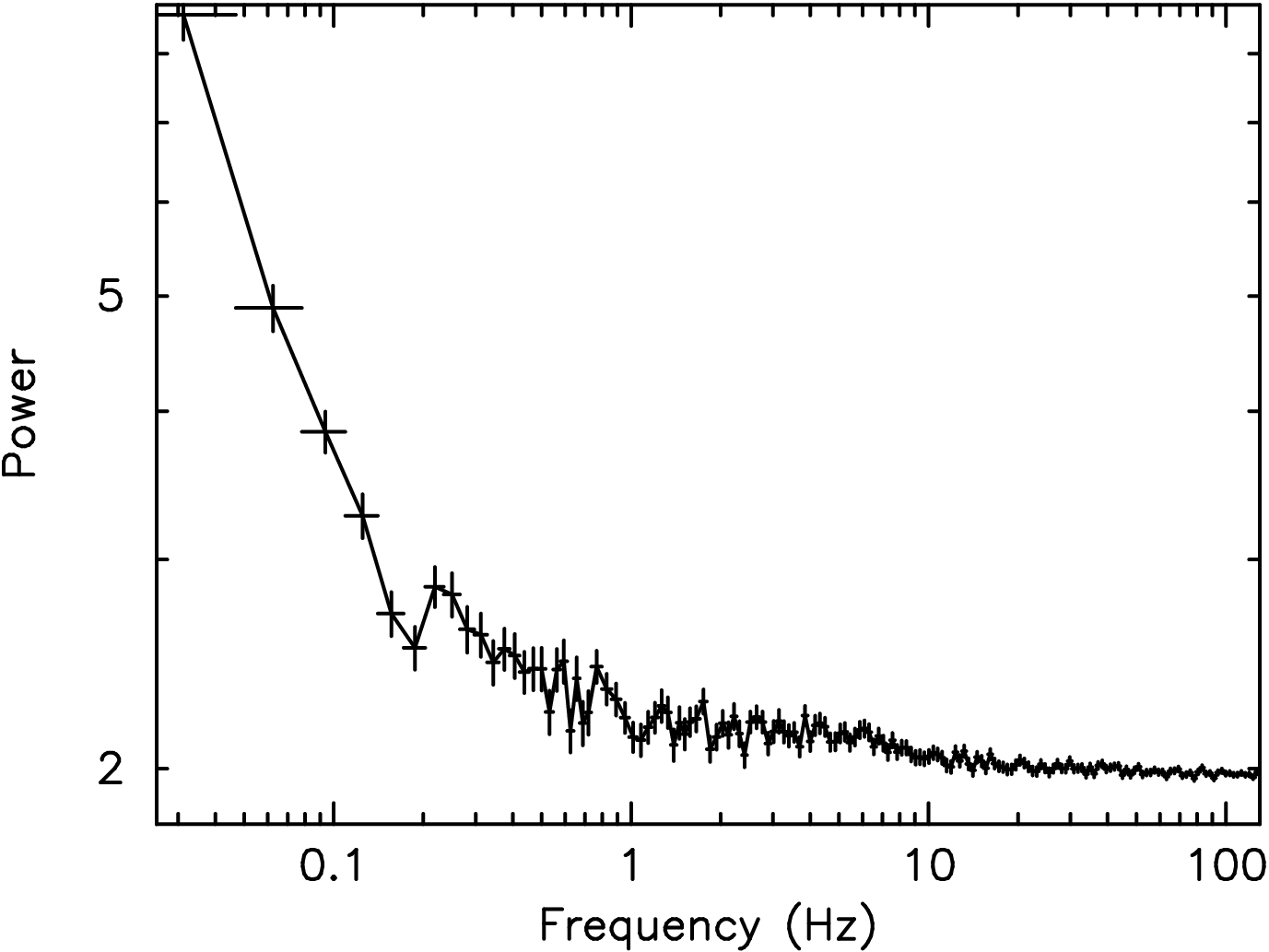}
\caption{
The left panel shows HBO with the centroid frequency $\sim$15-$\sim$50 Hz,  the middle
panel shows HBO with the centroid frequency $\gtrsim$50 Hz,
and the right one shows no QPO during the lower part of NB and FB.
}
\label{fig1}
\end{figure*}

\begin{figure*}
\includegraphics[width=3.1cm,angle=270,clip]{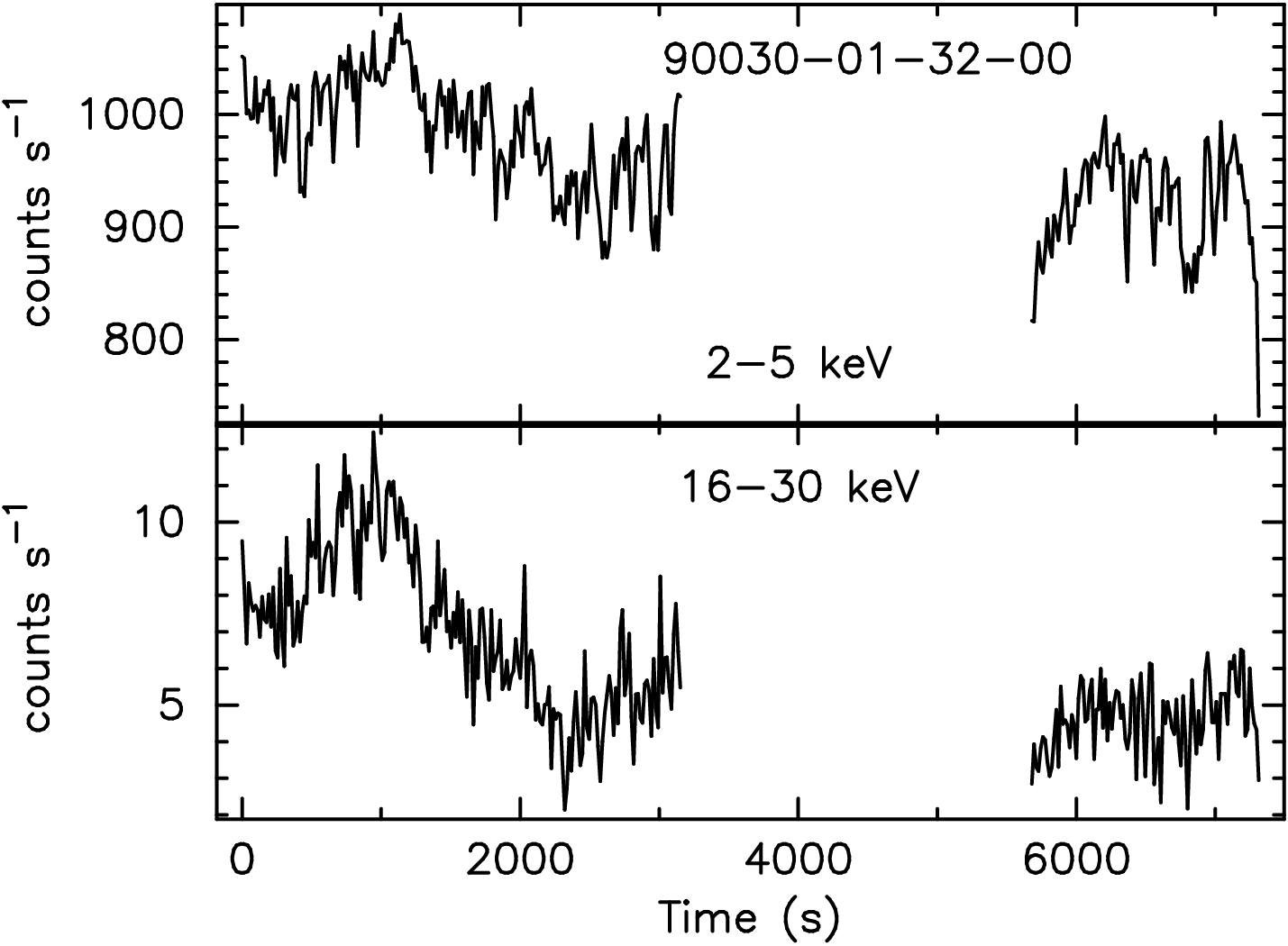}
\includegraphics[width=3.1cm,angle=270,clip]{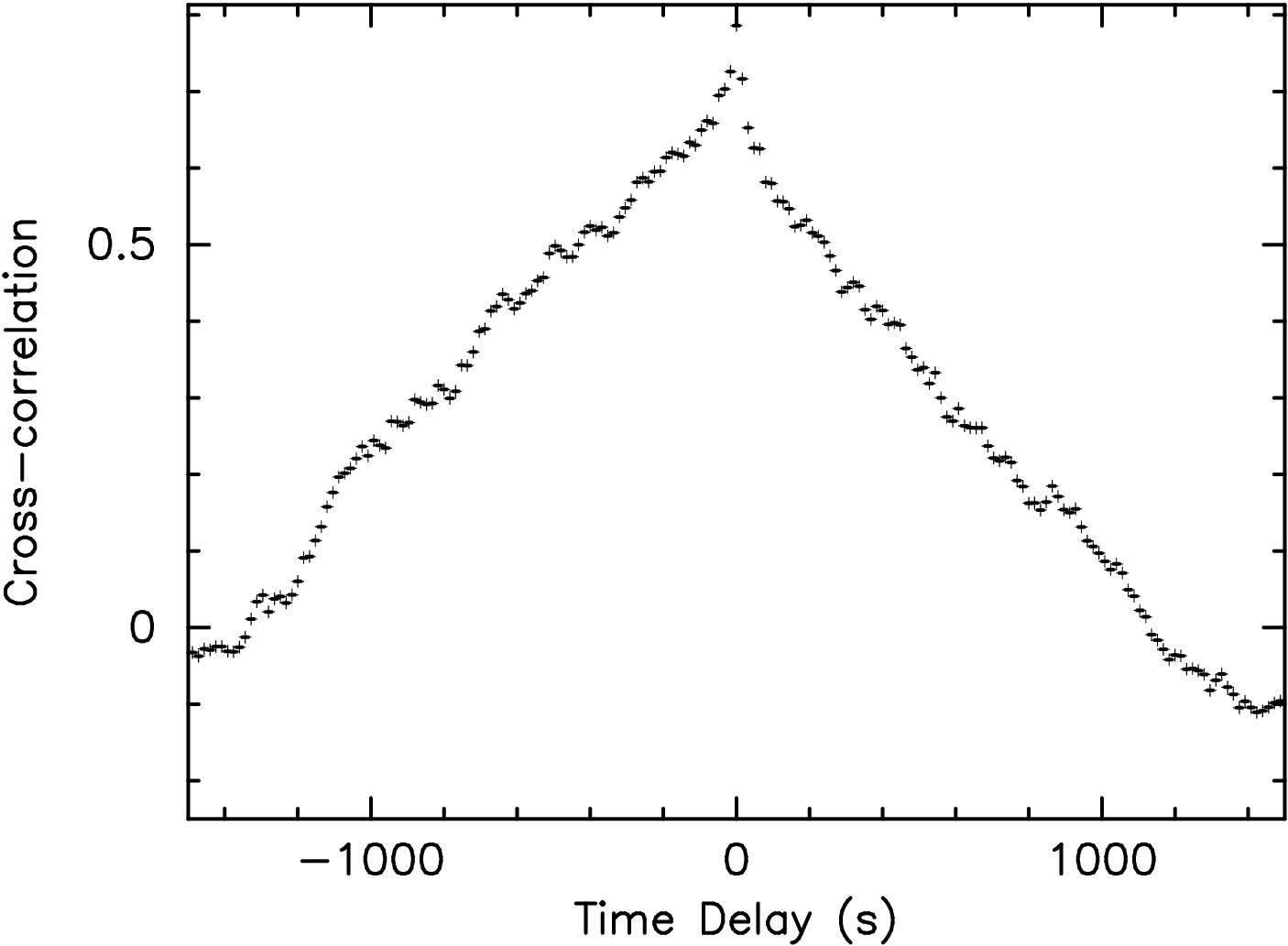}
\includegraphics[width=3.1cm,angle=270,clip]{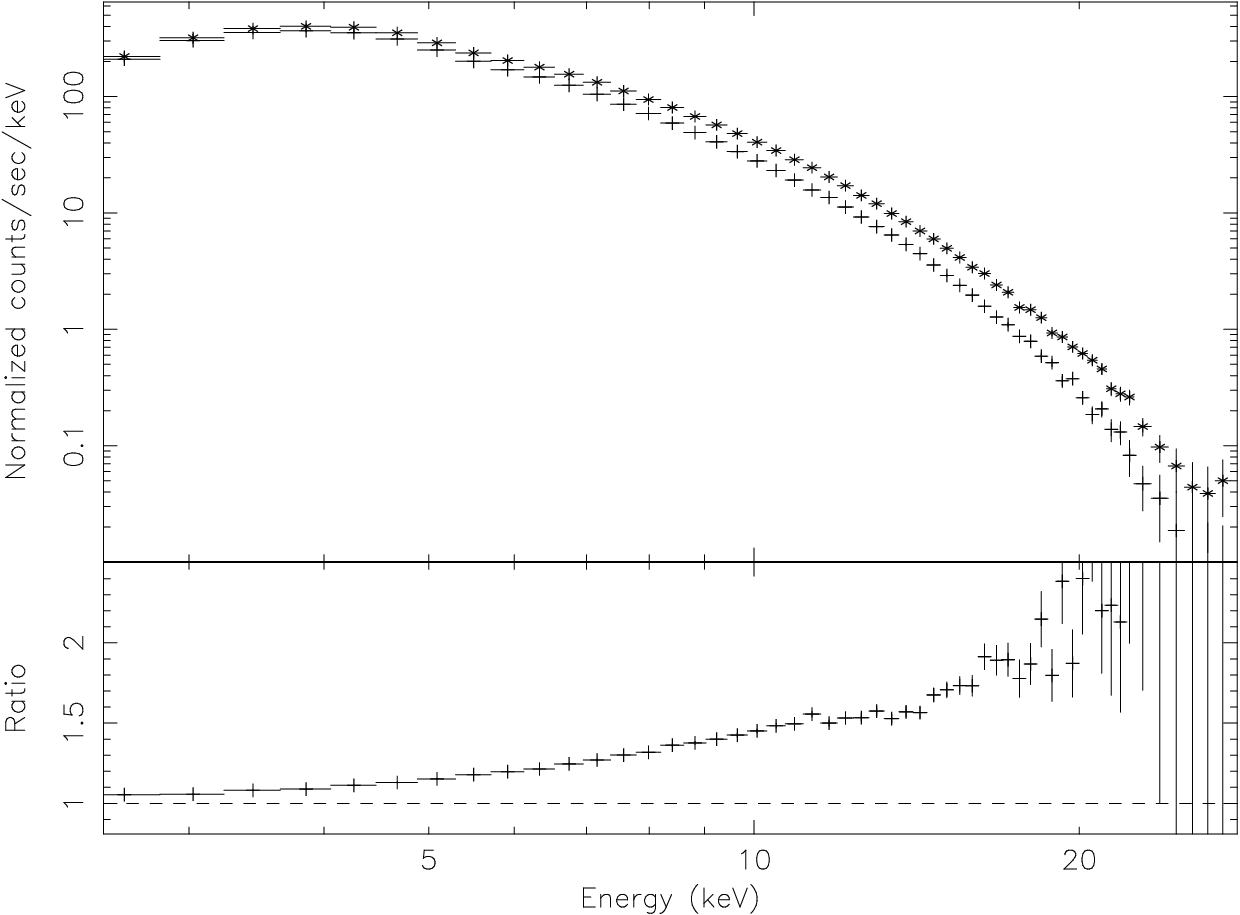}
\caption{
The left panel shows the lightcurves of ObsID 90030-01-32-00 and the middle
panel shows the cross-correlation function where a typical positive correlation is observed.
The right one shows the X-ray spectra of the ObsID
for hard and soft regions of the lightcurve (upper)
and the ratio of the hard and soft spectra (lower),
 where no pivoting is shown.
}
\label{fig2}
\end{figure*}

\begin{figure*}
\includegraphics[width=3.1cm,angle=270,clip]{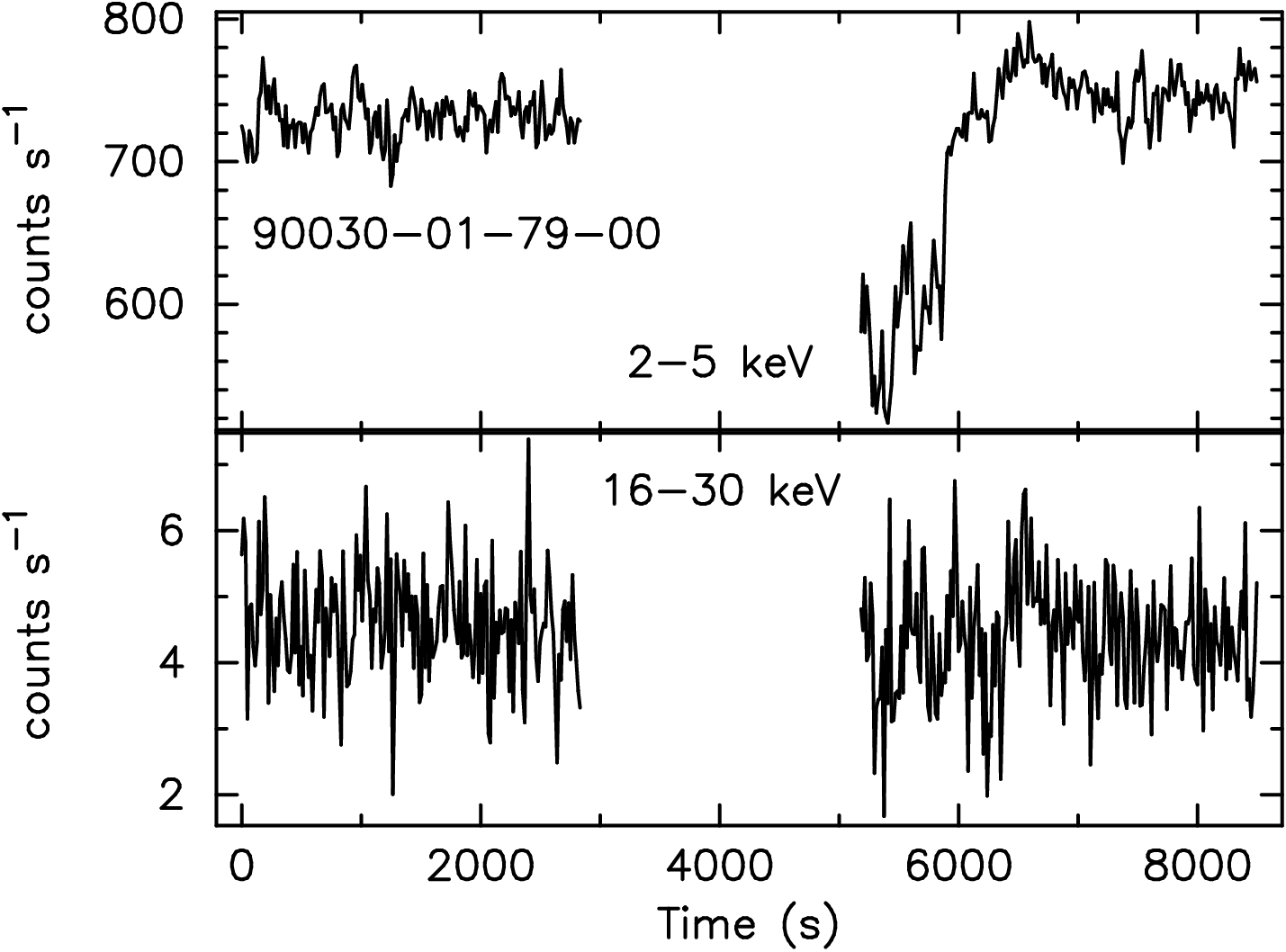}
\includegraphics[width=3.1cm,angle=270,clip]{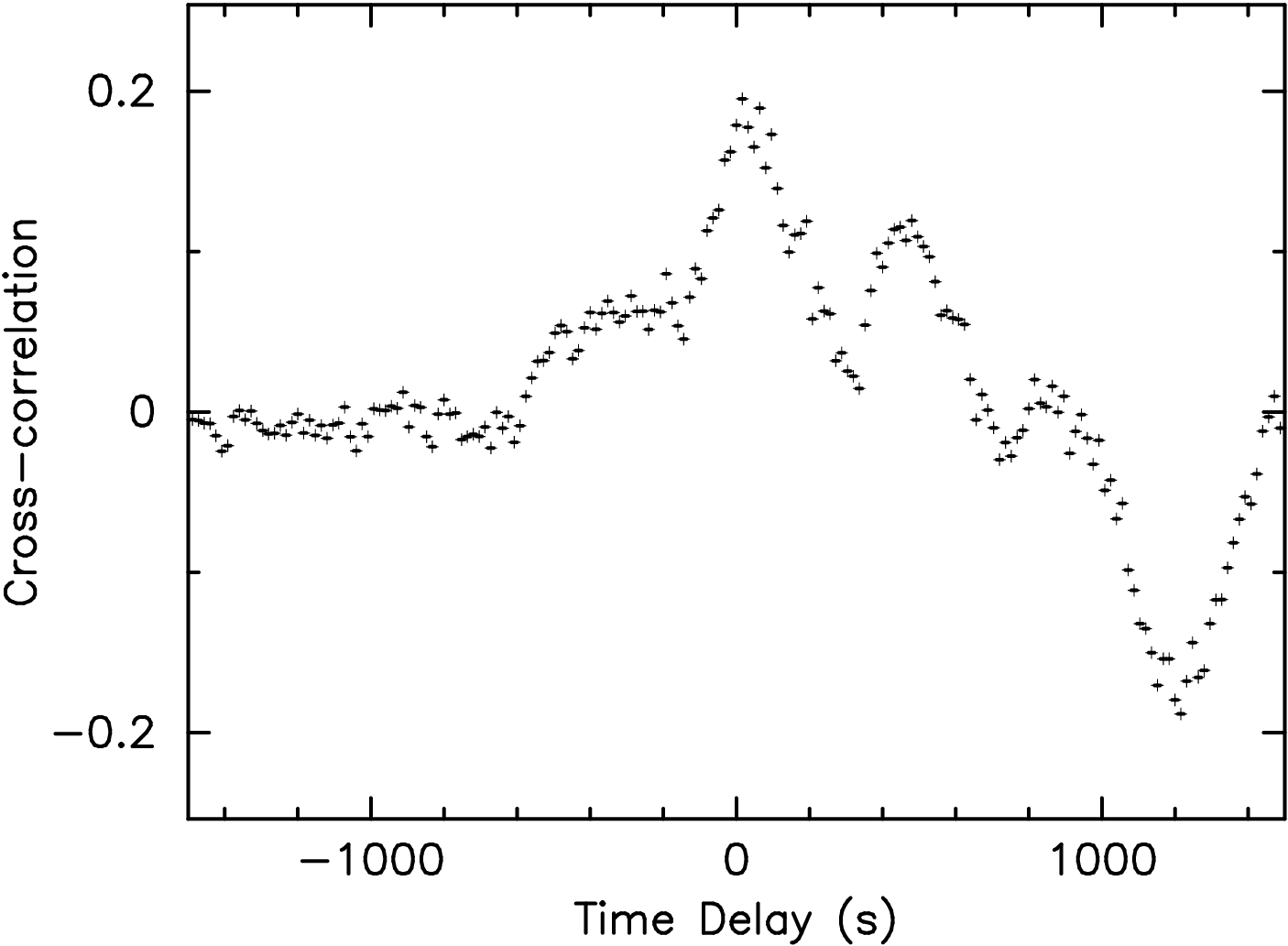}
\includegraphics[width=3.1cm,angle=270,clip]{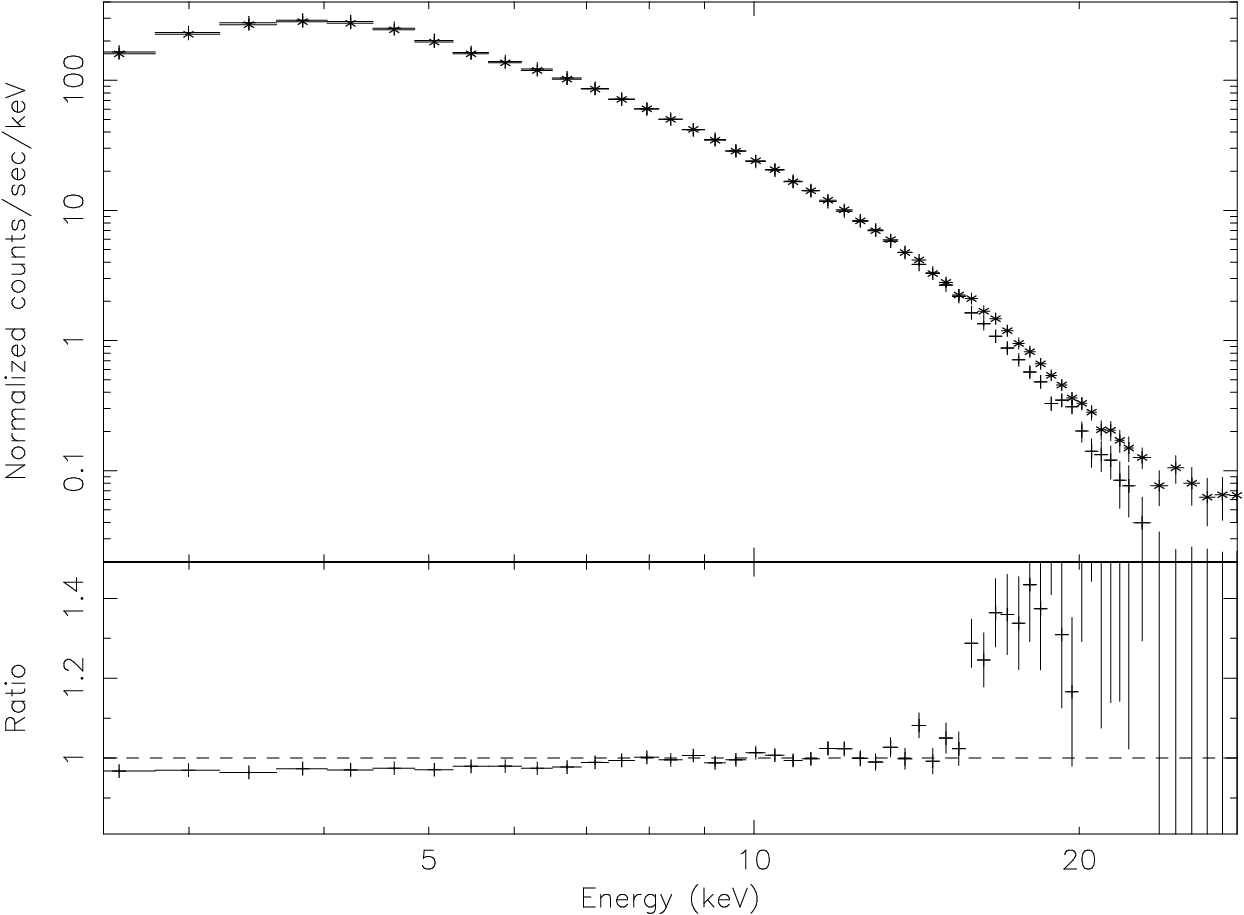}
\caption{
The left panel shows the lightcurves of ObsID 90030-01-79-00 and the middle
panel shows the cross-correlation function where no obvious correlation is observed.
The right one shows the X-ray spectra of the ObsID
for hard and soft regions of the lightcurve (upper)
and the ratio of the hard and soft spectra (lower),
 where no obvious pivoting is shown.
}
\label{fig3}
\end{figure*}

\clearpage
\begin{figure*}
\begin{center}
\includegraphics[width=3.1cm,angle=270,clip]{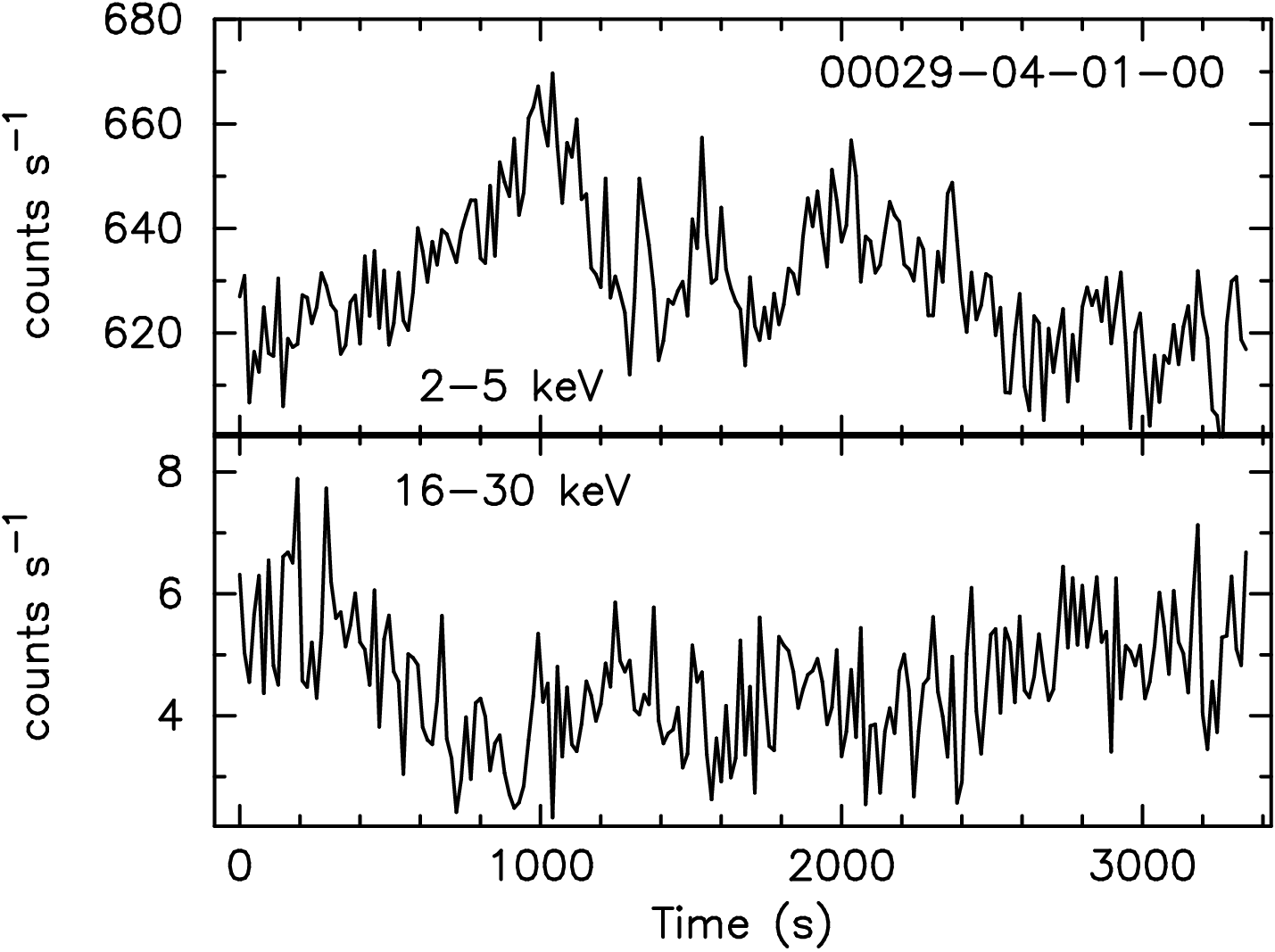}
\includegraphics[width=3.1cm,angle=270,clip]{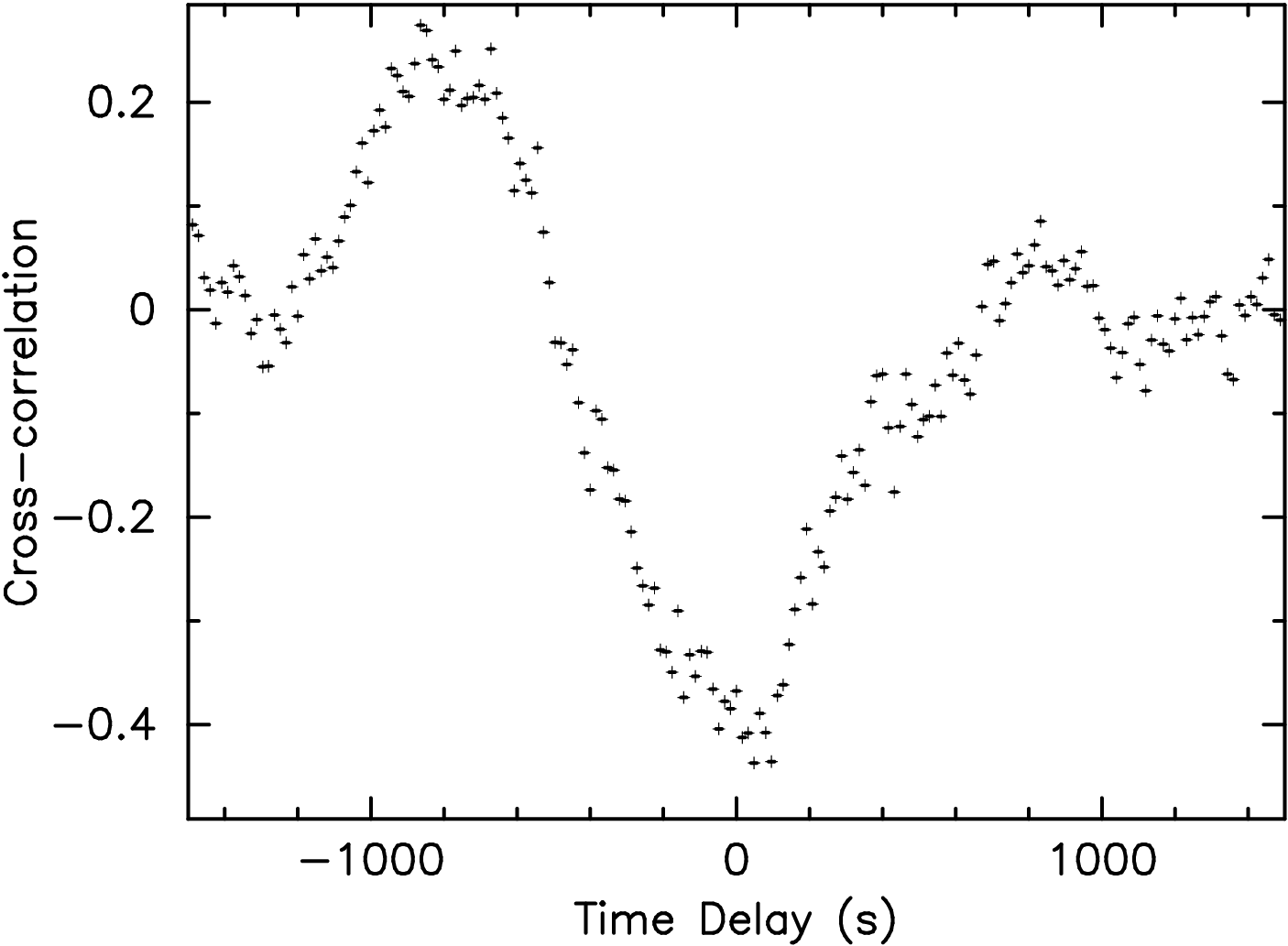}
\includegraphics[width=3.1cm,angle=270,clip]{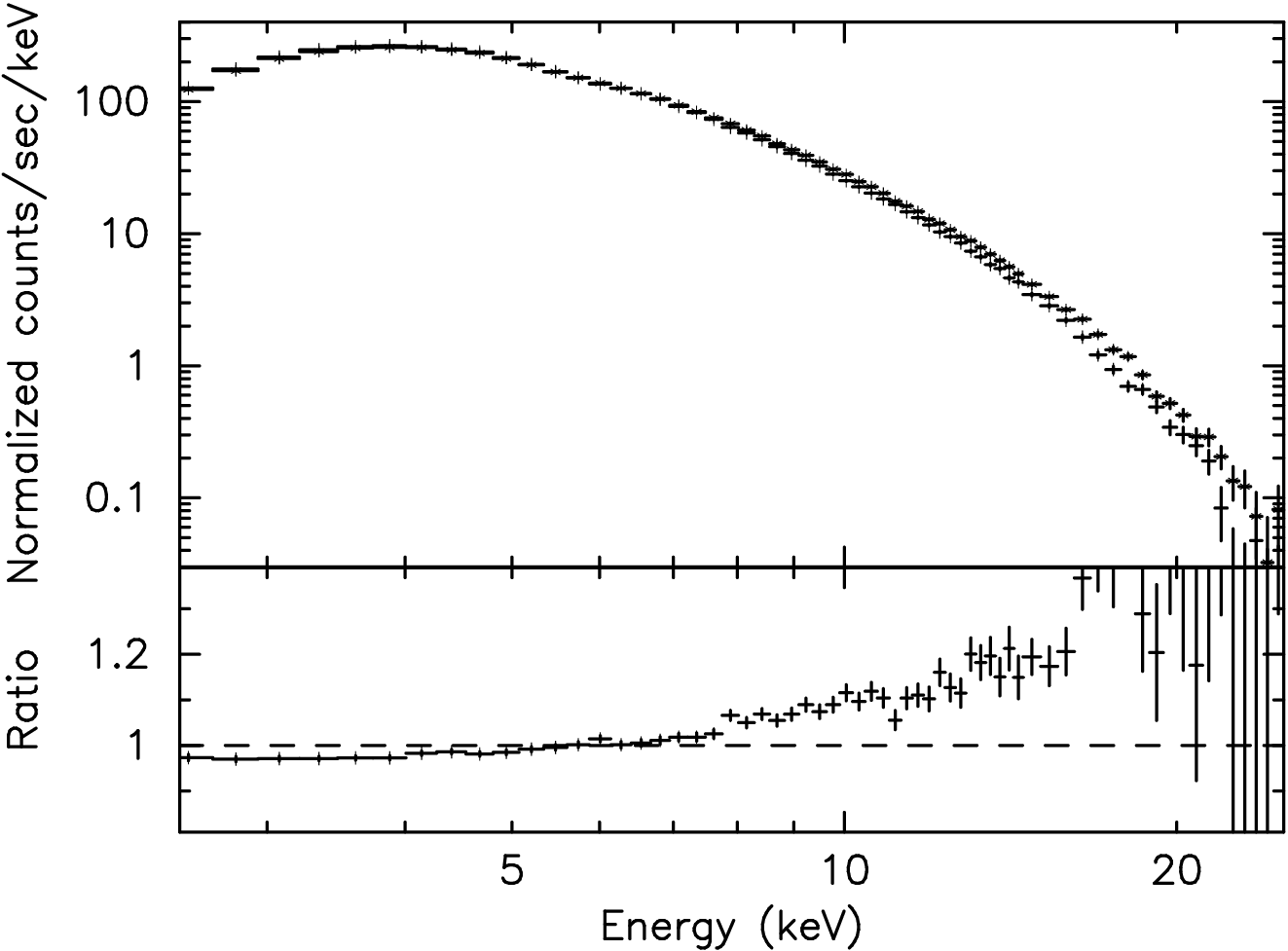}
\includegraphics[width=3.1cm,angle=270,clip]{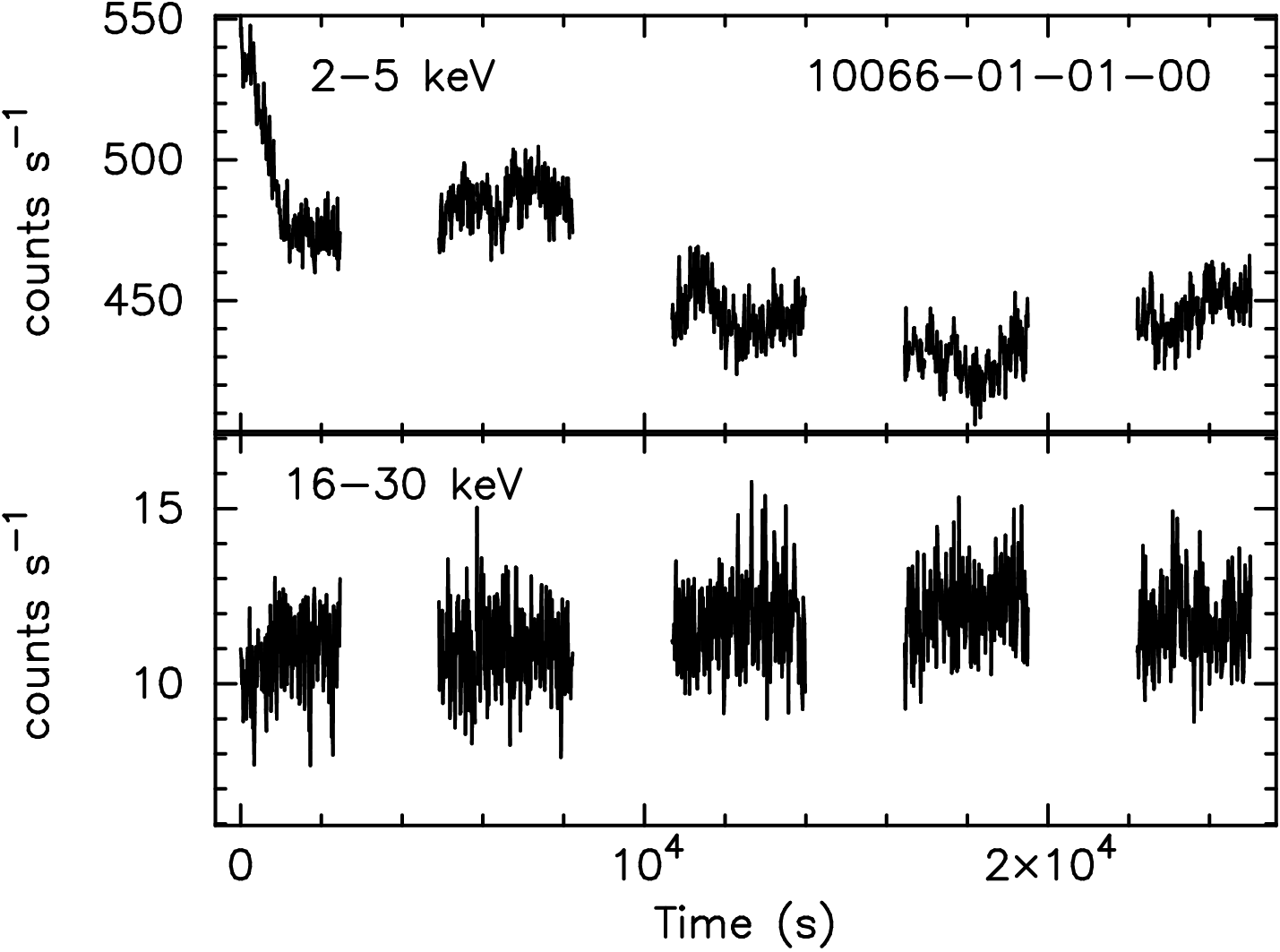}
\includegraphics[width=3.1cm,angle=270,clip]{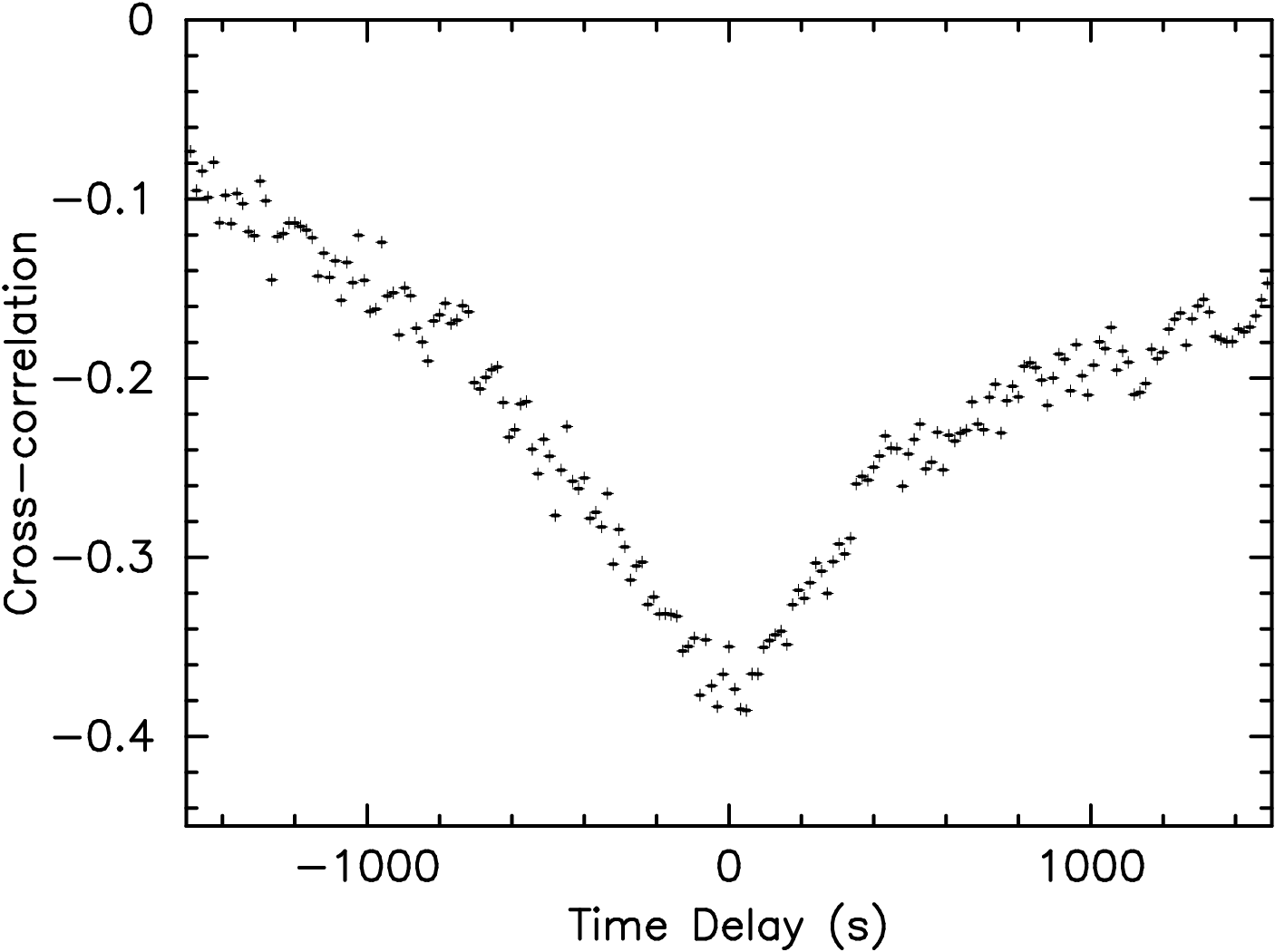}
\includegraphics[width=3.1cm,angle=270,clip]{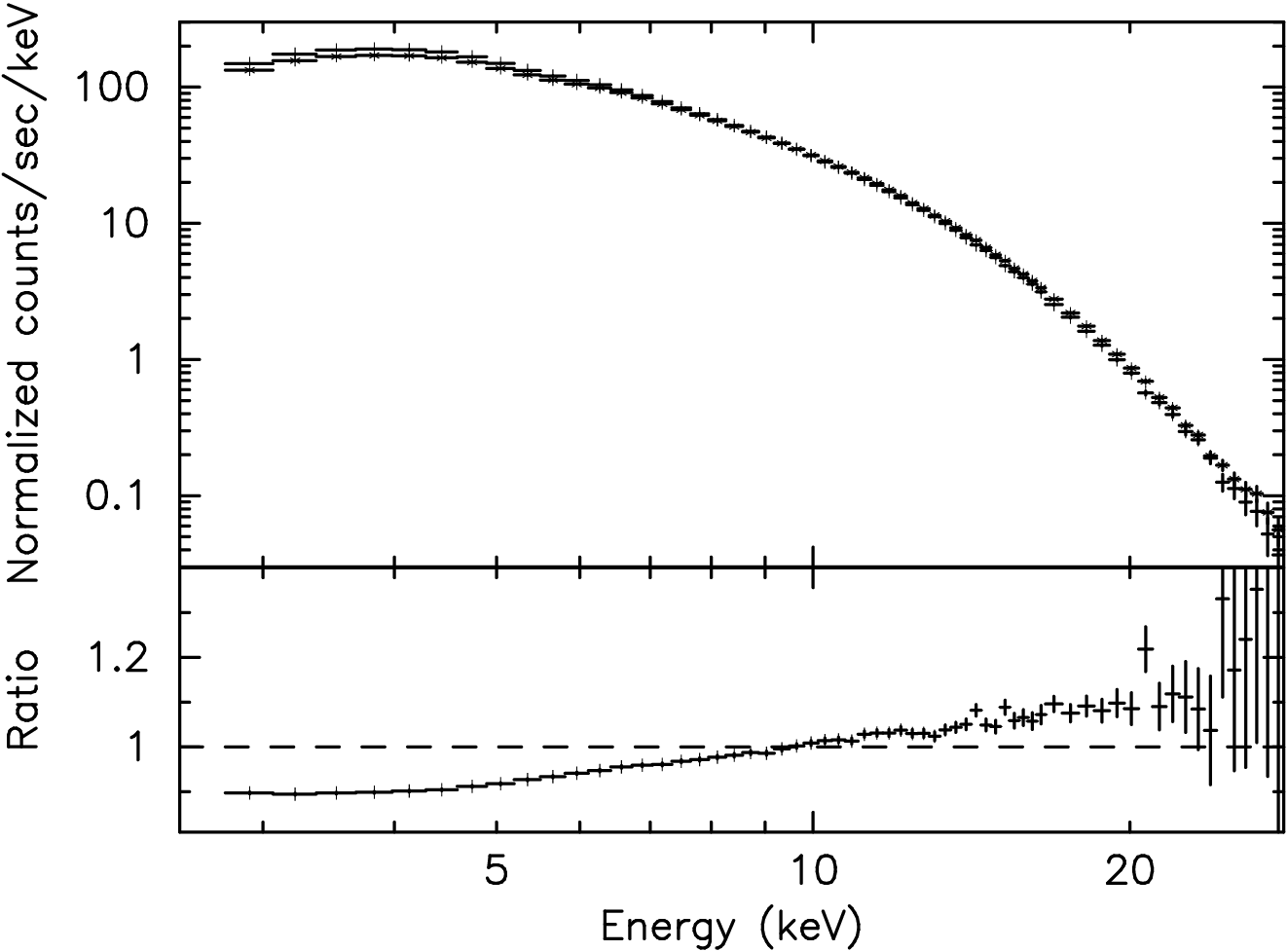}
\includegraphics[width=3.1cm,angle=270,clip]{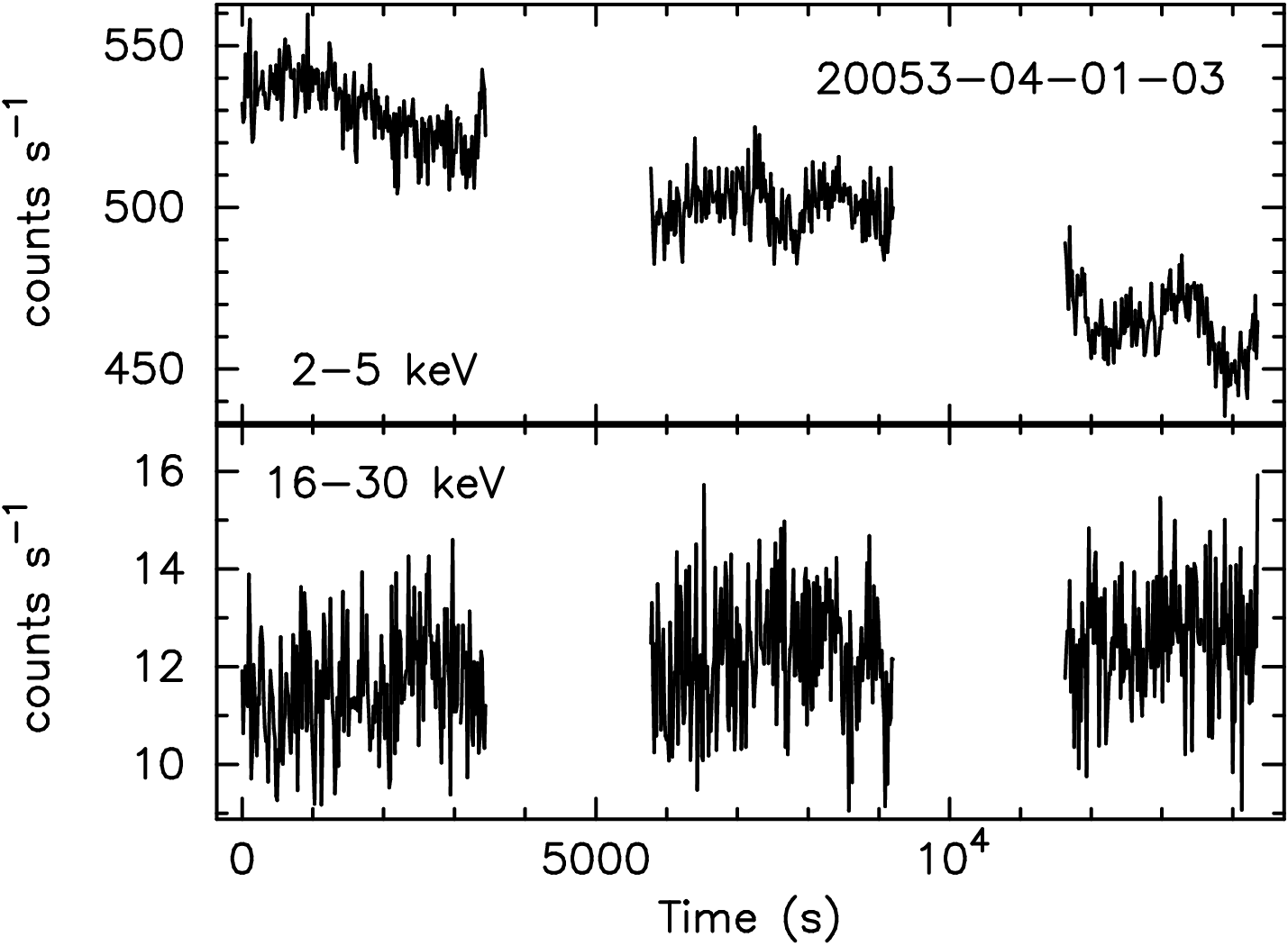}
\includegraphics[width=3.1cm,angle=270,clip]{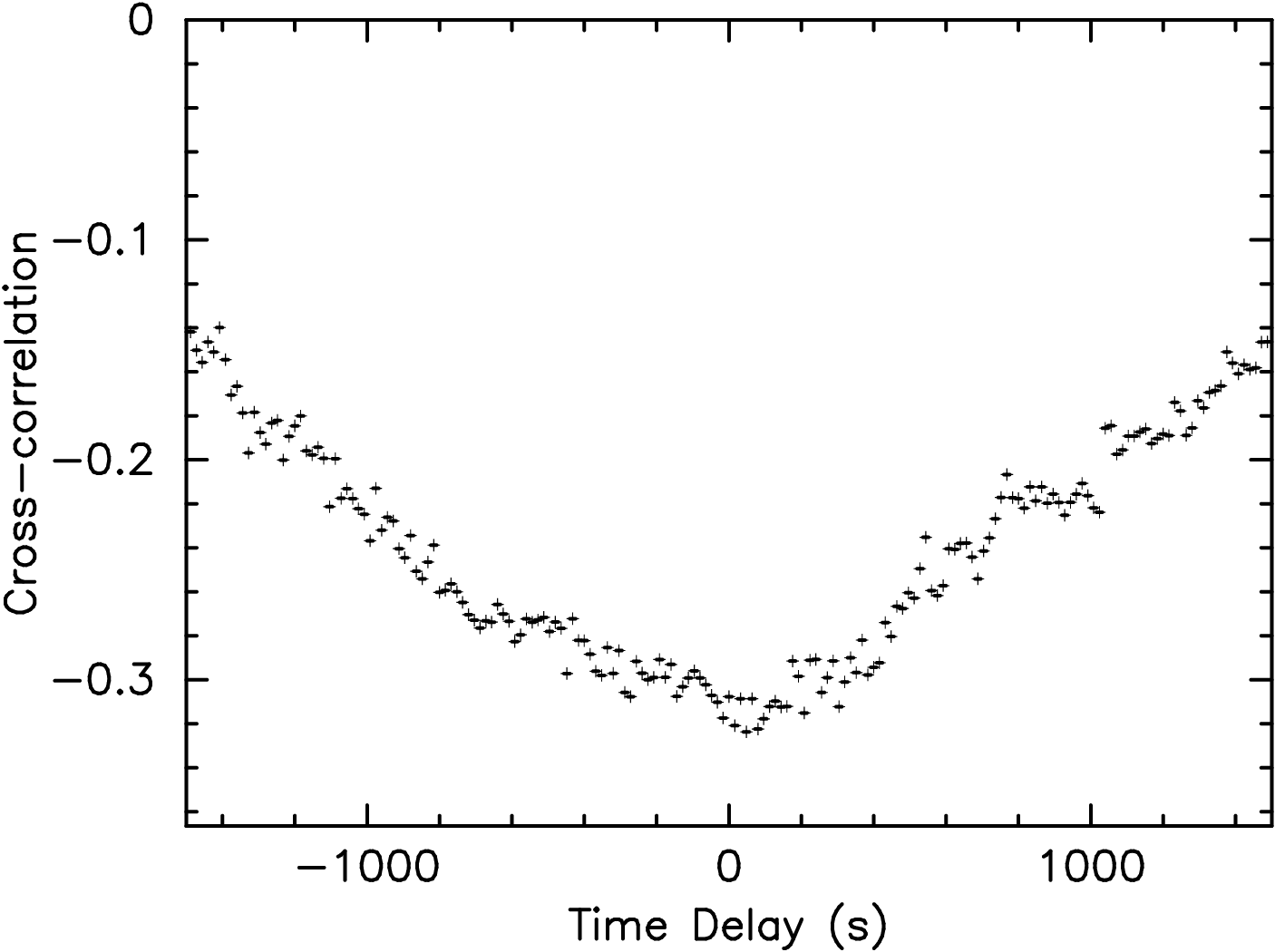}
\includegraphics[width=3.1cm,angle=270,clip]{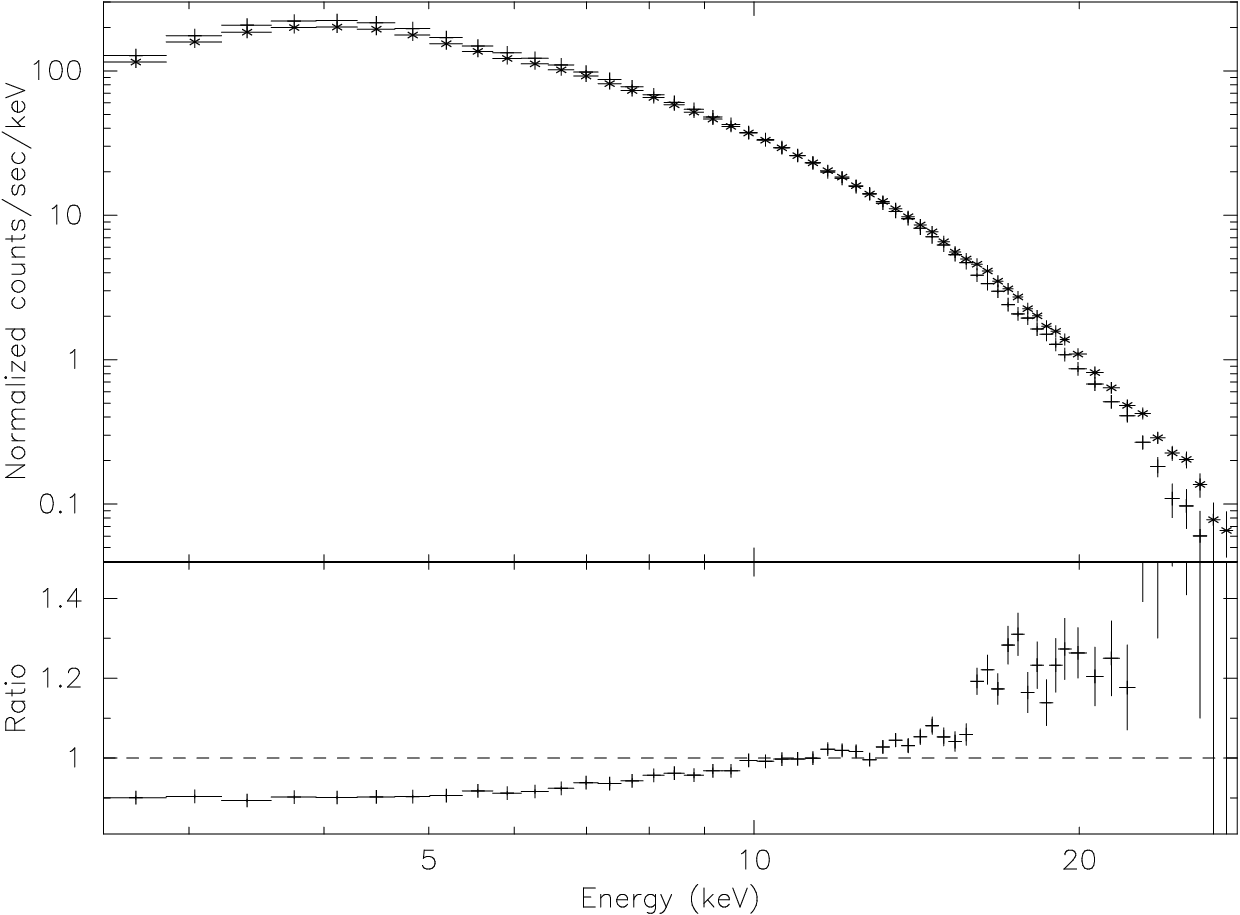}
\includegraphics[width=3.1cm,angle=270,clip]{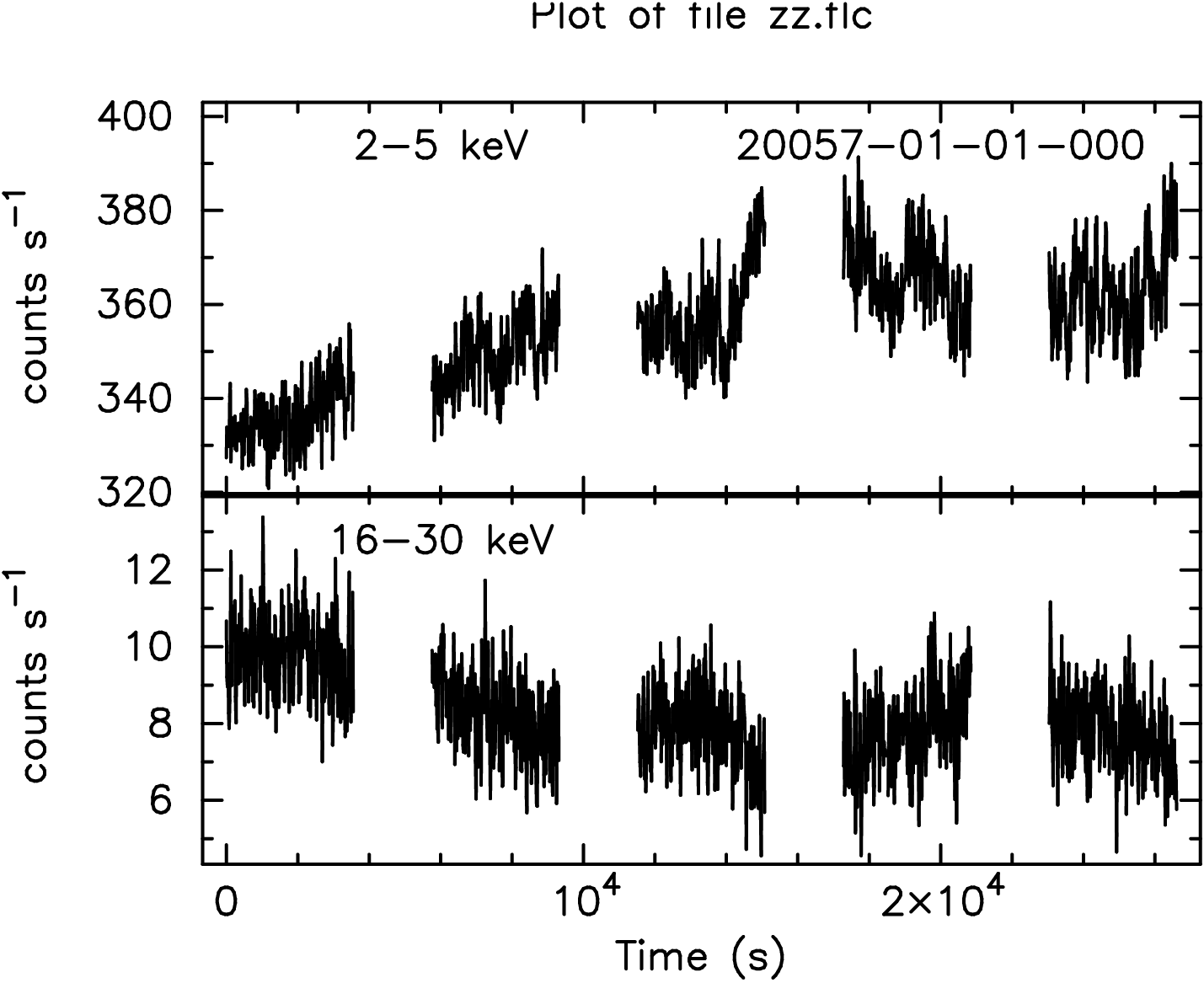}
\includegraphics[width=3.1cm,angle=270,clip]{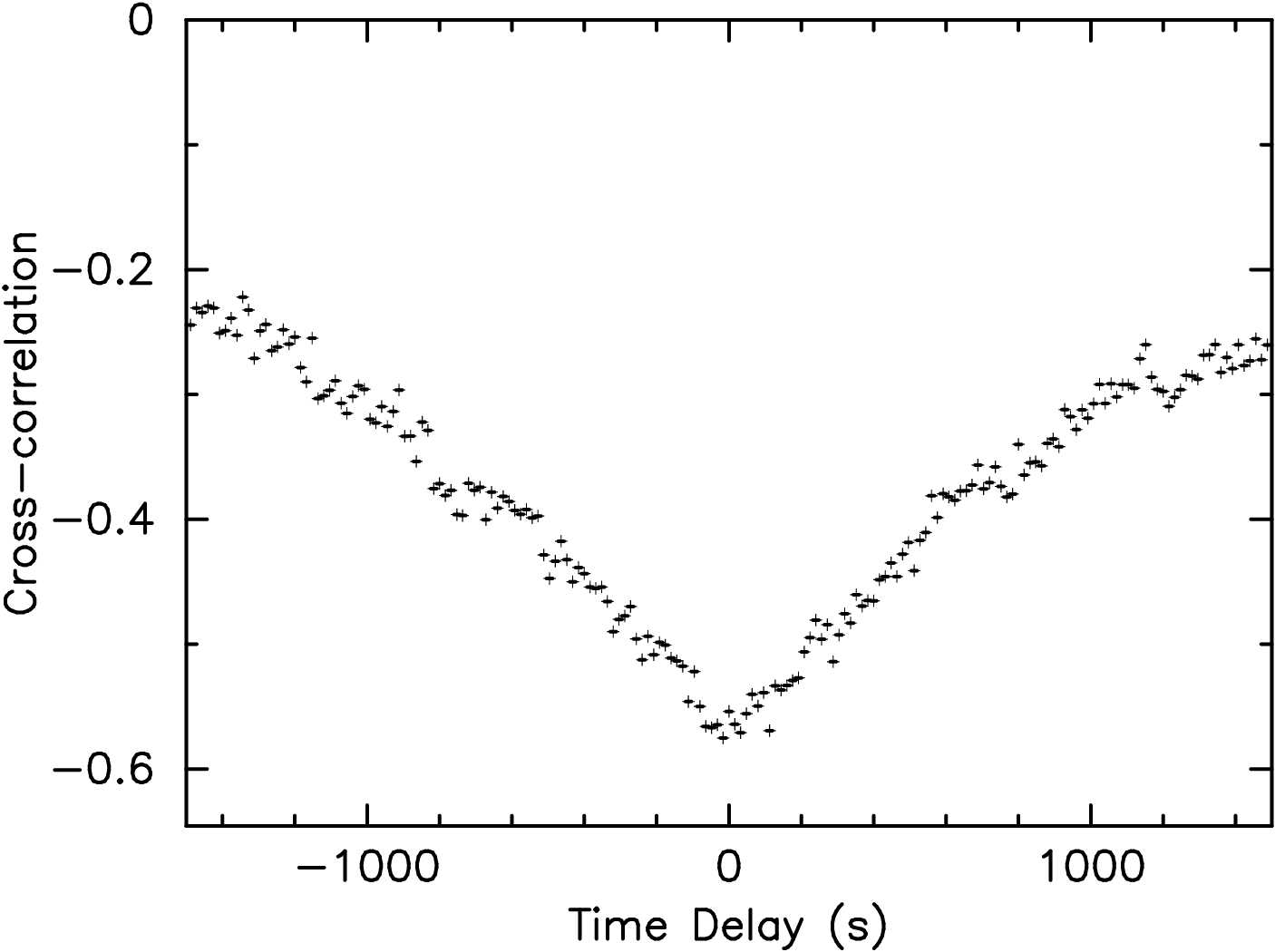}
\includegraphics[width=3.1cm,angle=270,clip]{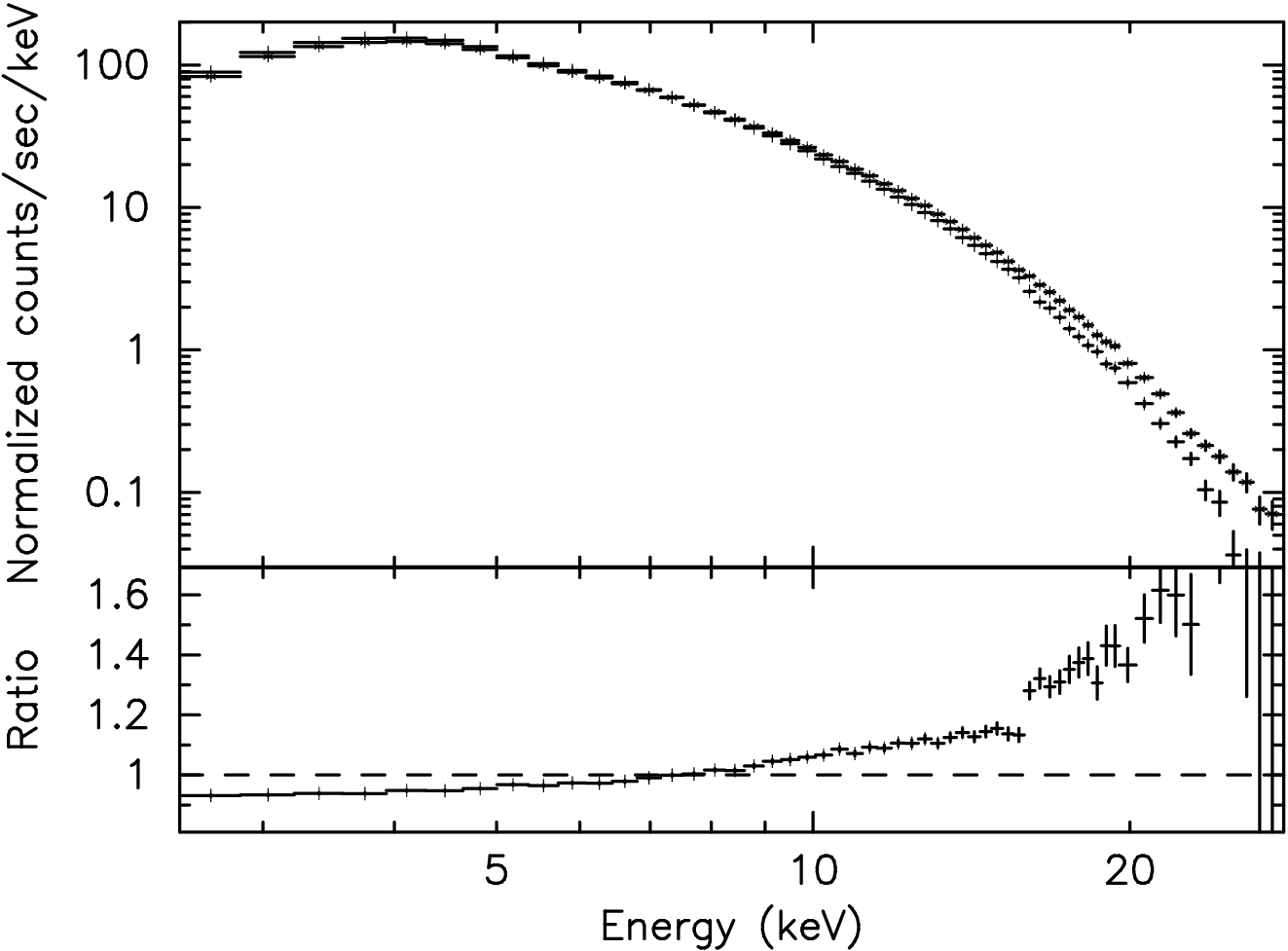}
\includegraphics[width=3.1cm,angle=270,clip]{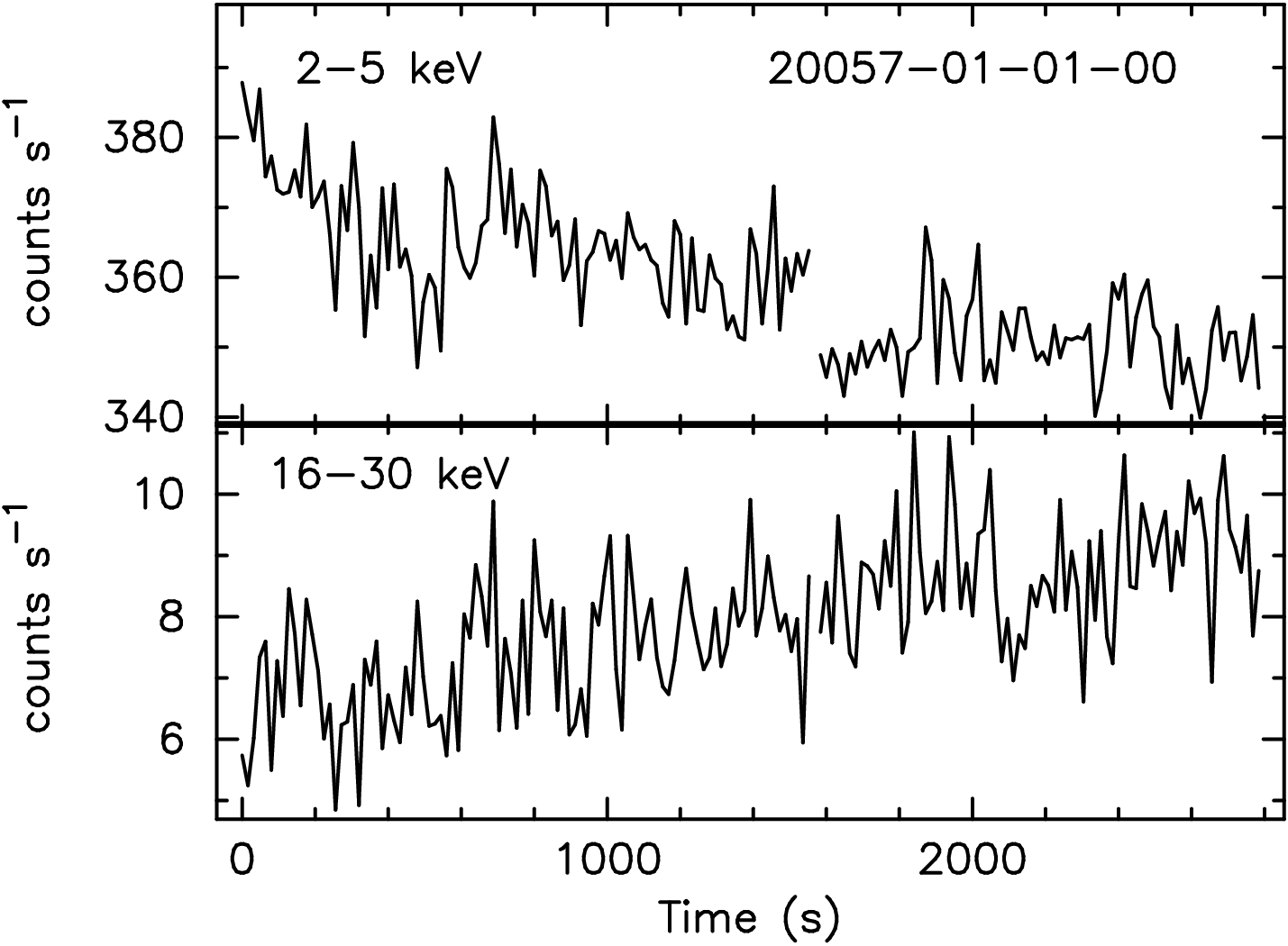}
\includegraphics[width=3.1cm,angle=270,clip]{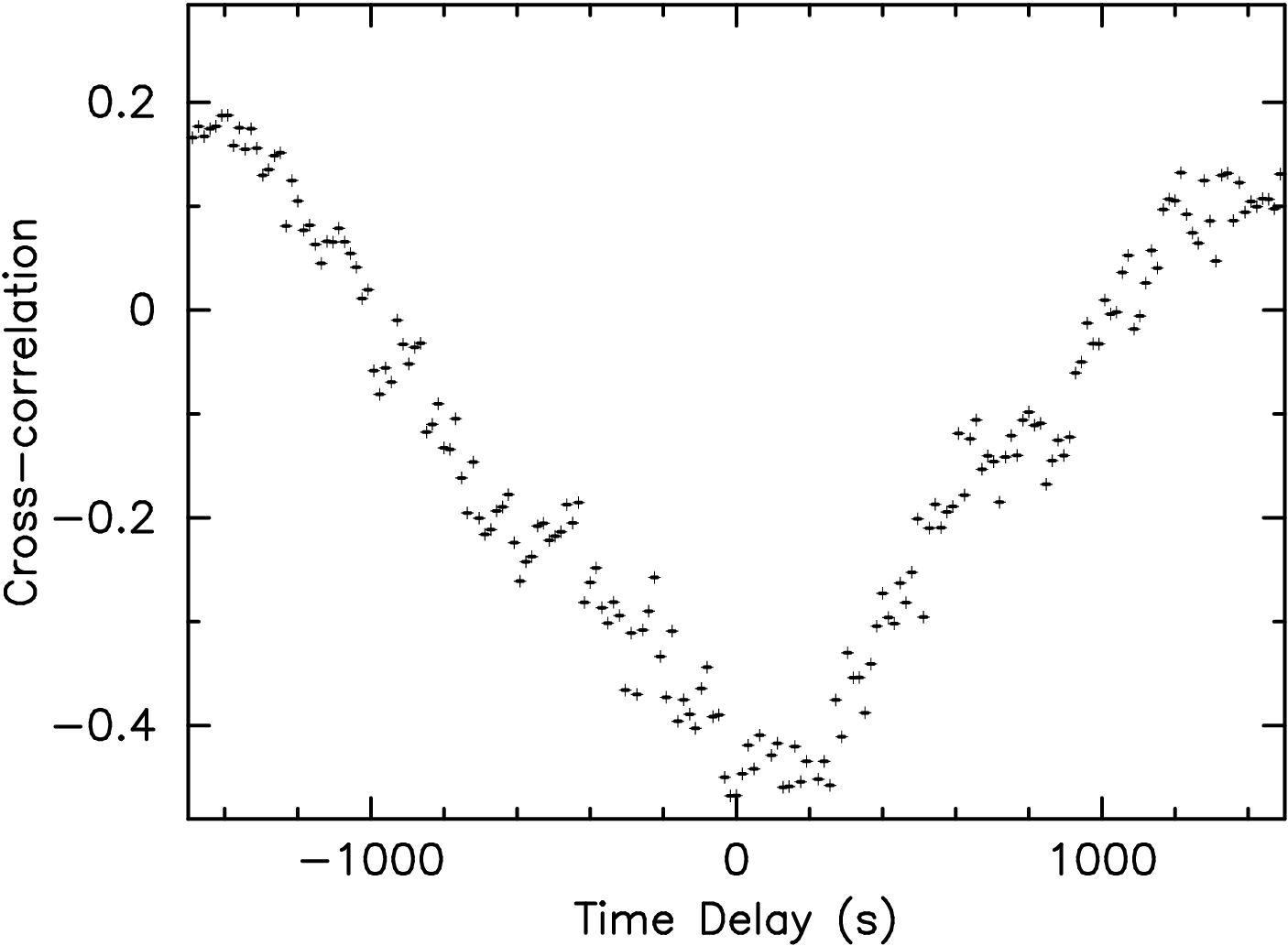}
\includegraphics[width=3.1cm,angle=270,clip]{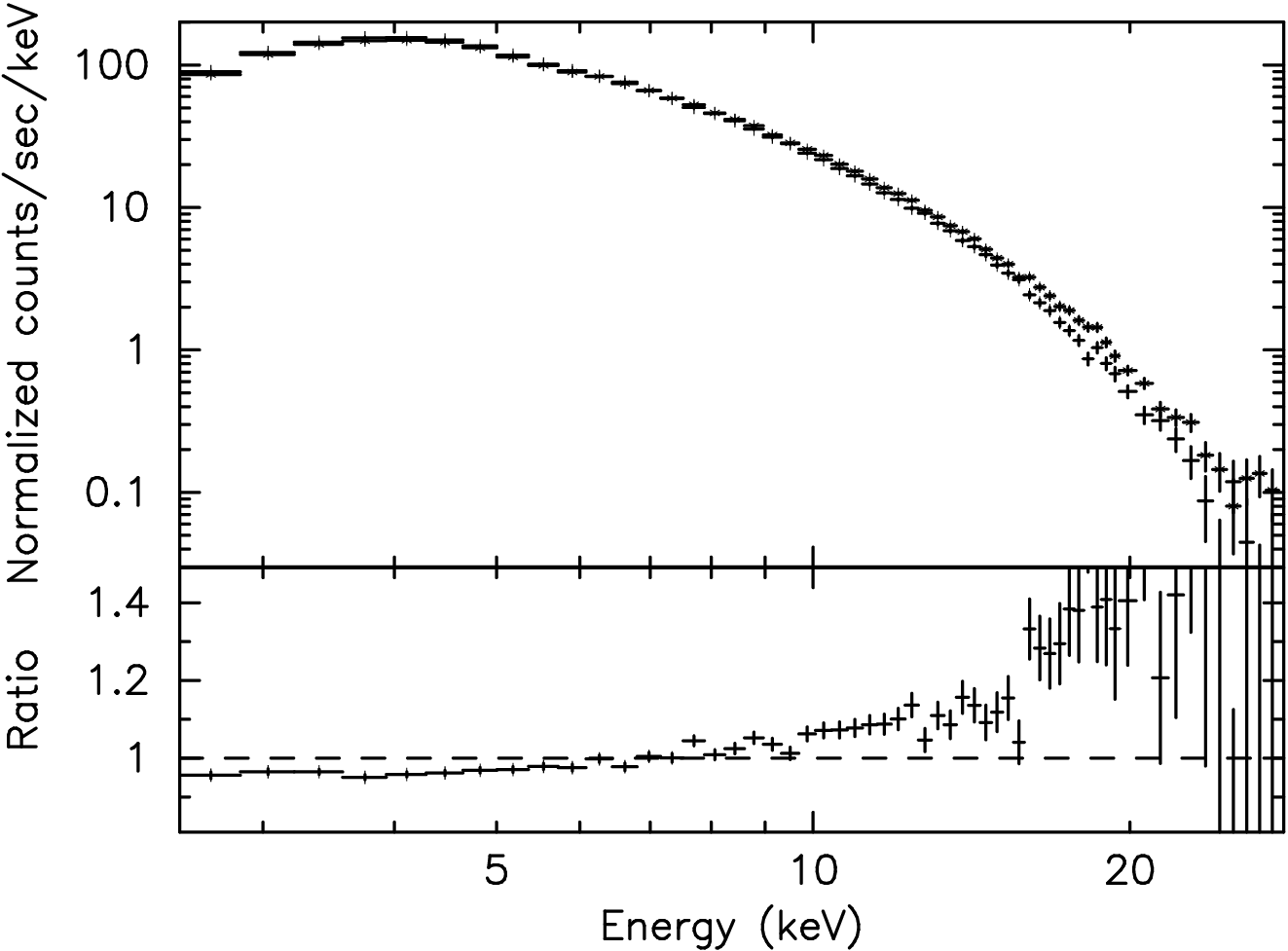}
\includegraphics[width=3.1cm,angle=270,clip]{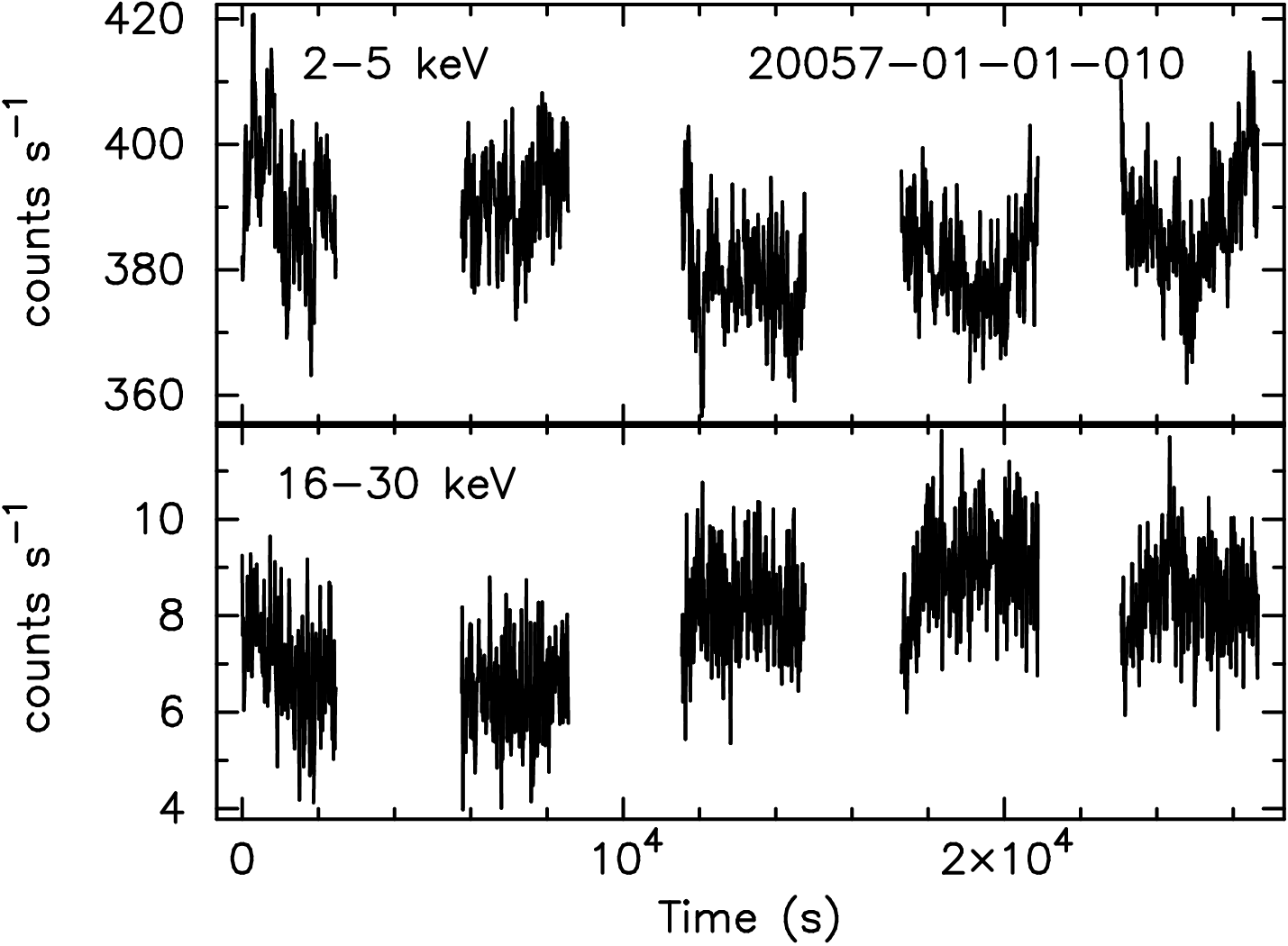}
\includegraphics[width=3.1cm,angle=270,clip]{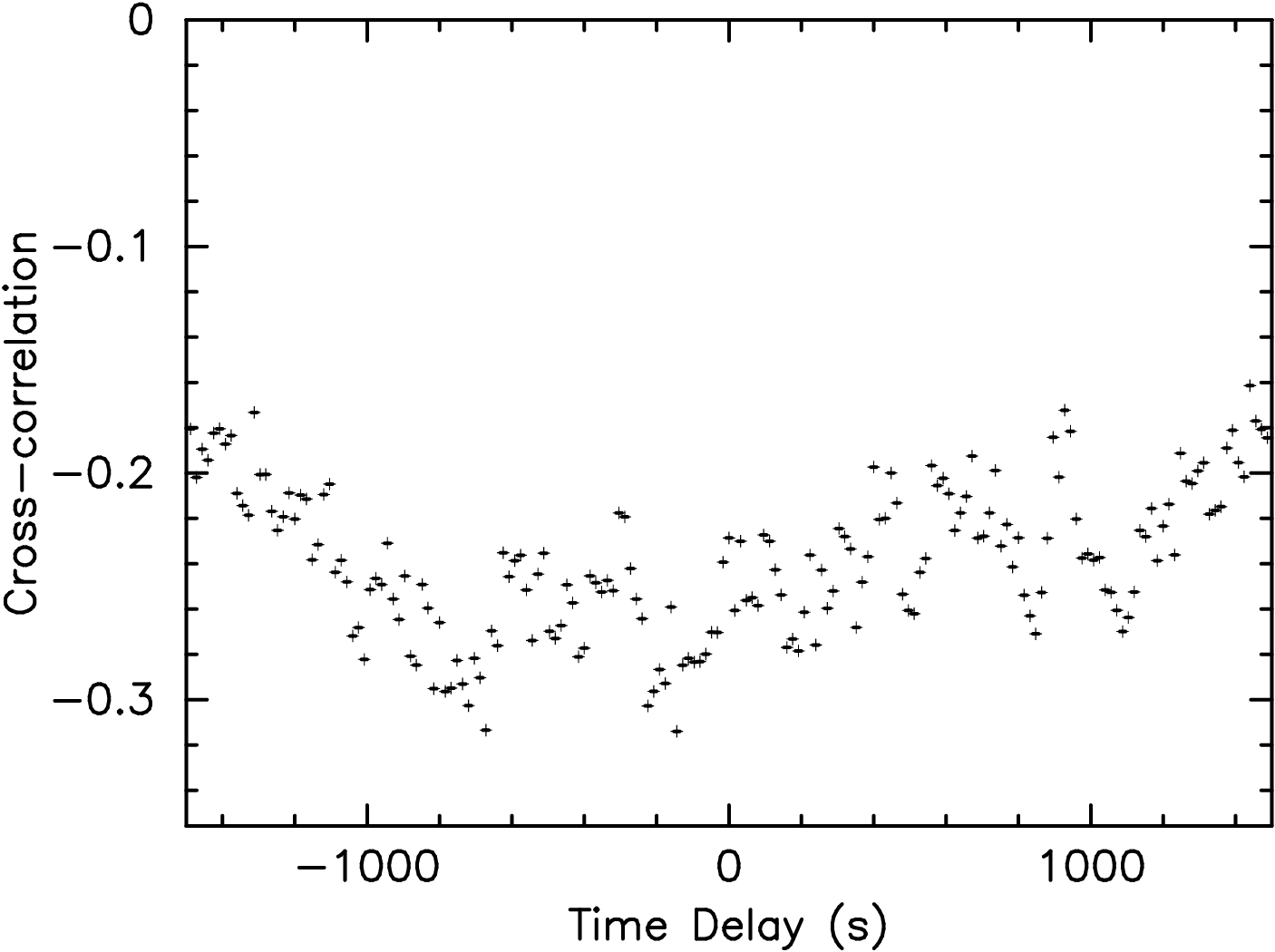}
\includegraphics[width=3.1cm,angle=270,clip]{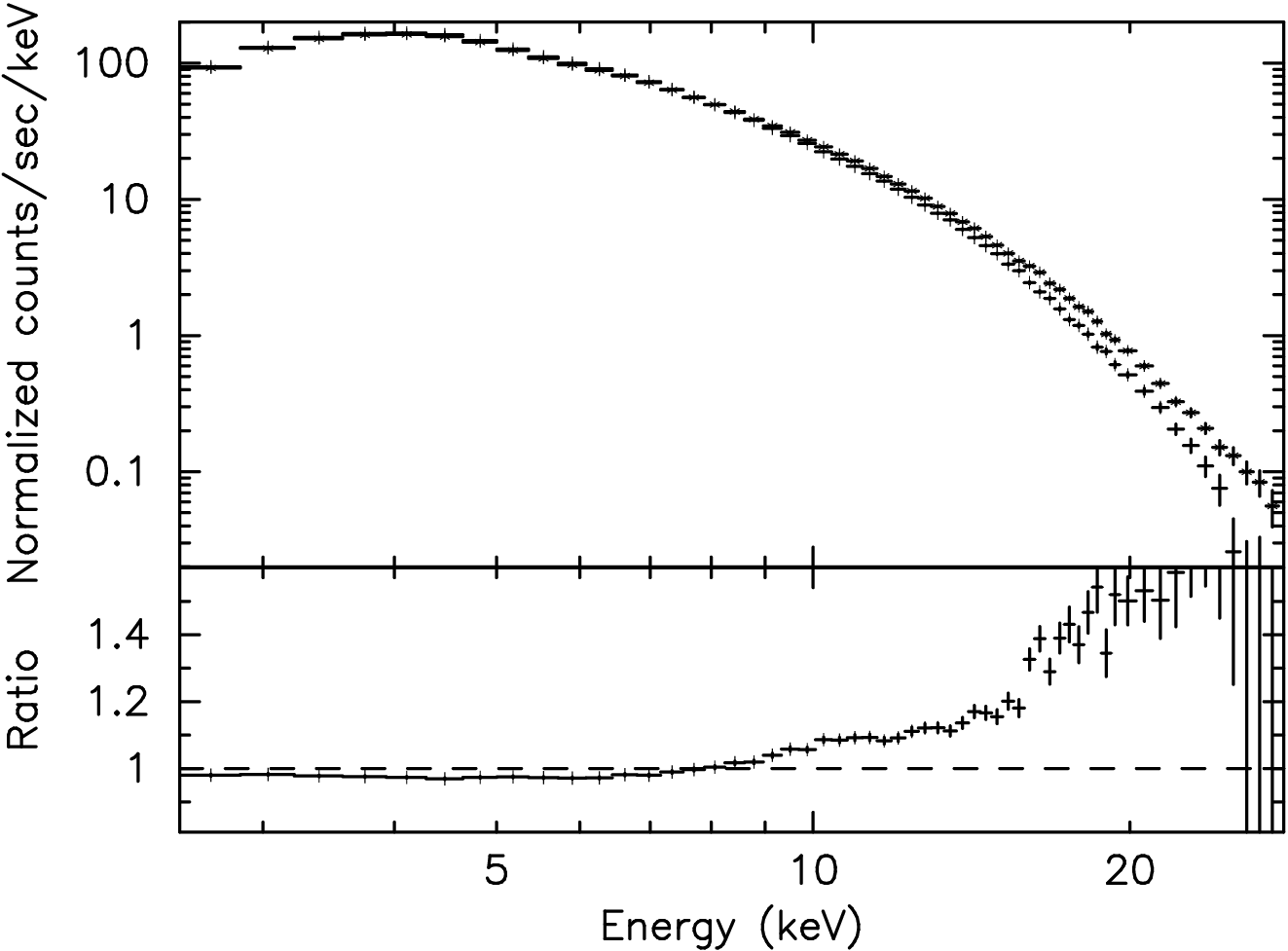}
\includegraphics[width=3.1cm,angle=270,clip]{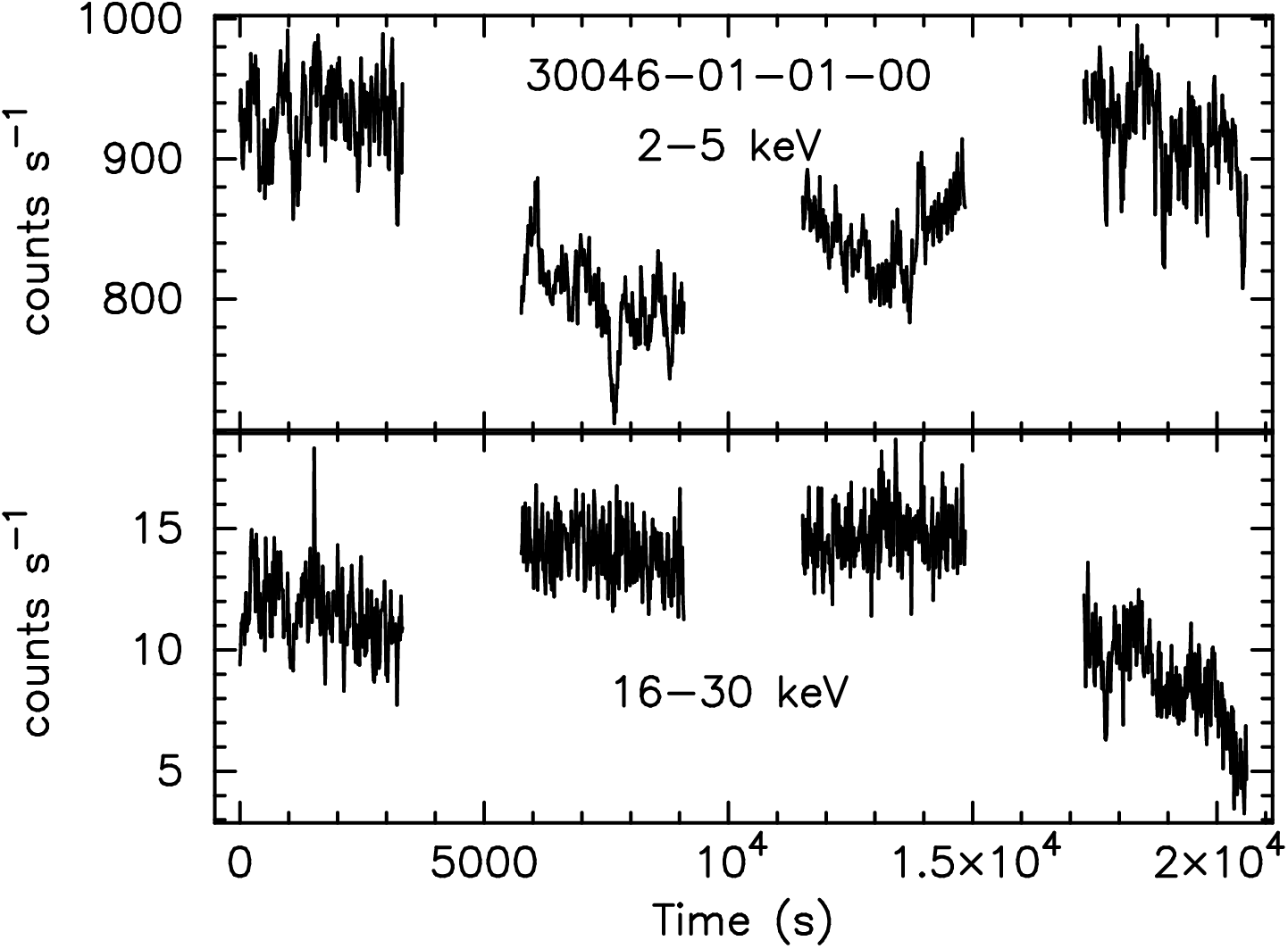}
\includegraphics[width=3.1cm,angle=270,clip]{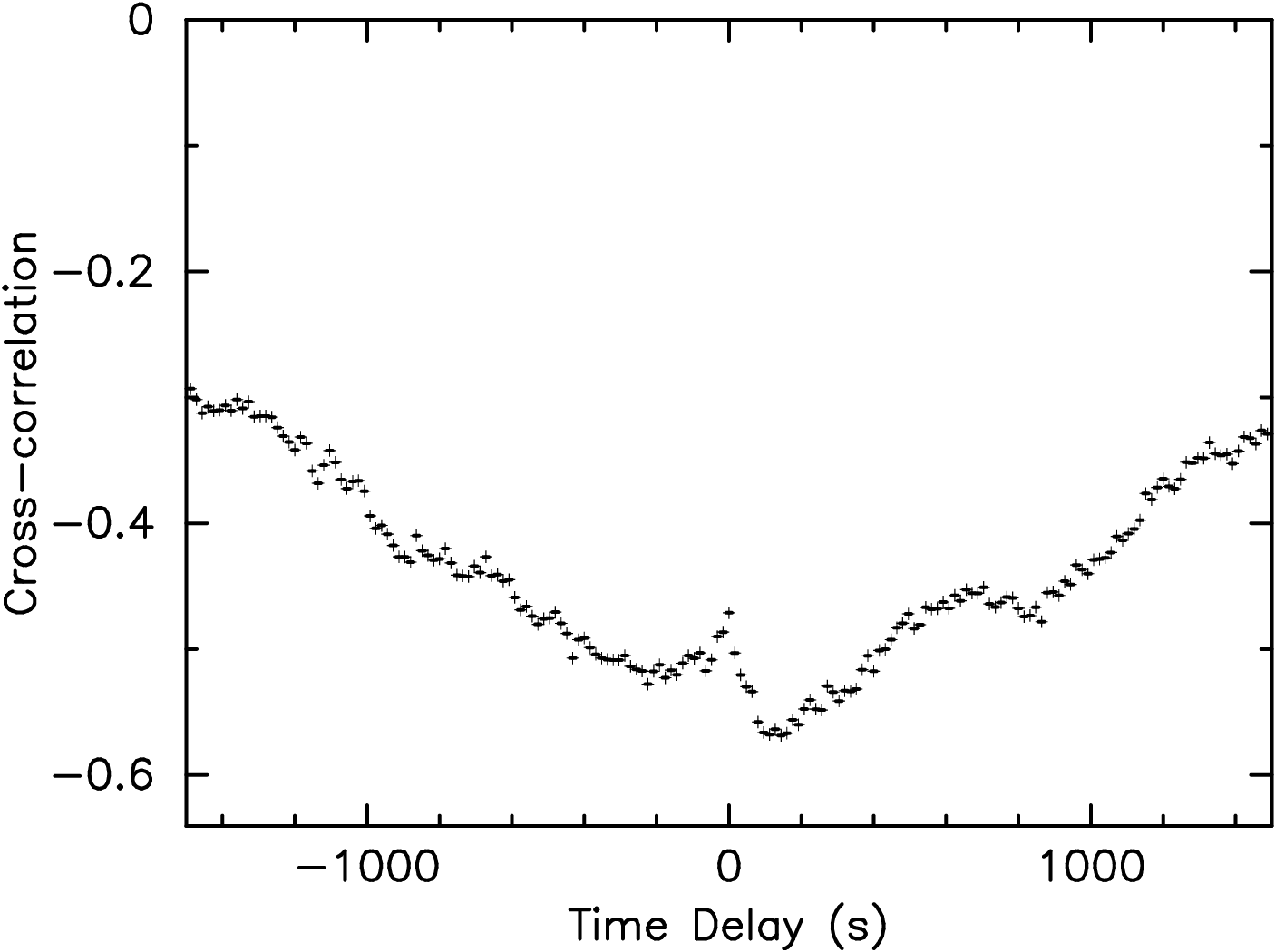}
\includegraphics[width=3.1cm,angle=270,clip]{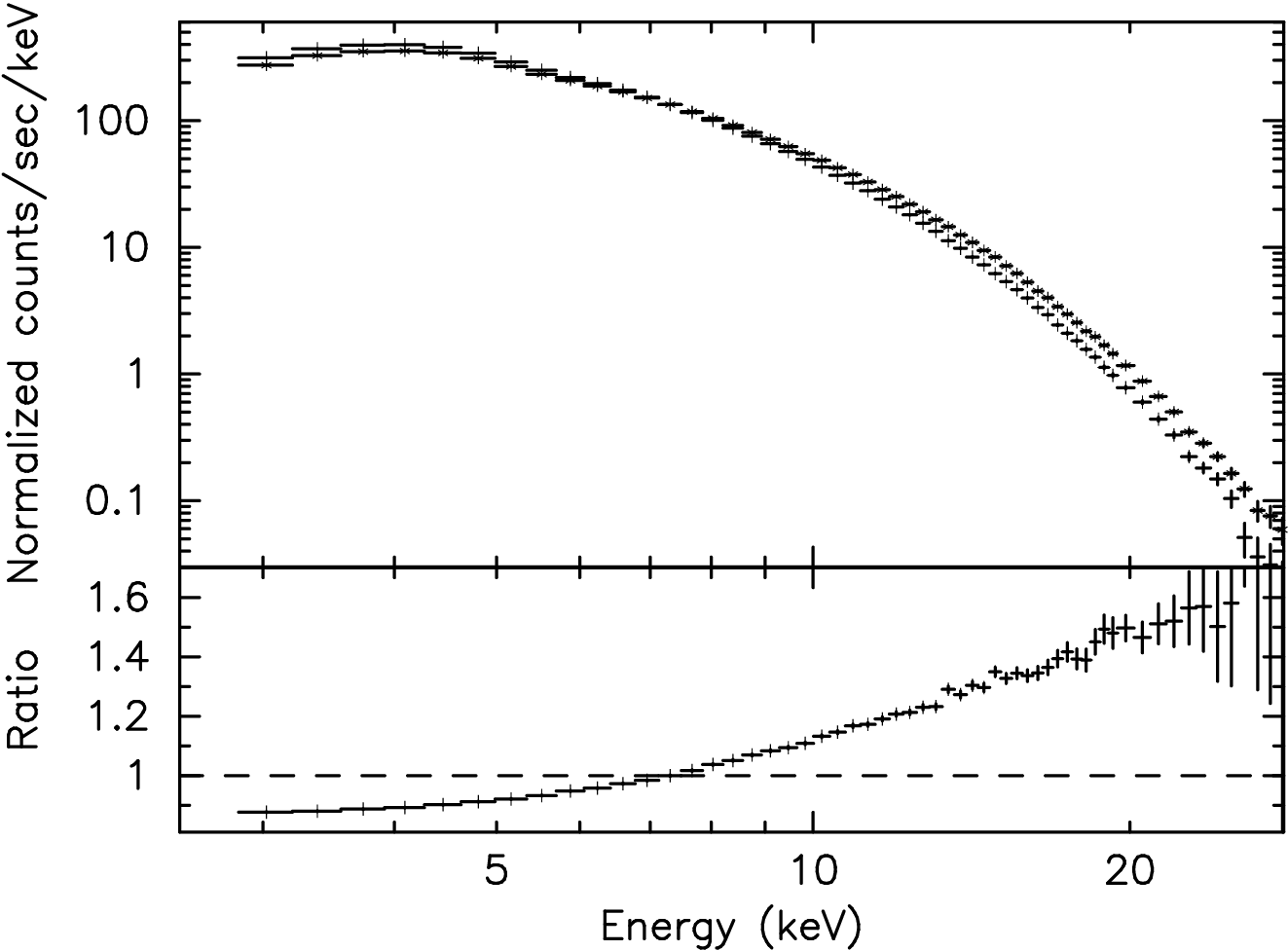}
\caption{The lightcurves, cross-correlations between the
soft (2-5 keV) and hard (16-30 keV) X-rays, the X-ray spectra of the ObsID
for hard and soft regions of the lightcurve are shown.
} \label{fig4}
\end{center}
\end{figure*}
\clearpage
\pagestyle{empty}
\begin{center}
\vspace*{-27mm}
\includegraphics[width=3.1cm,angle=270,clip]{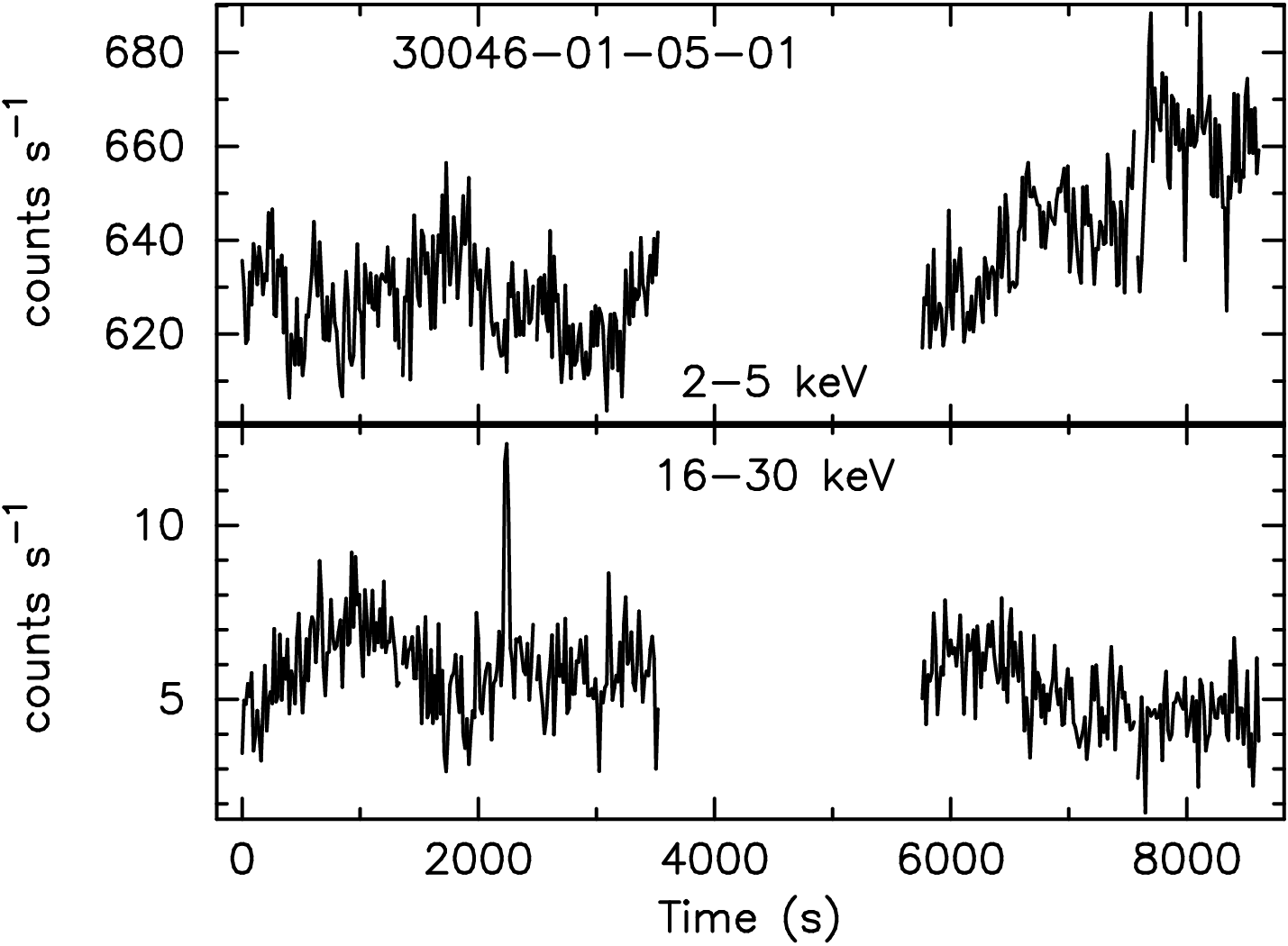}
\includegraphics[width=3.1cm,angle=270,clip]{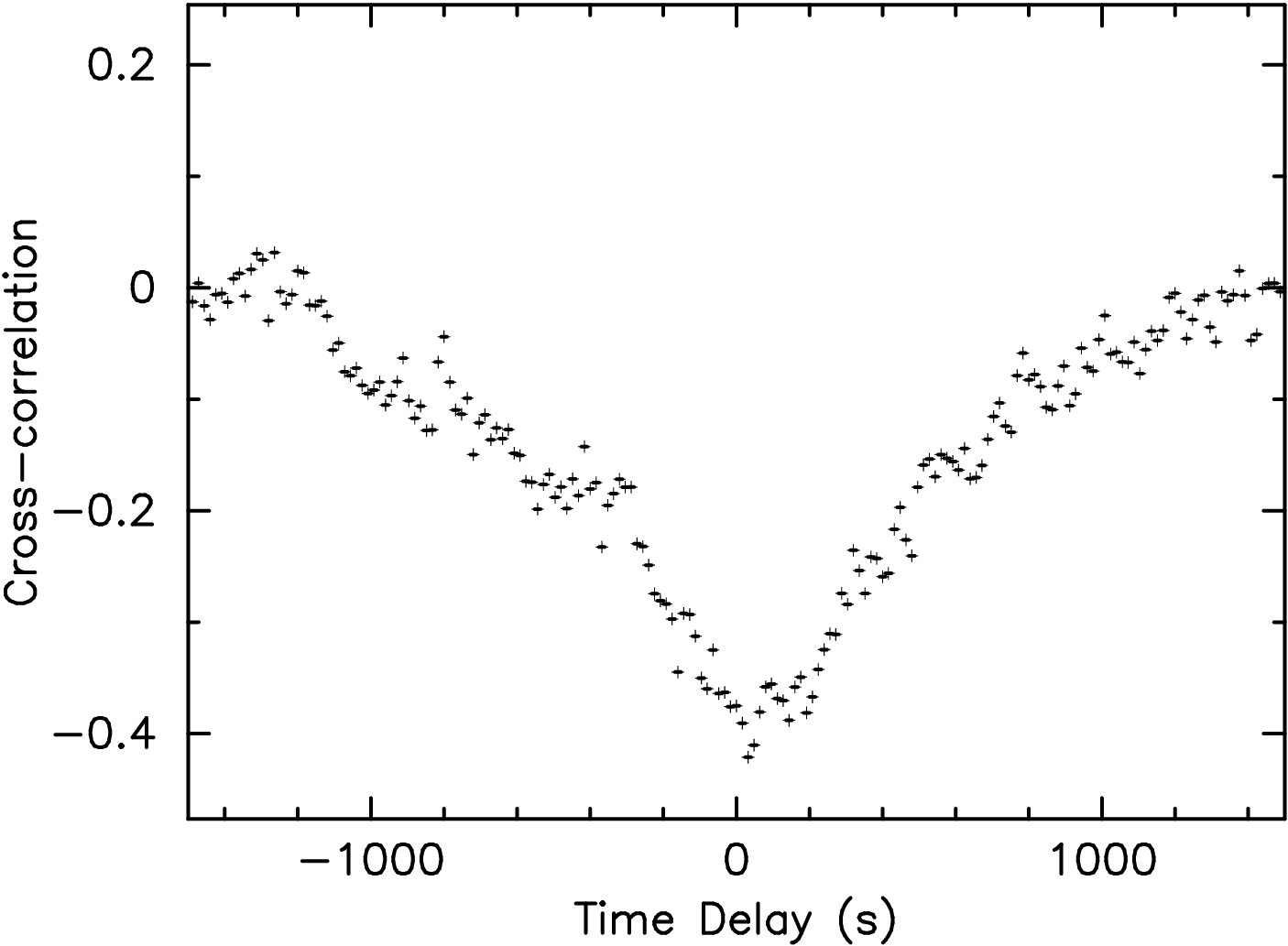}
\includegraphics[width=3.1cm,angle=270,clip]{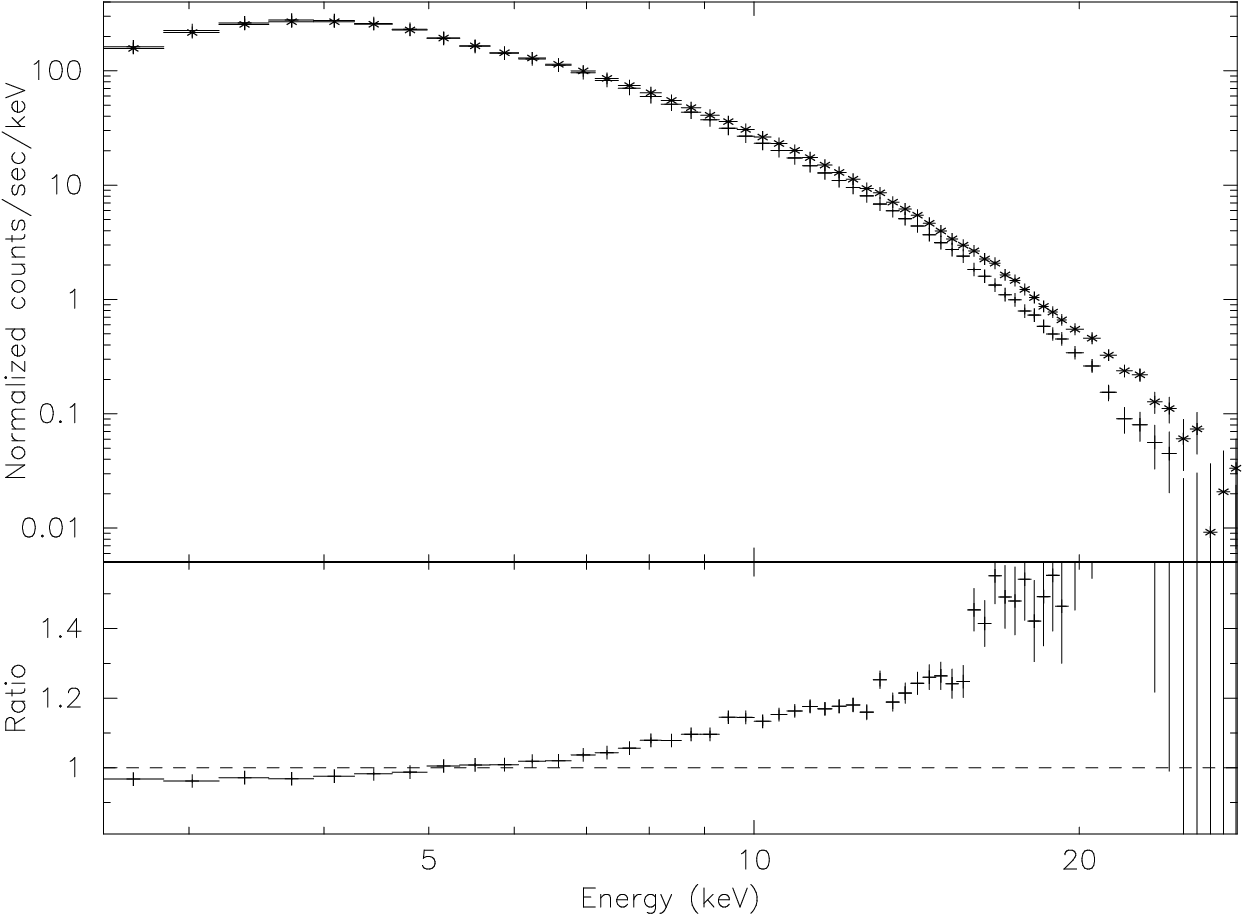}\\
\includegraphics[width=3.1cm,angle=270,clip]{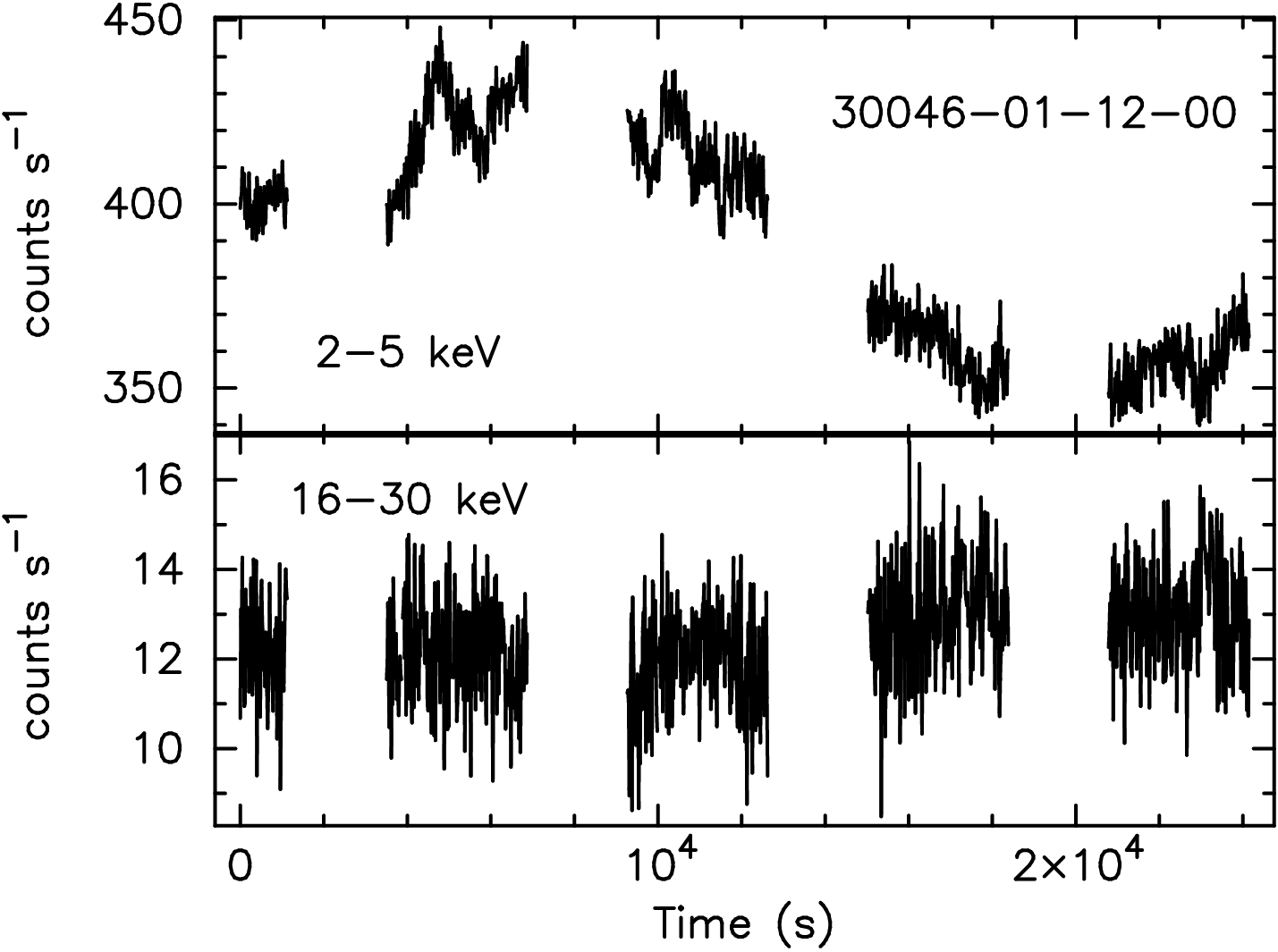}
\includegraphics[width=3.1cm,angle=270,clip]{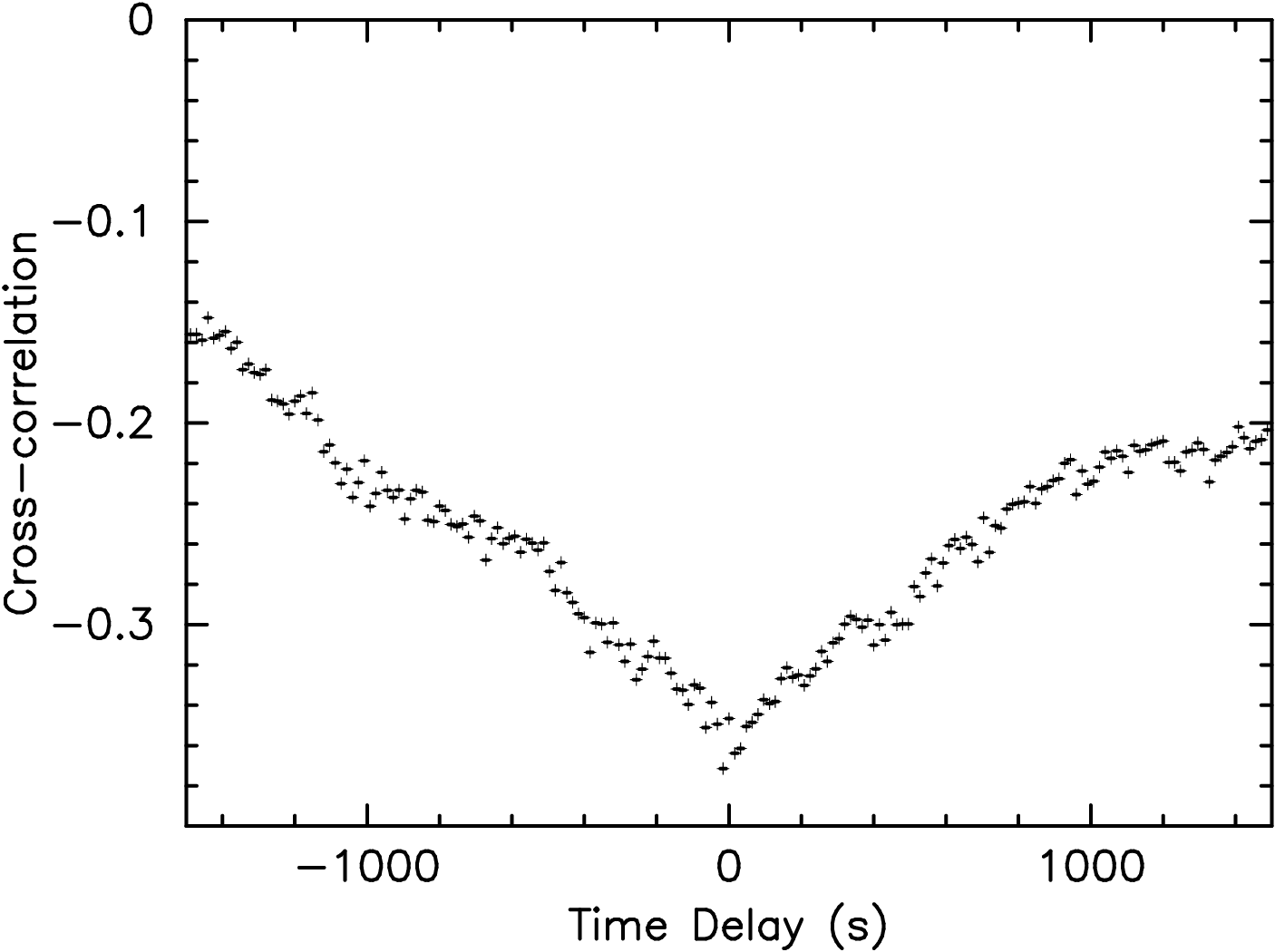}
\includegraphics[width=3.1cm,angle=270,clip]{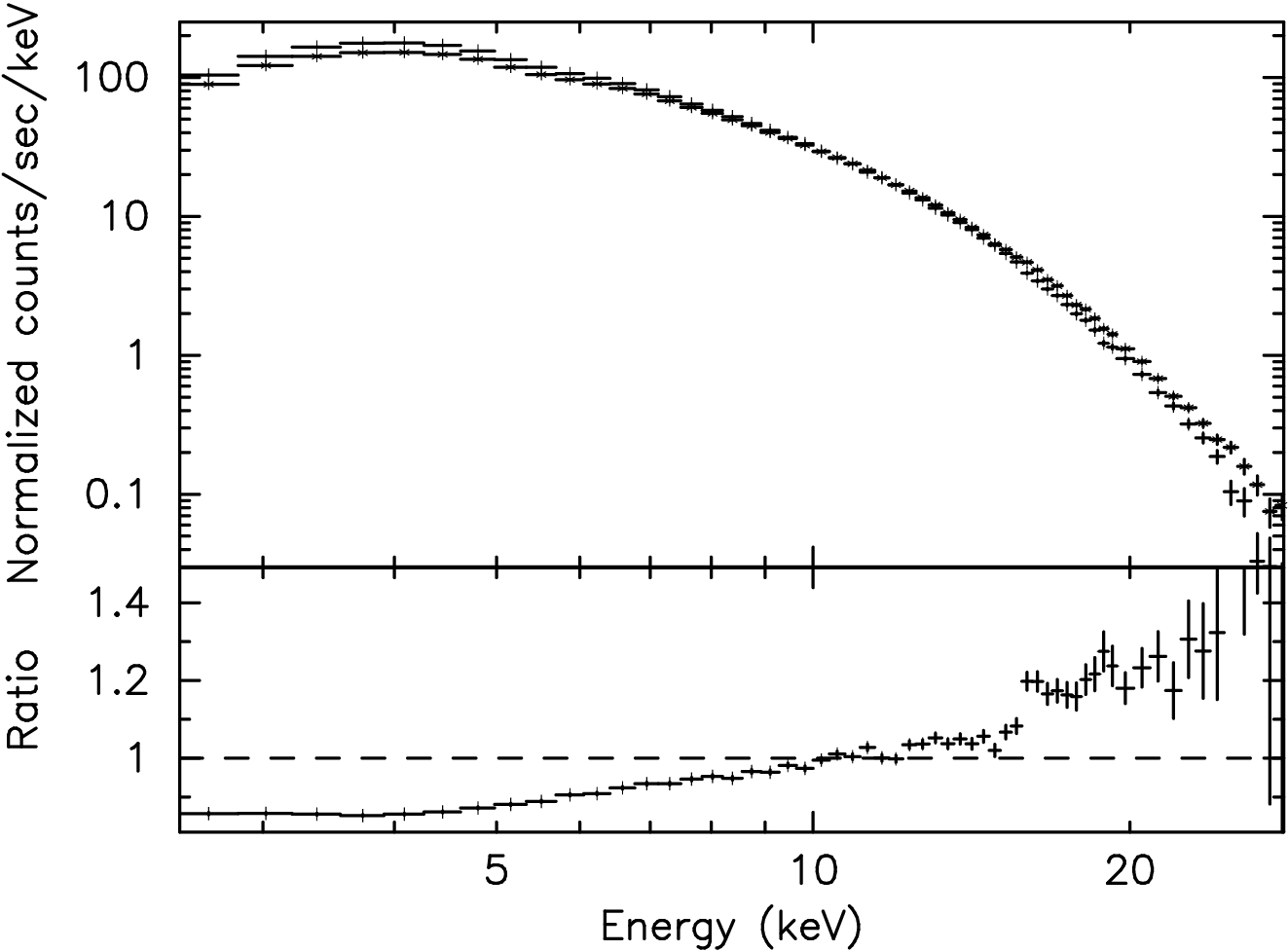}\\
\includegraphics[width=3.1cm,angle=270,clip]{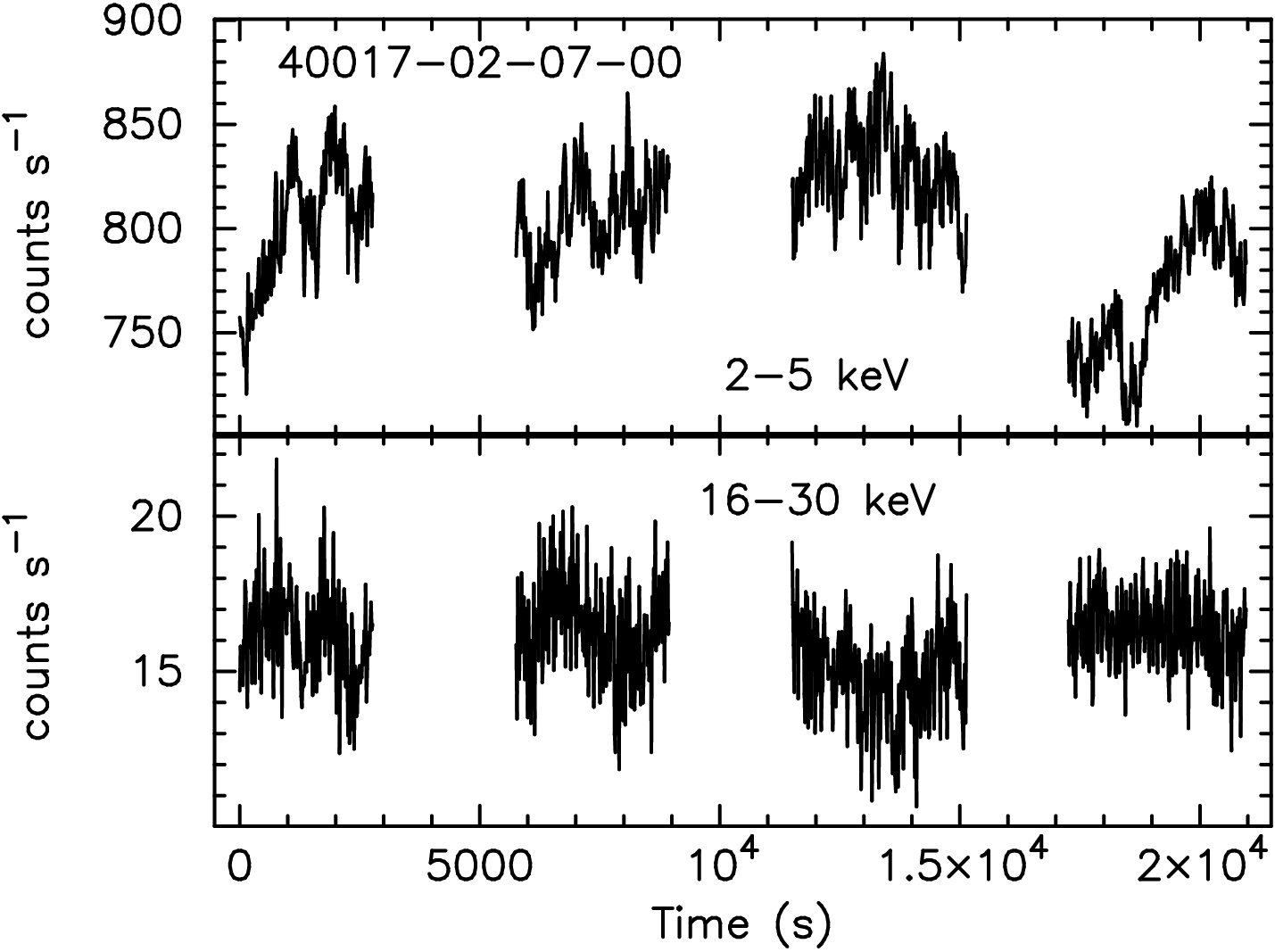}
\includegraphics[width=3.1cm,angle=270,clip]{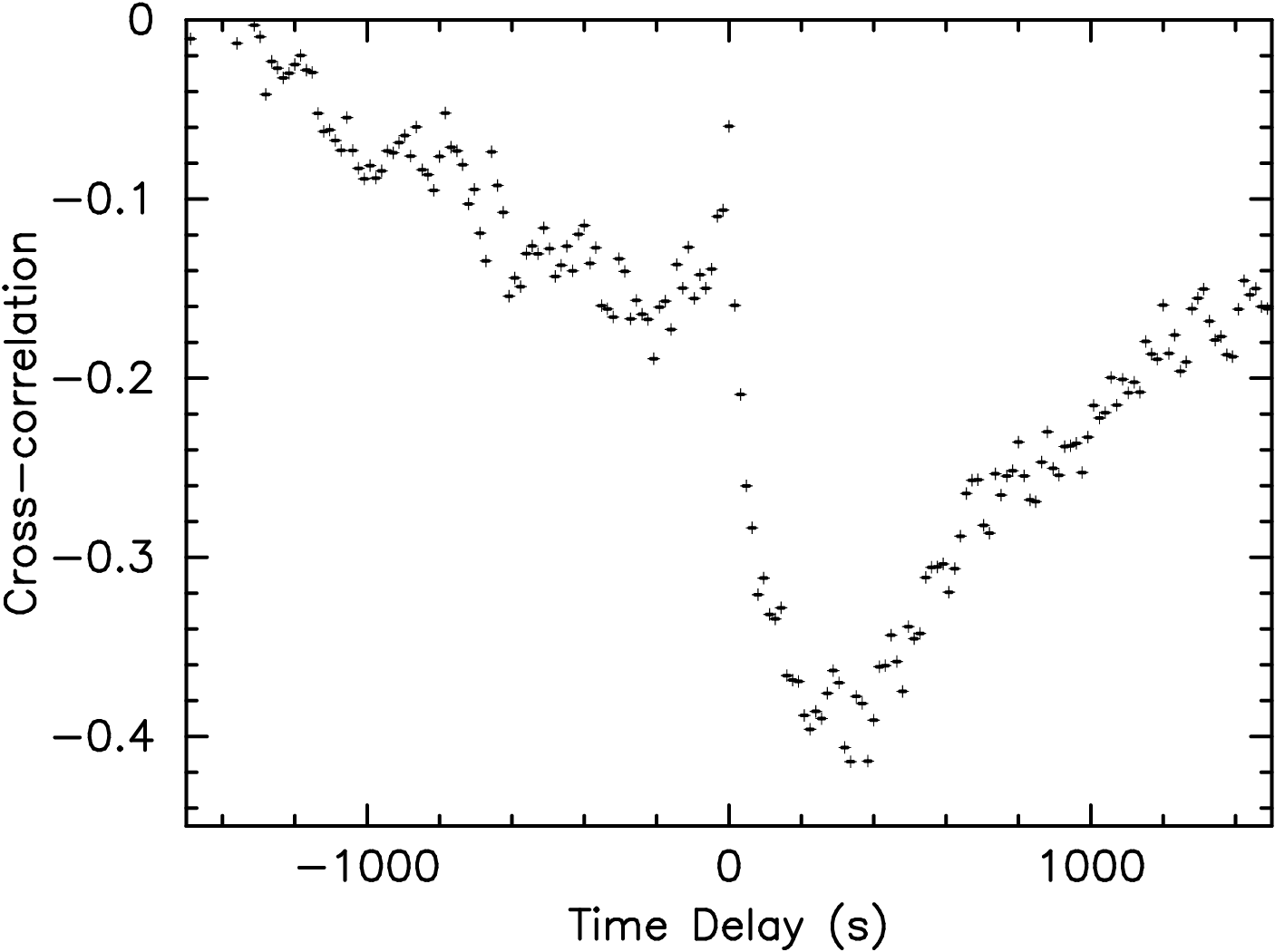}
\includegraphics[width=3.1cm,angle=270,clip]{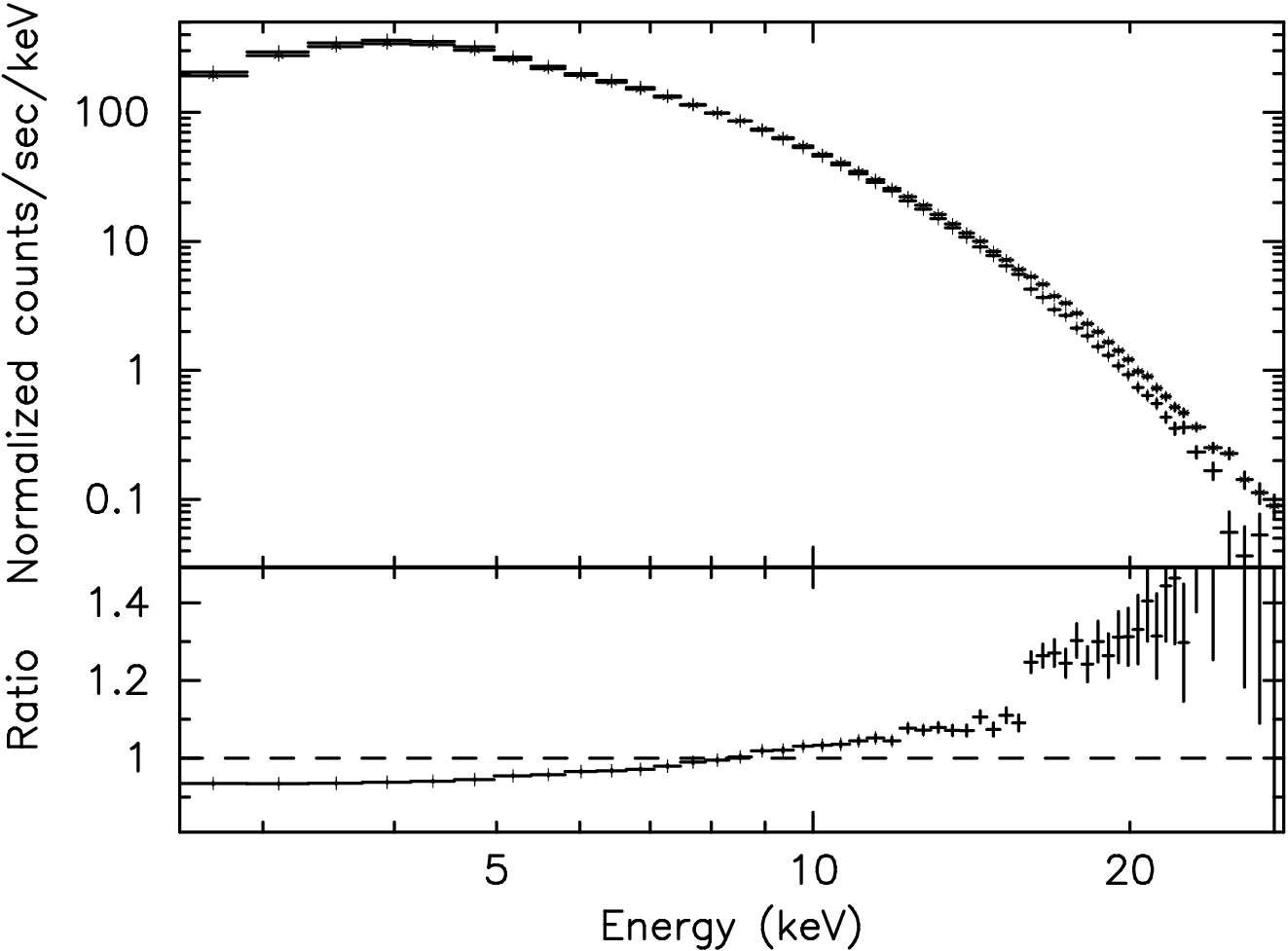}\\
\includegraphics[width=3.1cm,angle=270,clip]{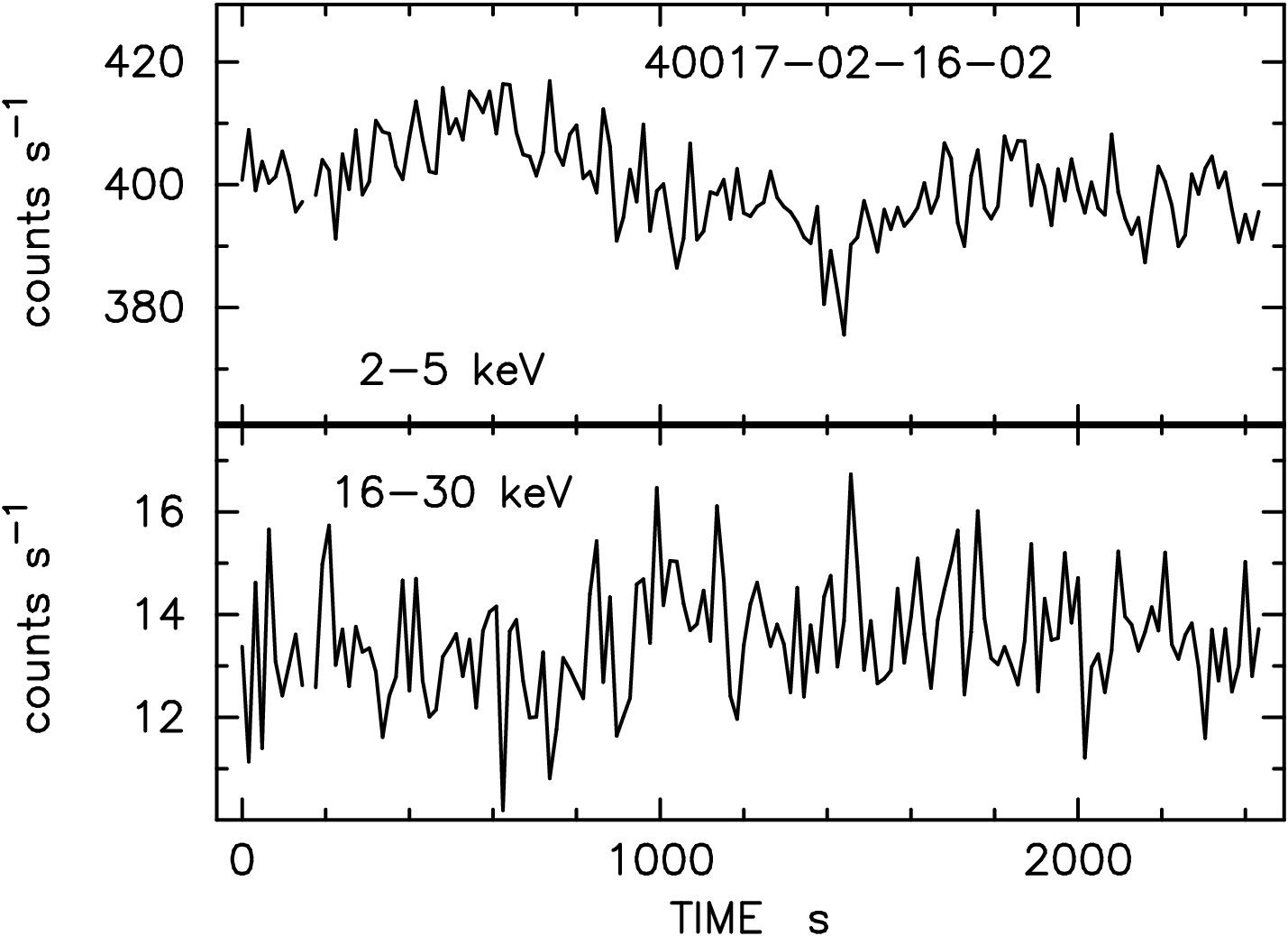}
\includegraphics[width=3.1cm,angle=270,clip]{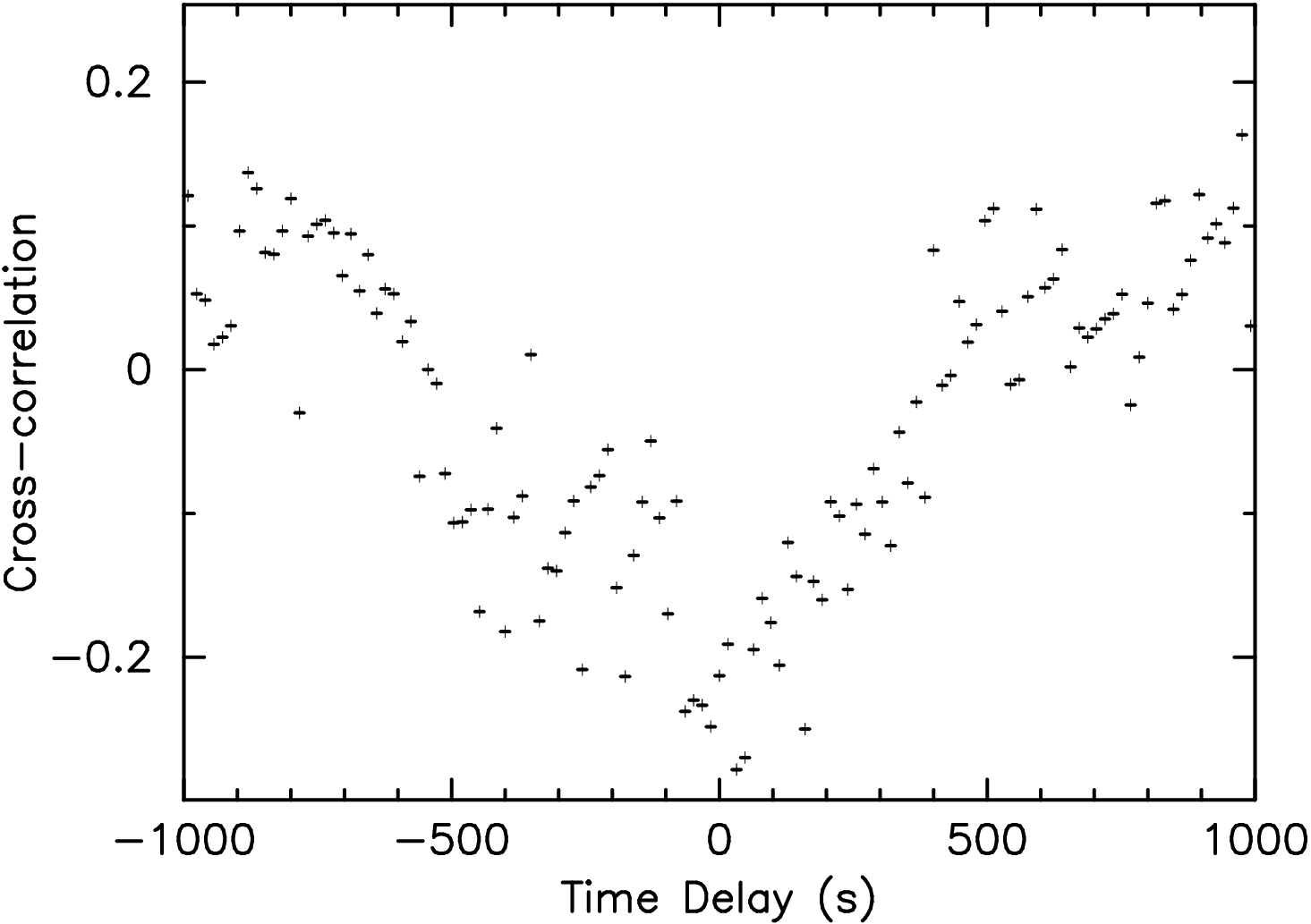}
\includegraphics[width=3.1cm,angle=270,clip]{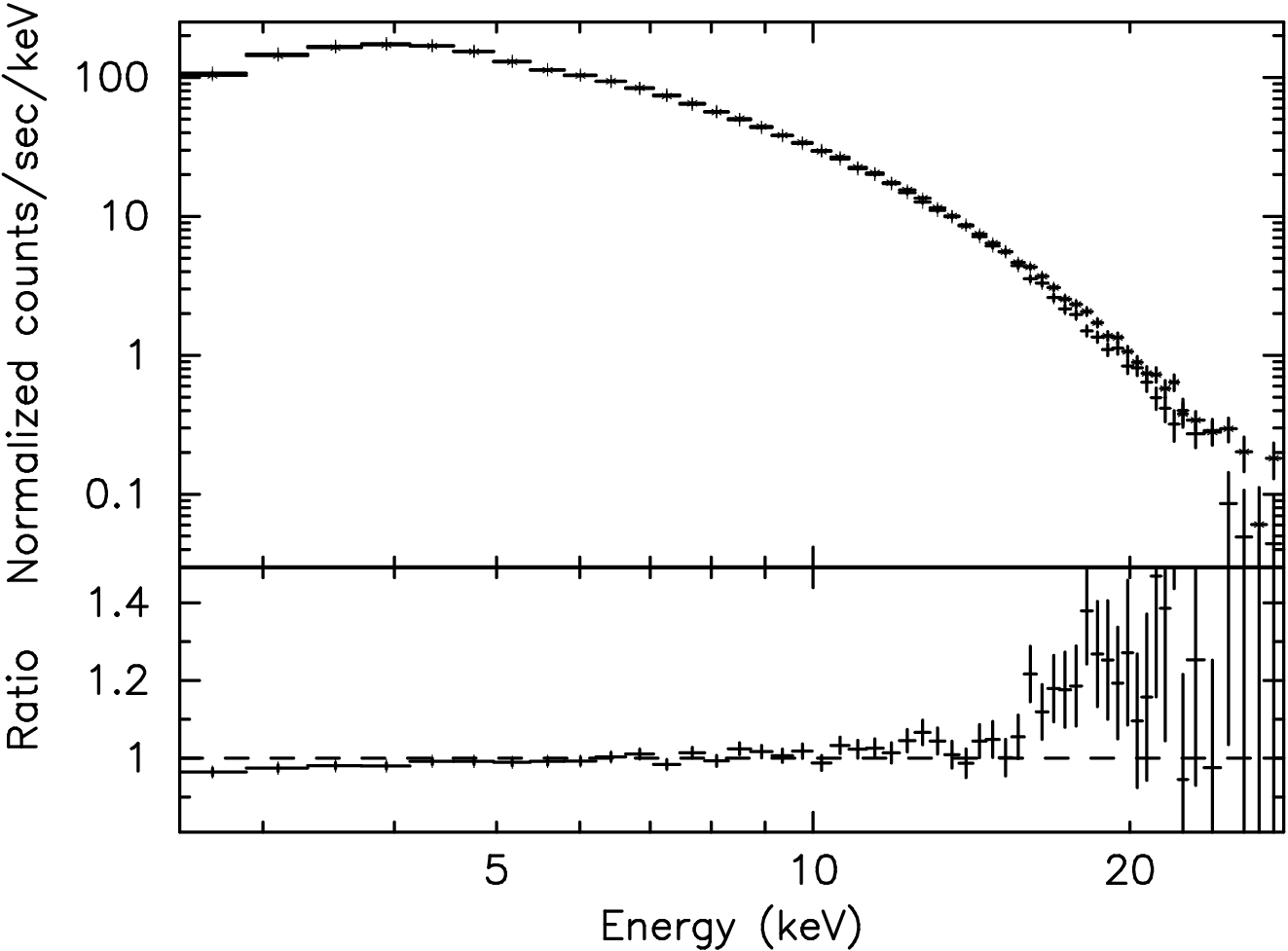}\\
\includegraphics[width=3.1cm,angle=270,clip]{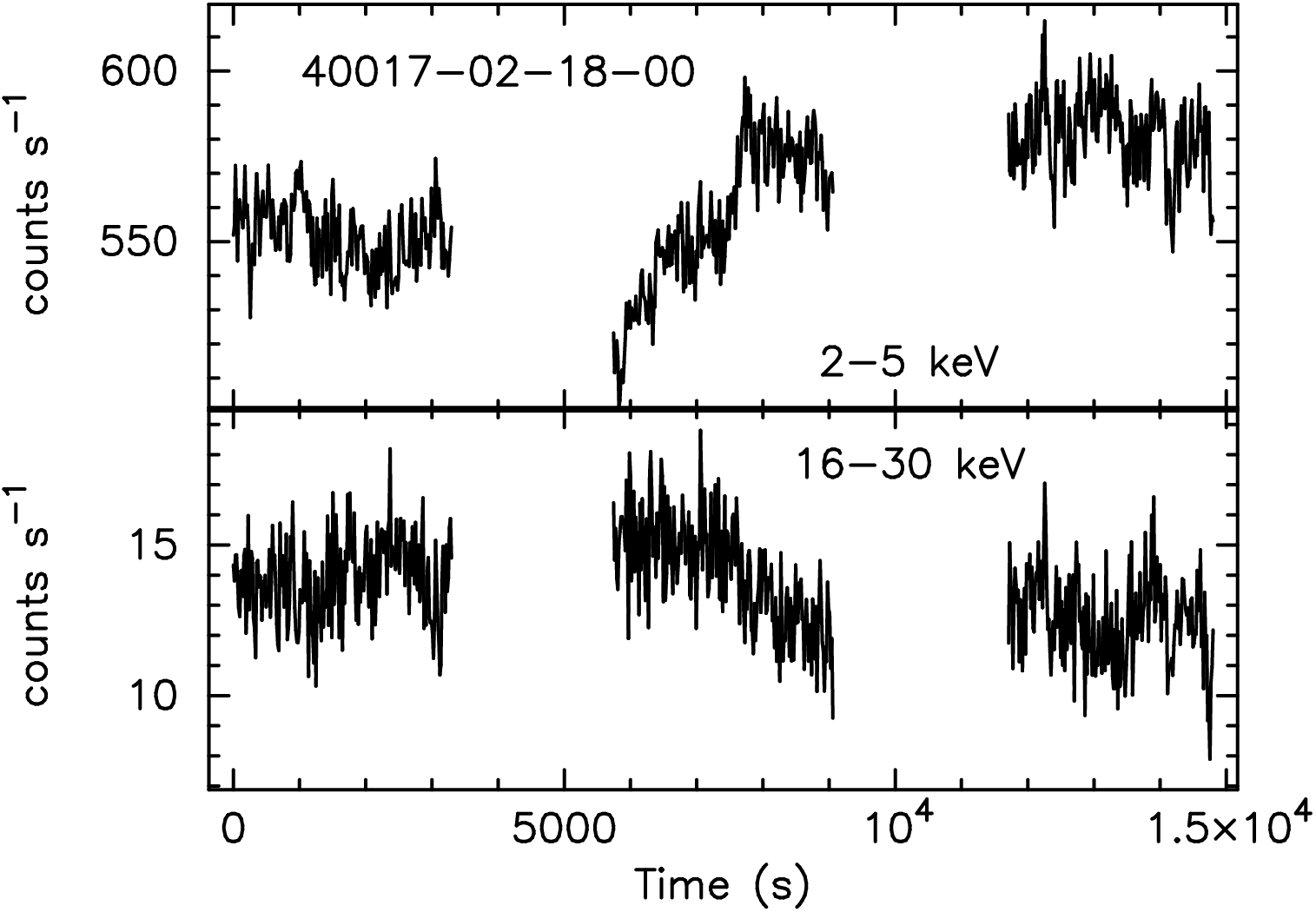}
\includegraphics[width=3.1cm,angle=270,clip]{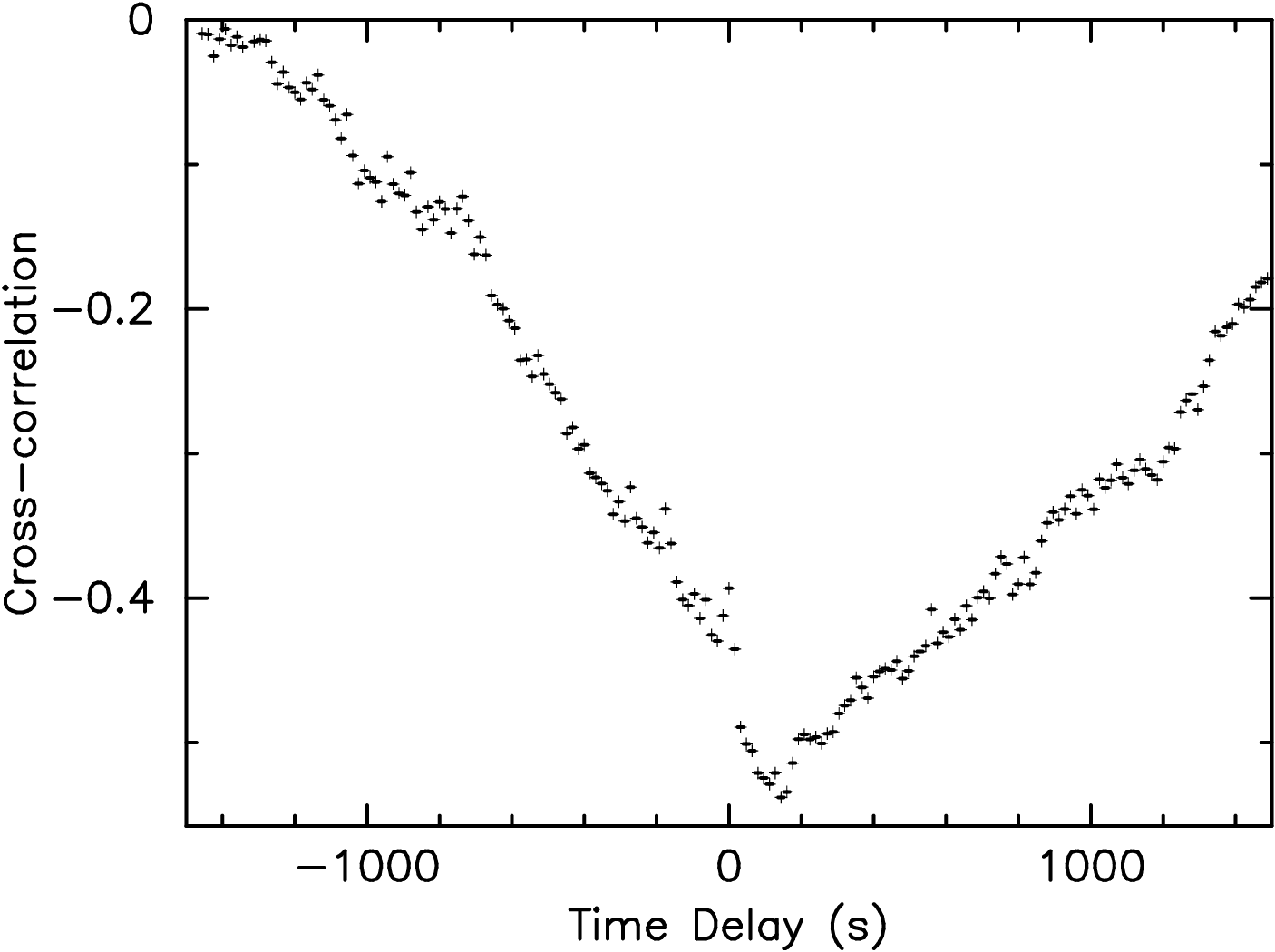}
\includegraphics[width=3.1cm,angle=270,clip]{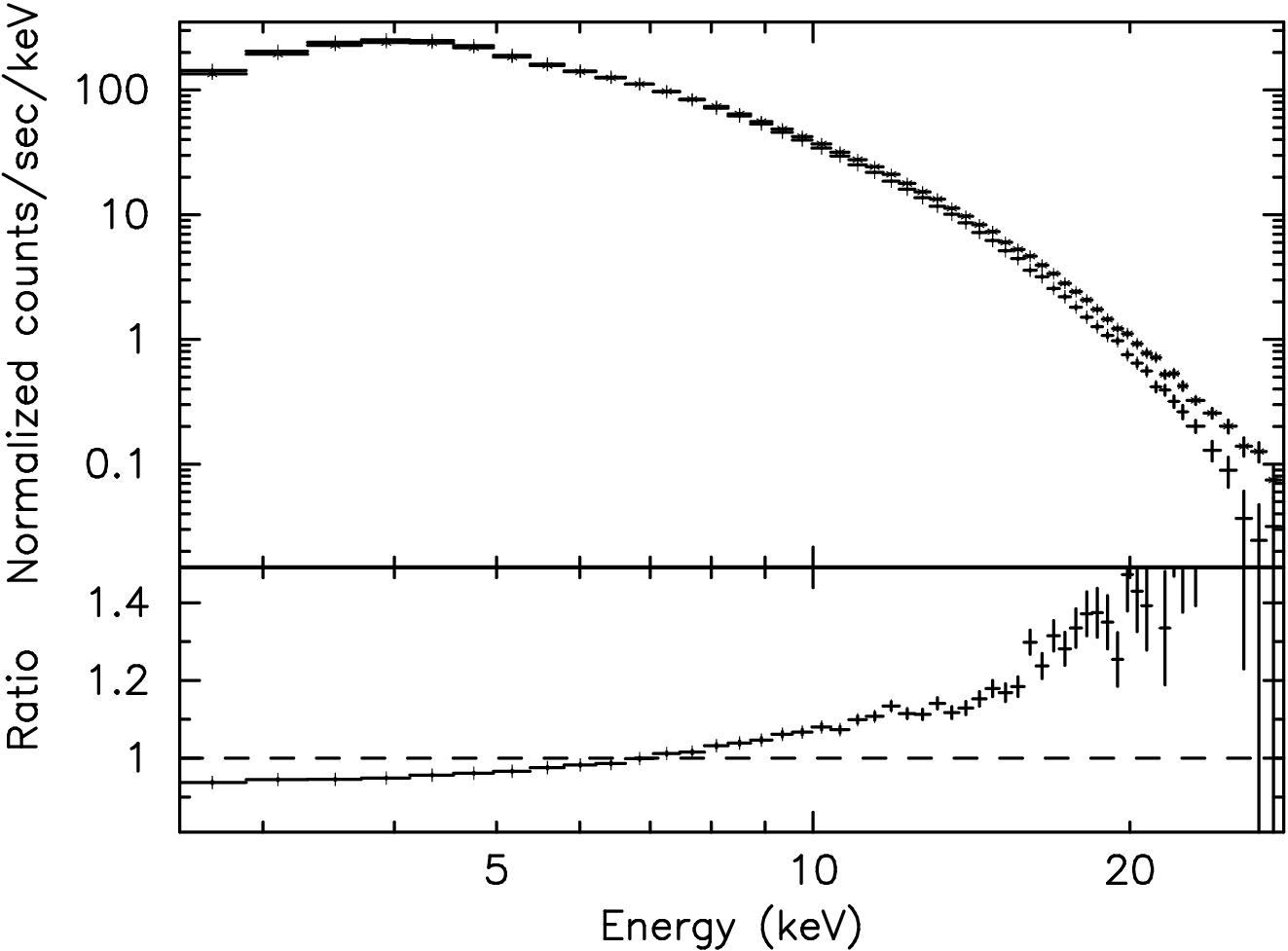}\\
\includegraphics[width=3.1cm,angle=270,clip]{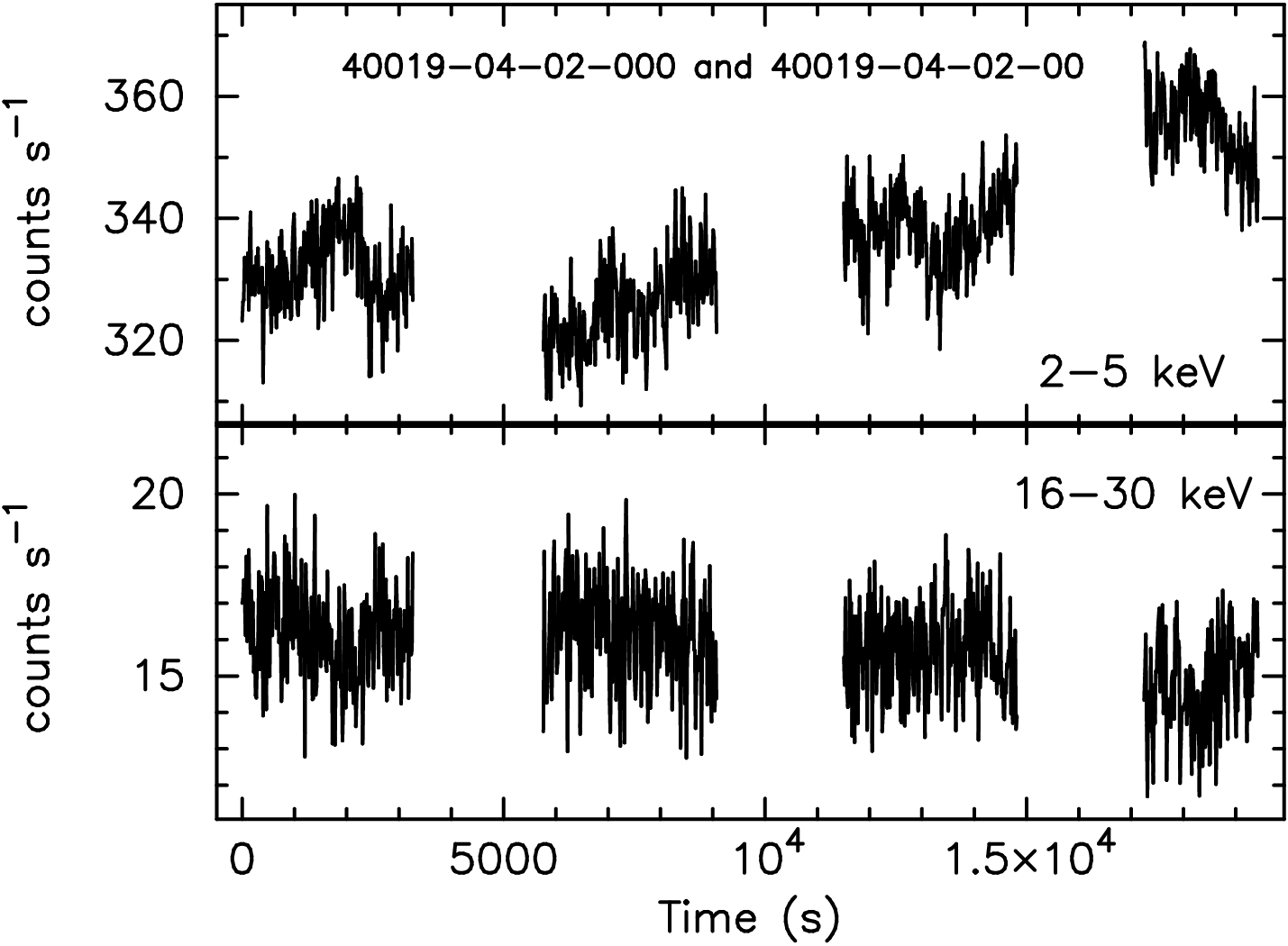}
\includegraphics[width=3.1cm,angle=270,clip]{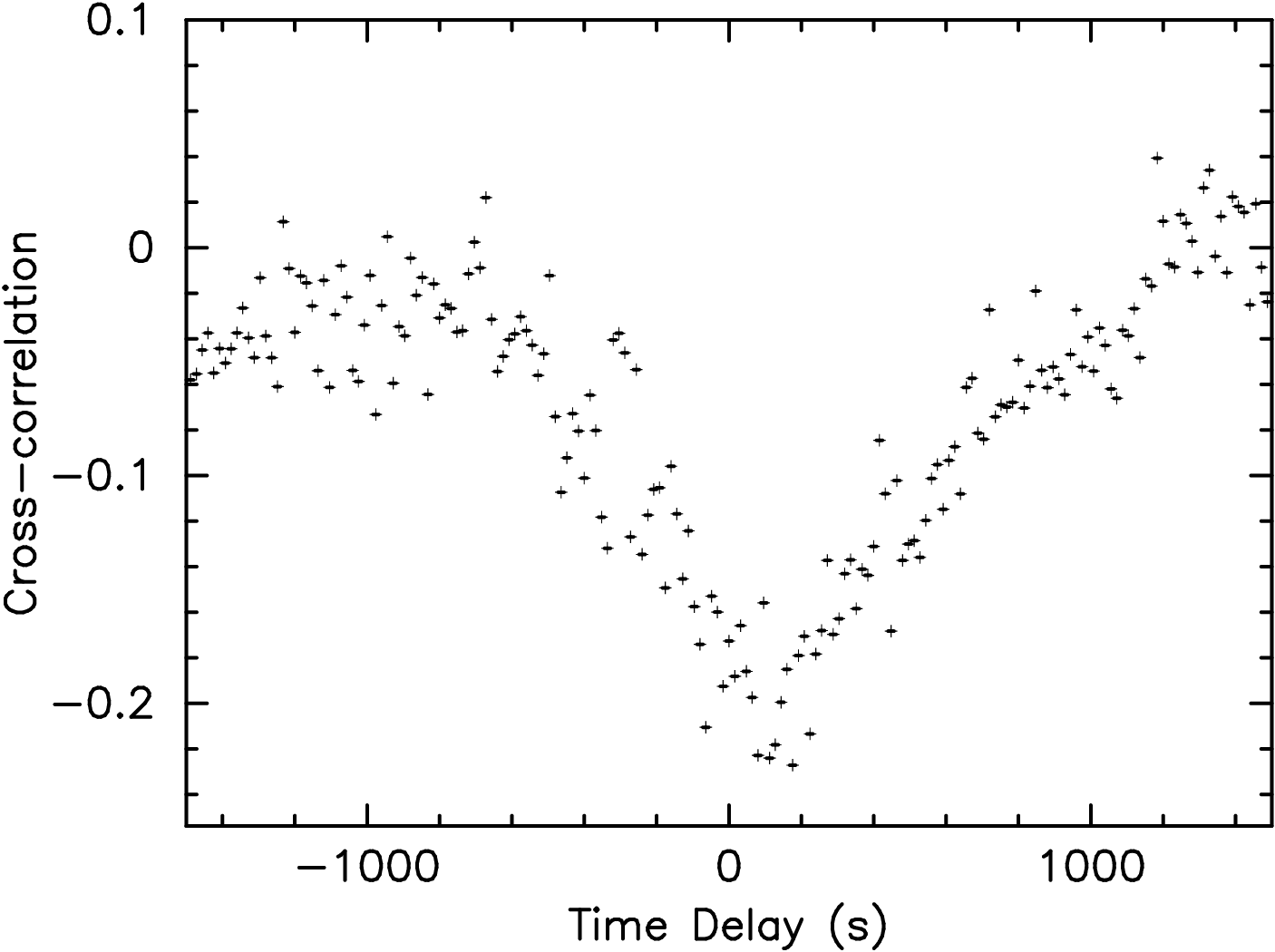}
\includegraphics[width=3.1cm,angle=270,clip]{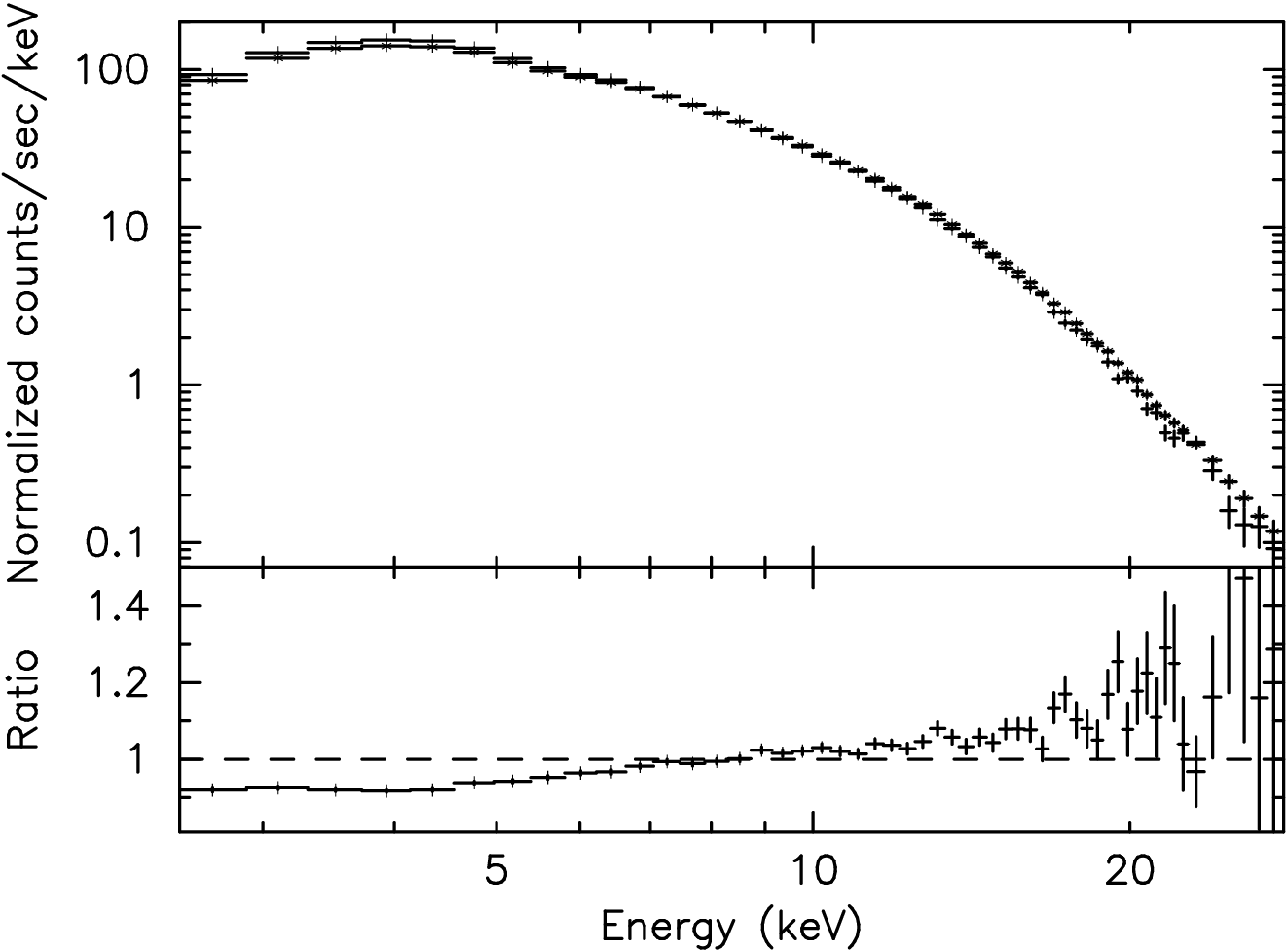}\\
\includegraphics[width=3.1cm,angle=270,clip]{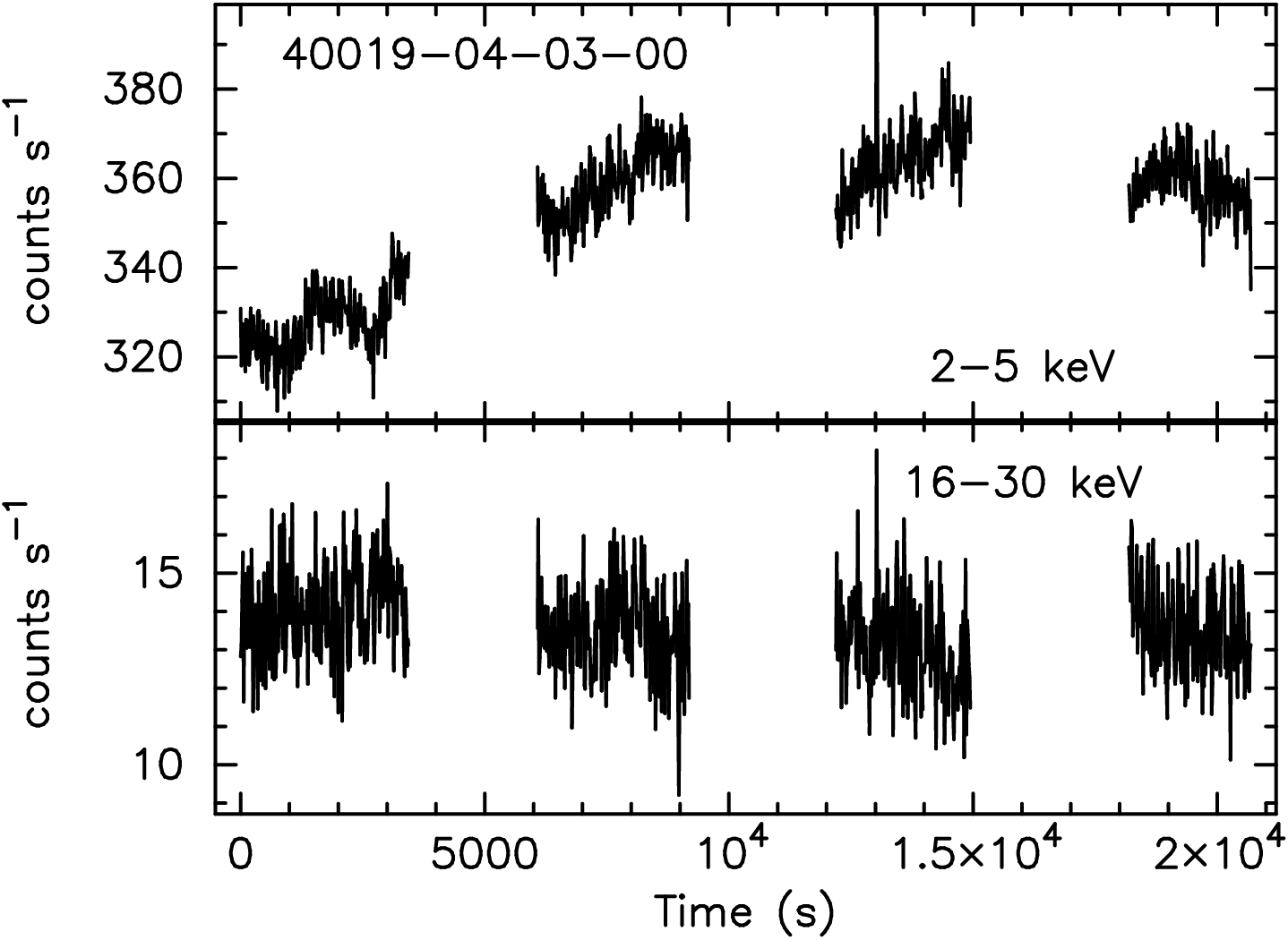}
\includegraphics[width=3.1cm,angle=270,clip]{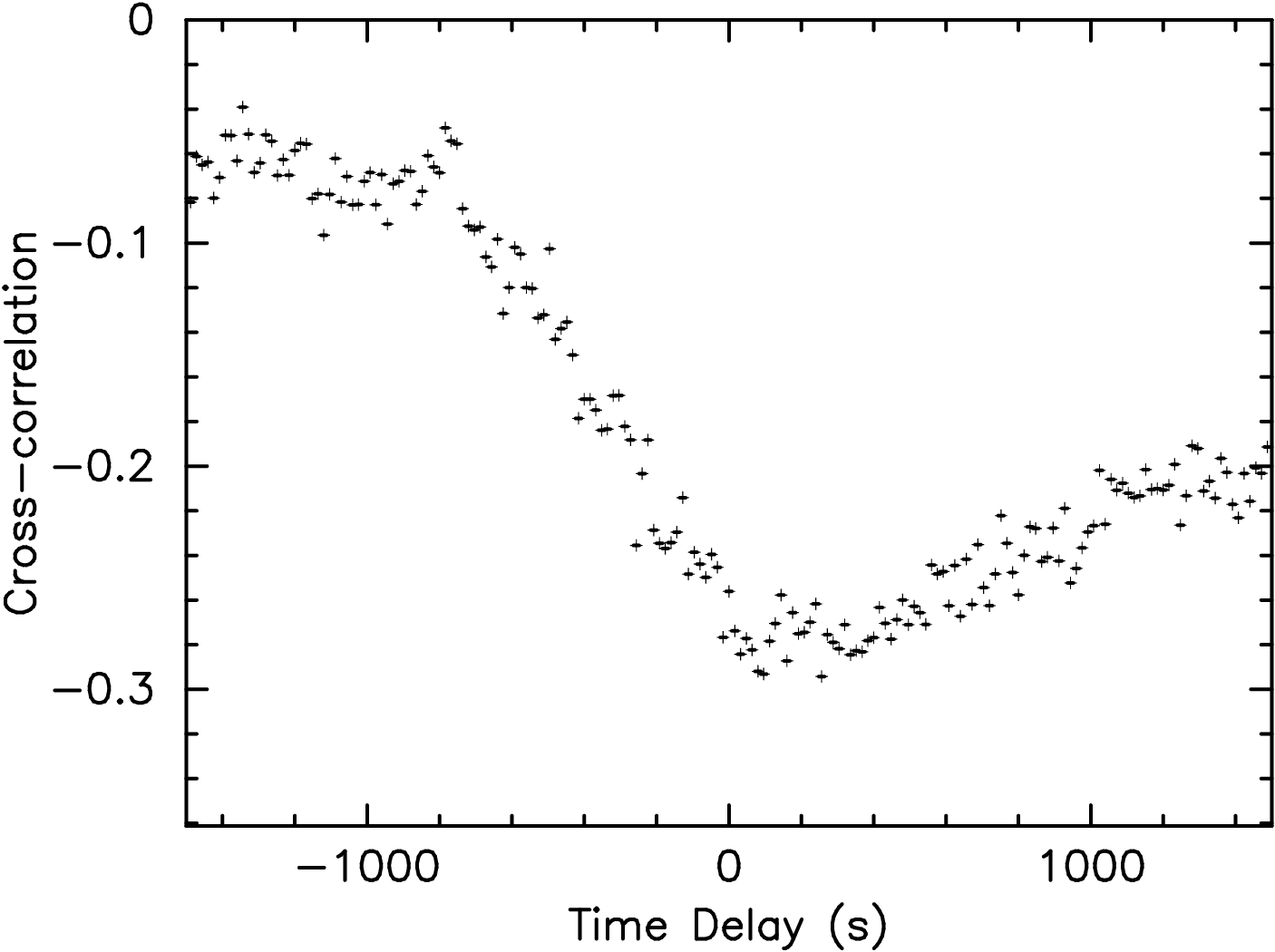}
\includegraphics[width=3.1cm,angle=270,clip]{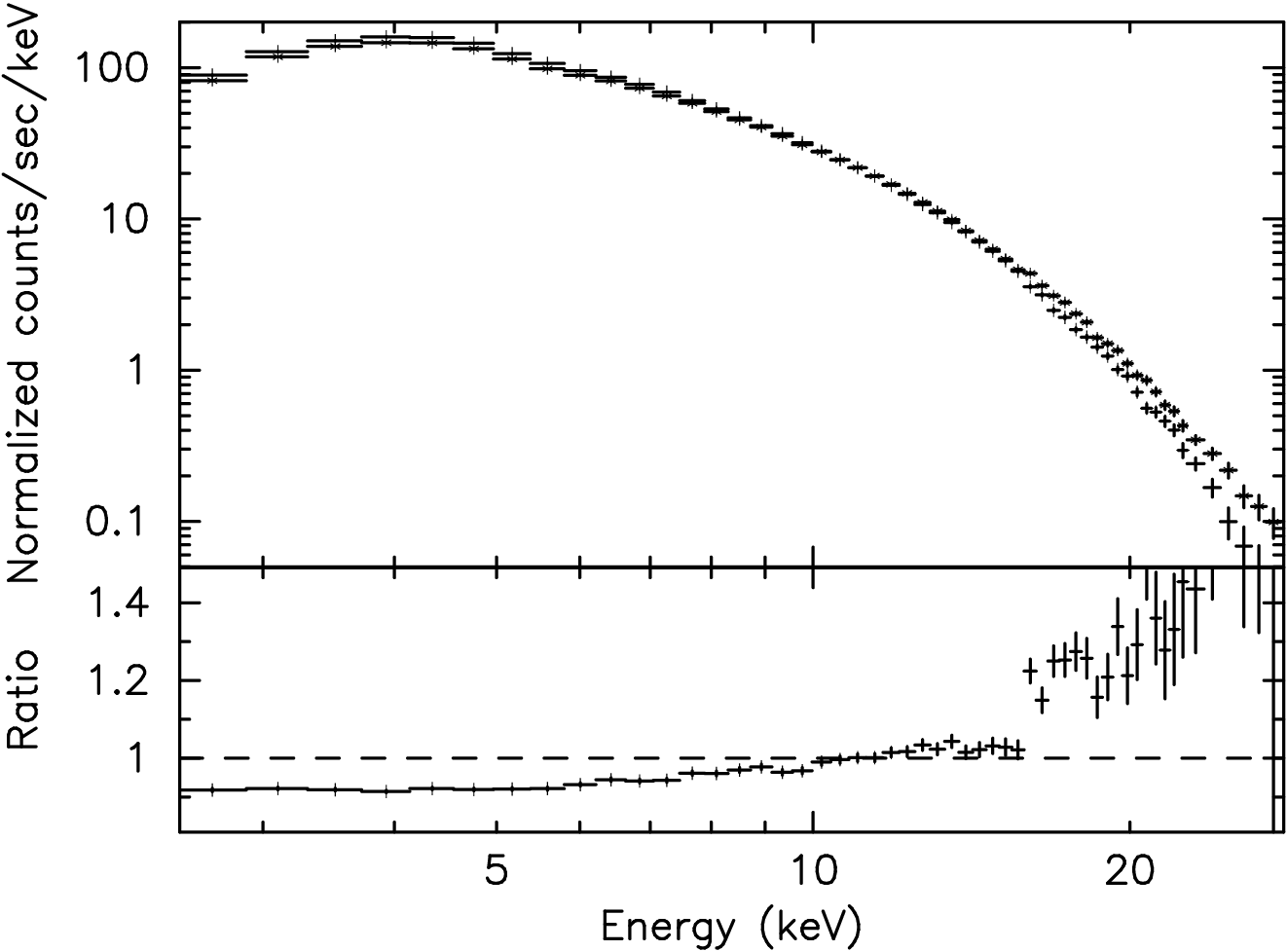}\\[5mm]
\centerline{Fig. 4. --- Continued.}
\end{center}
\clearpage
\begin{center}
\vspace*{-27mm}
\includegraphics[width=3.1cm,angle=270,clip]{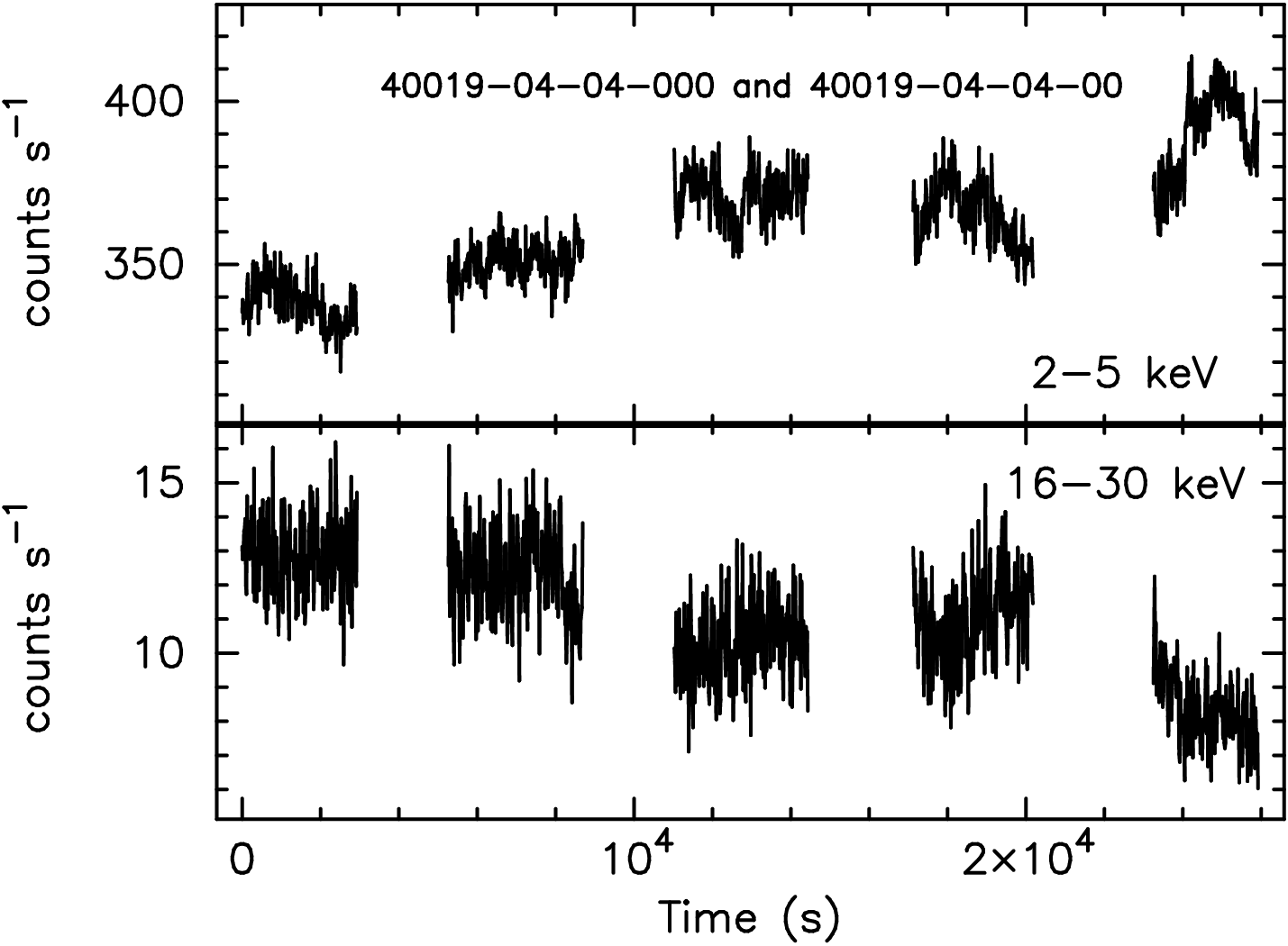}
\includegraphics[width=3.1cm,angle=270,clip]{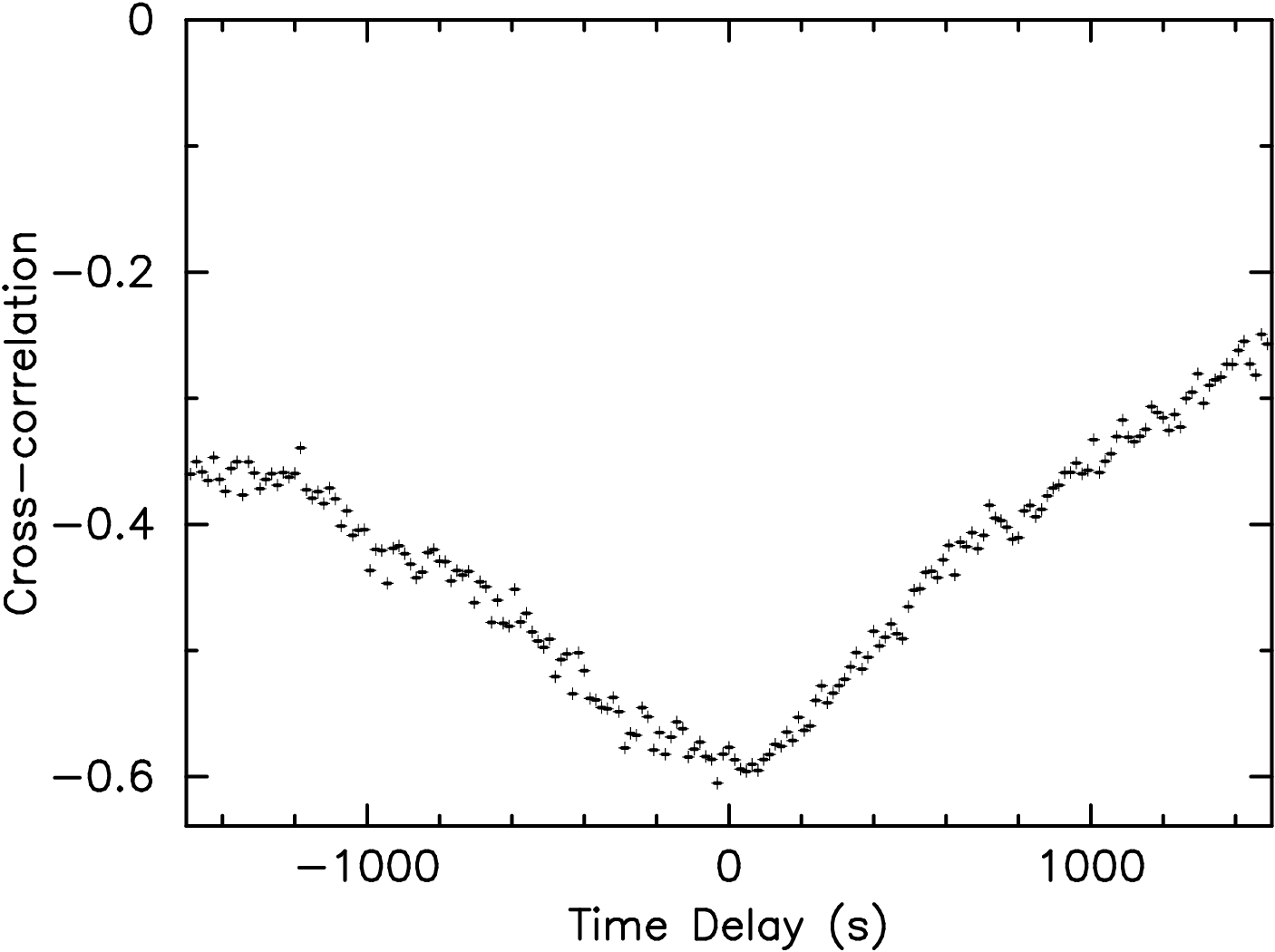}
\includegraphics[width=3.1cm,angle=270,clip]{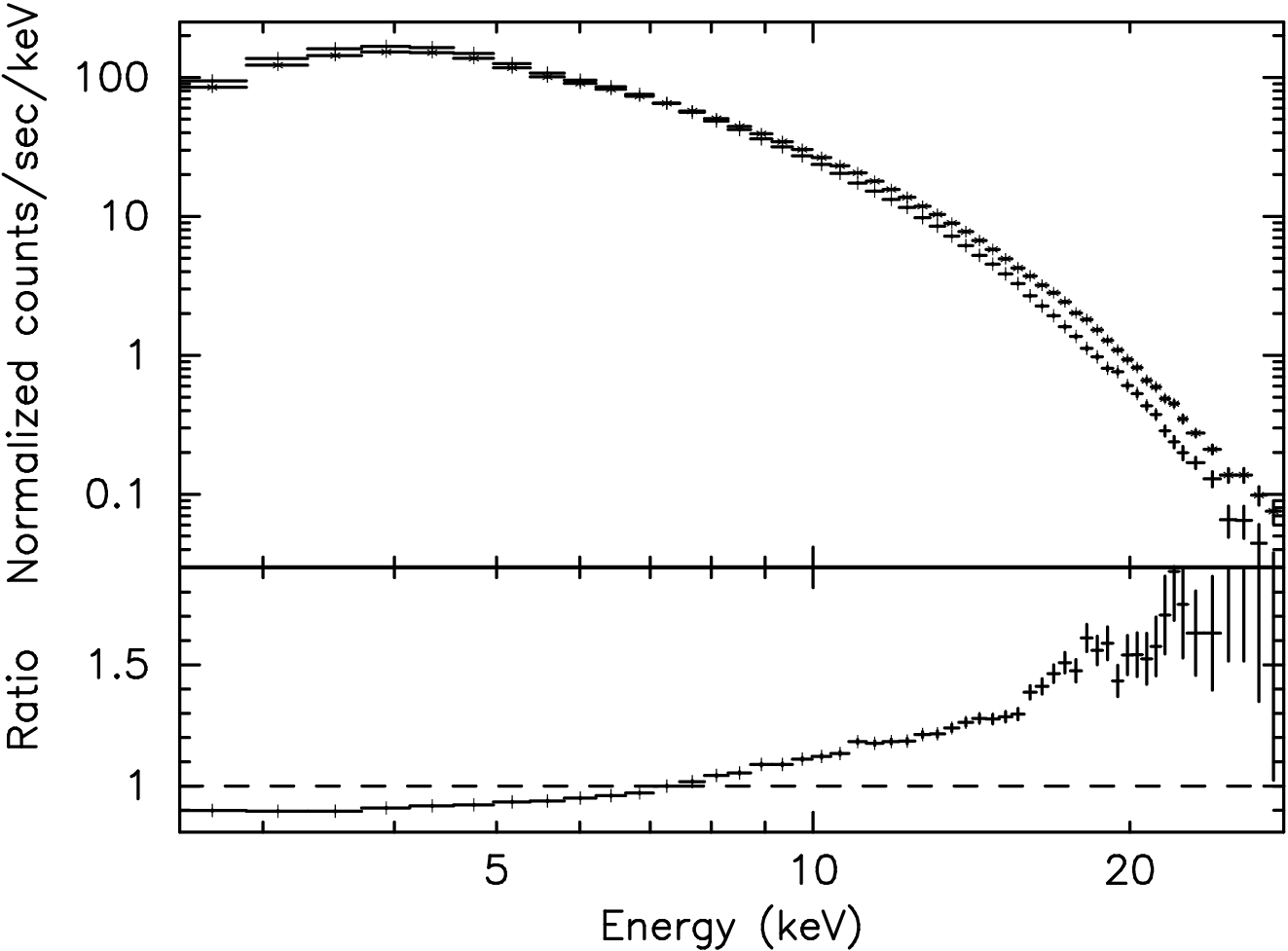}\\
\includegraphics[width=3.1cm,angle=270,clip]{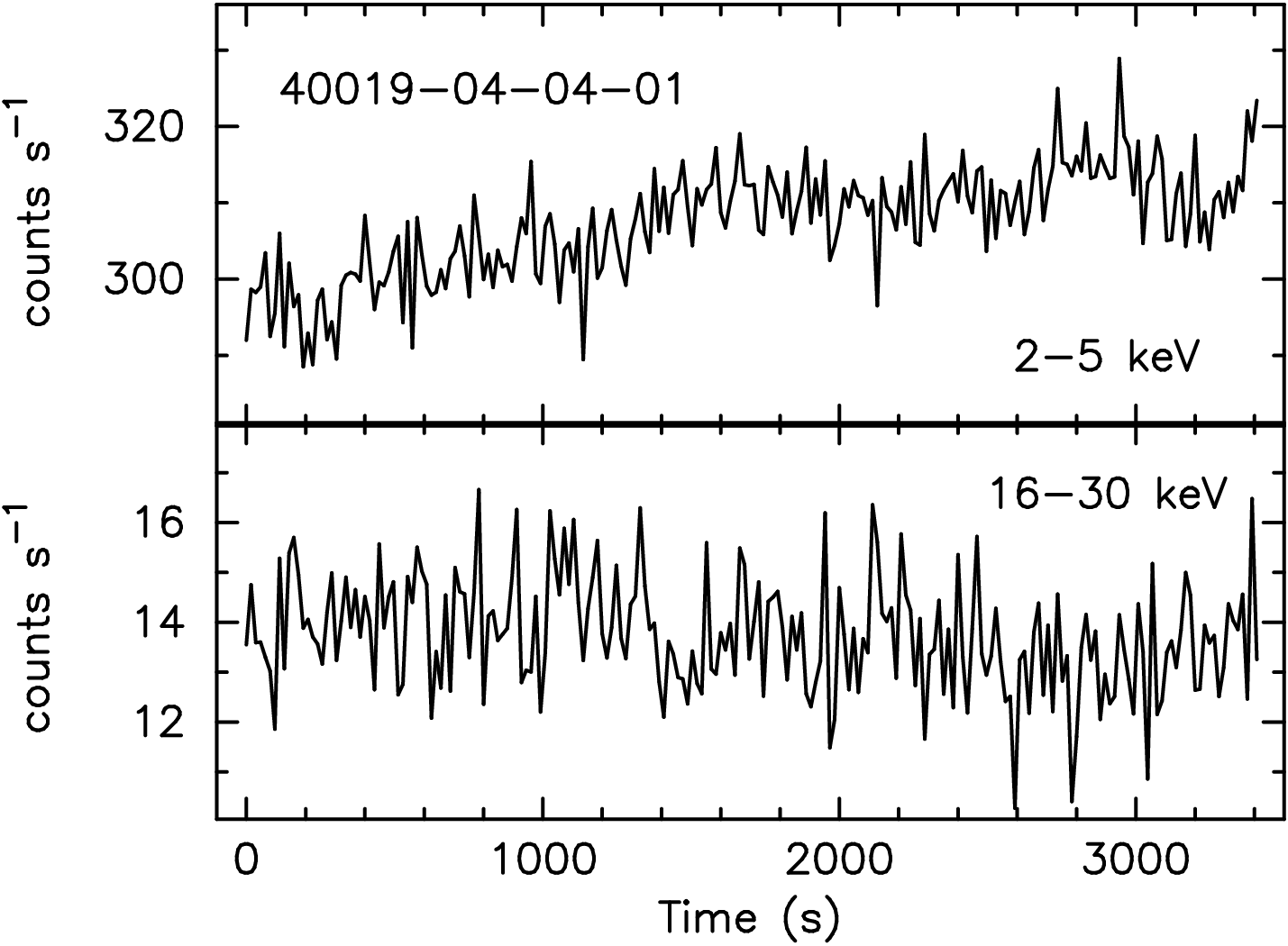}
\includegraphics[width=3.1cm,angle=270,clip]{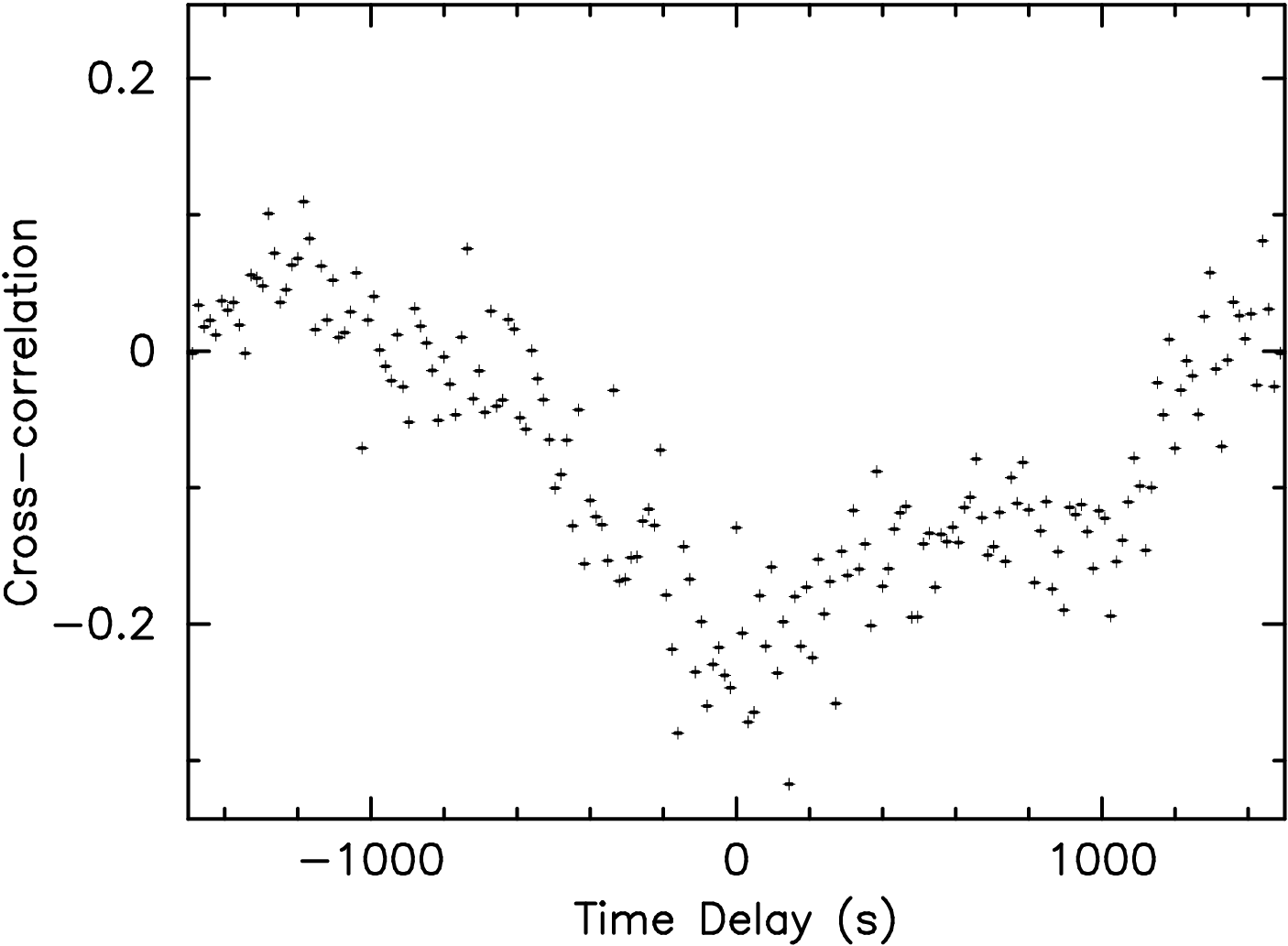}
\includegraphics[width=3.1cm,angle=270,clip]{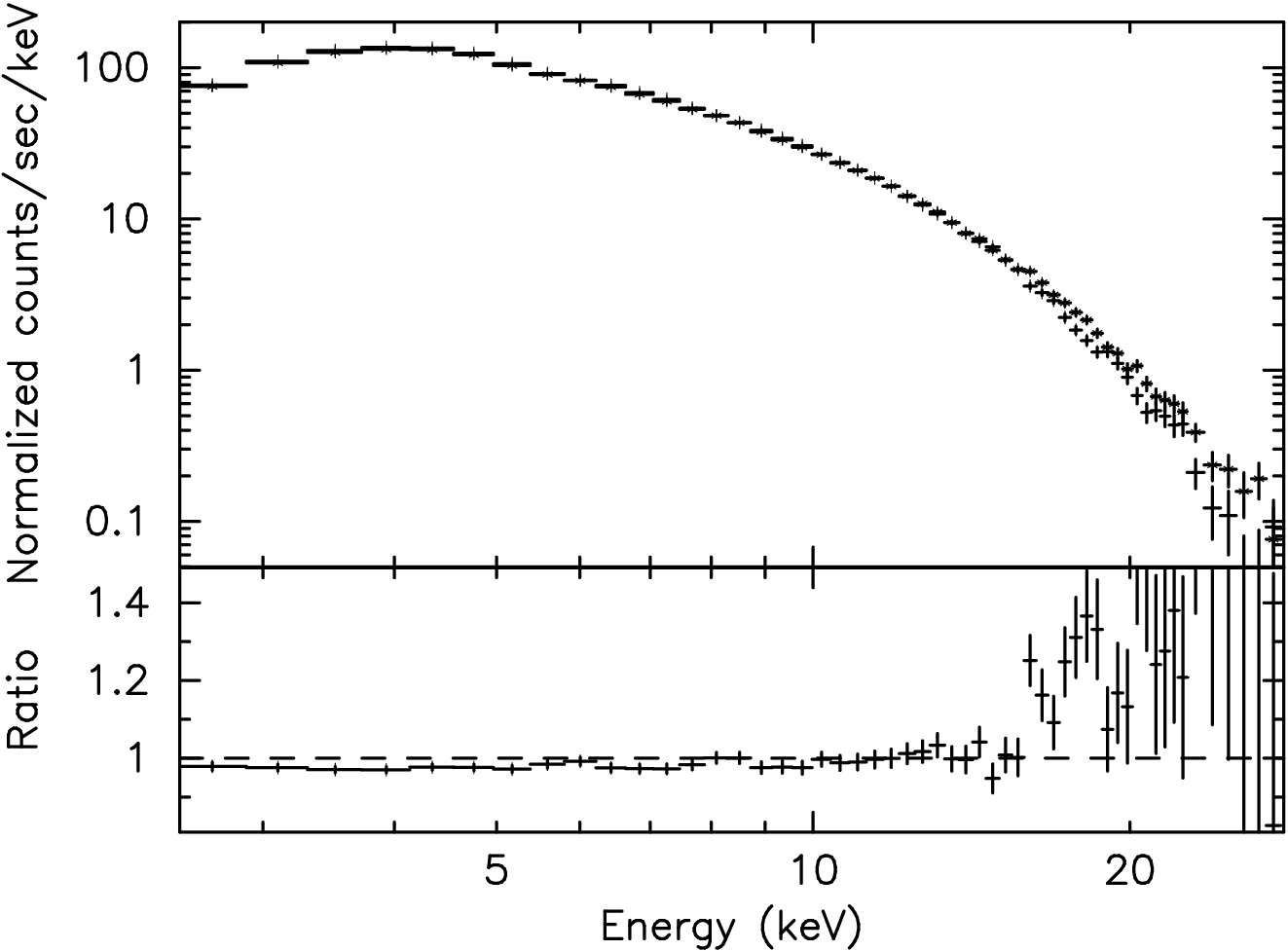}\\
\includegraphics[width=3.1cm,angle=270,clip]{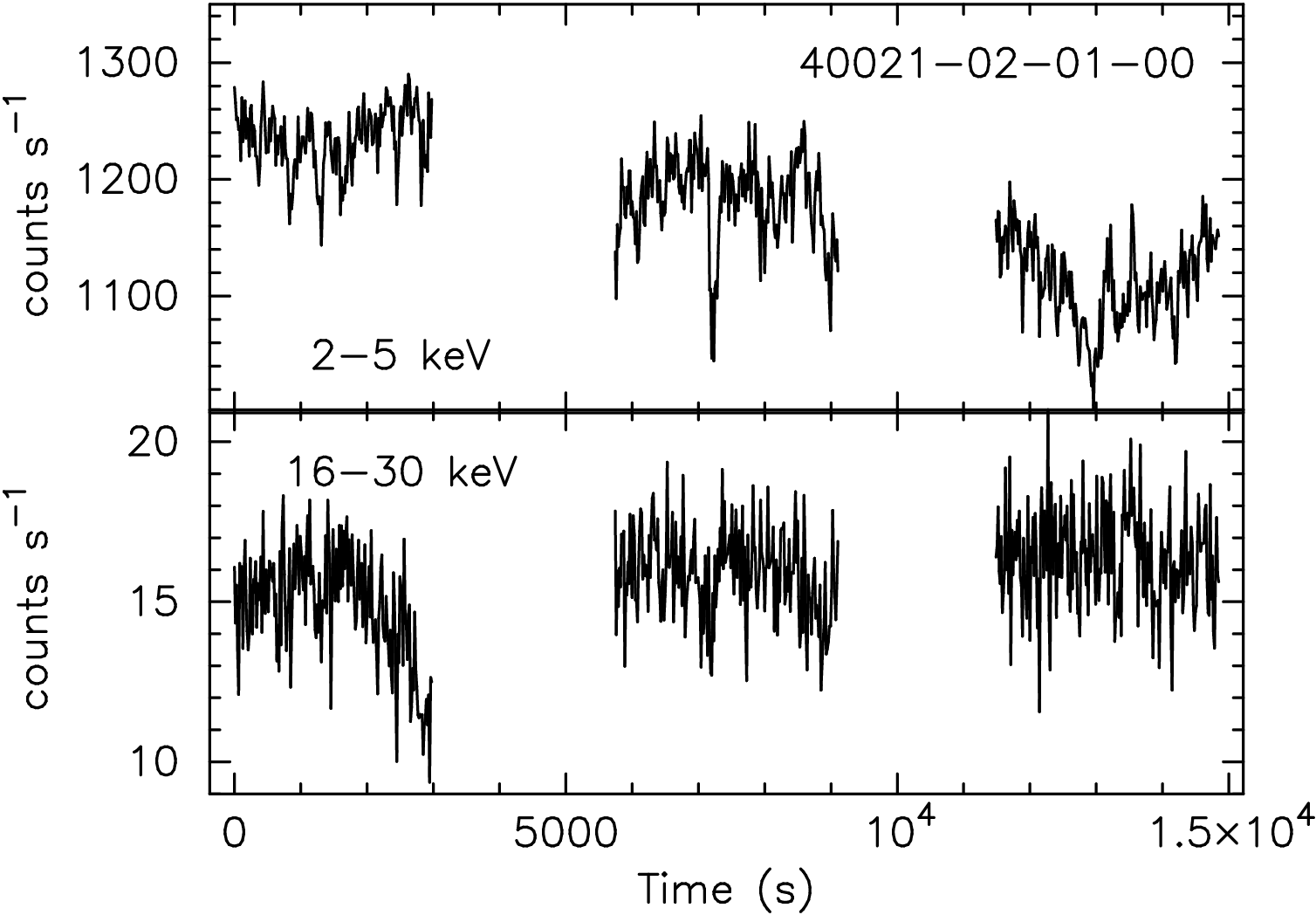}
\includegraphics[width=3.1cm,angle=270,clip]{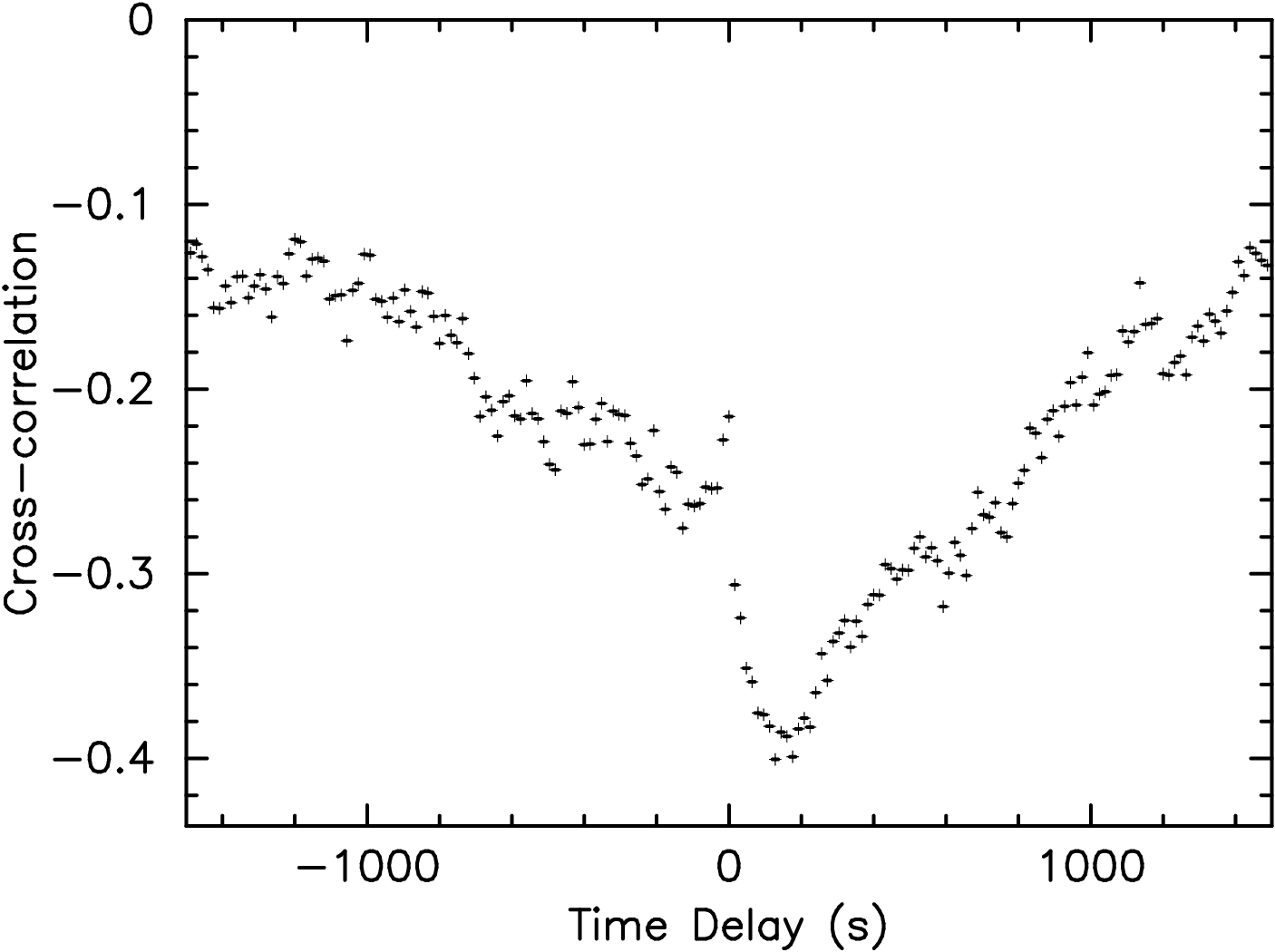}
\includegraphics[width=3.1cm,angle=270,clip]{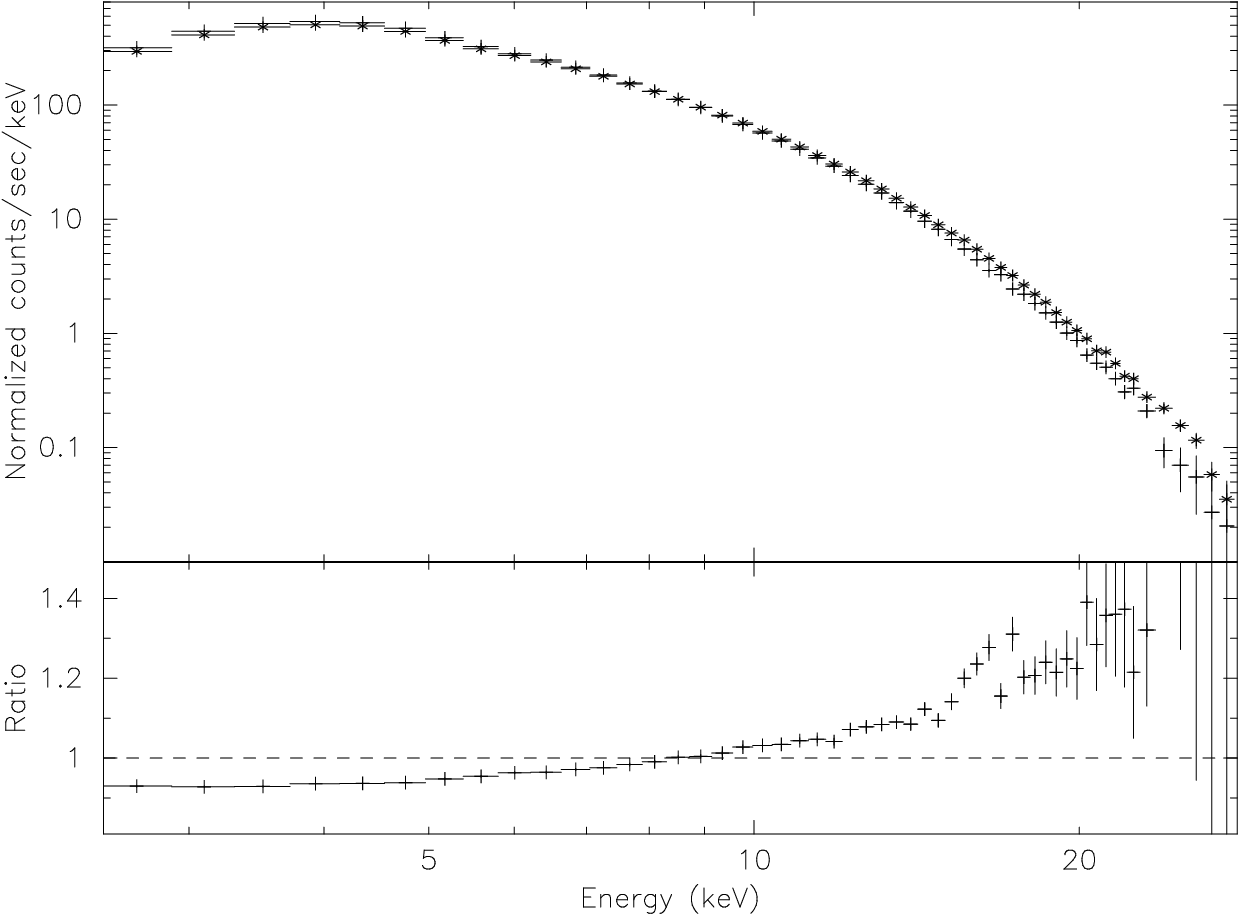}\\
\includegraphics[width=3.1cm,angle=270,clip]{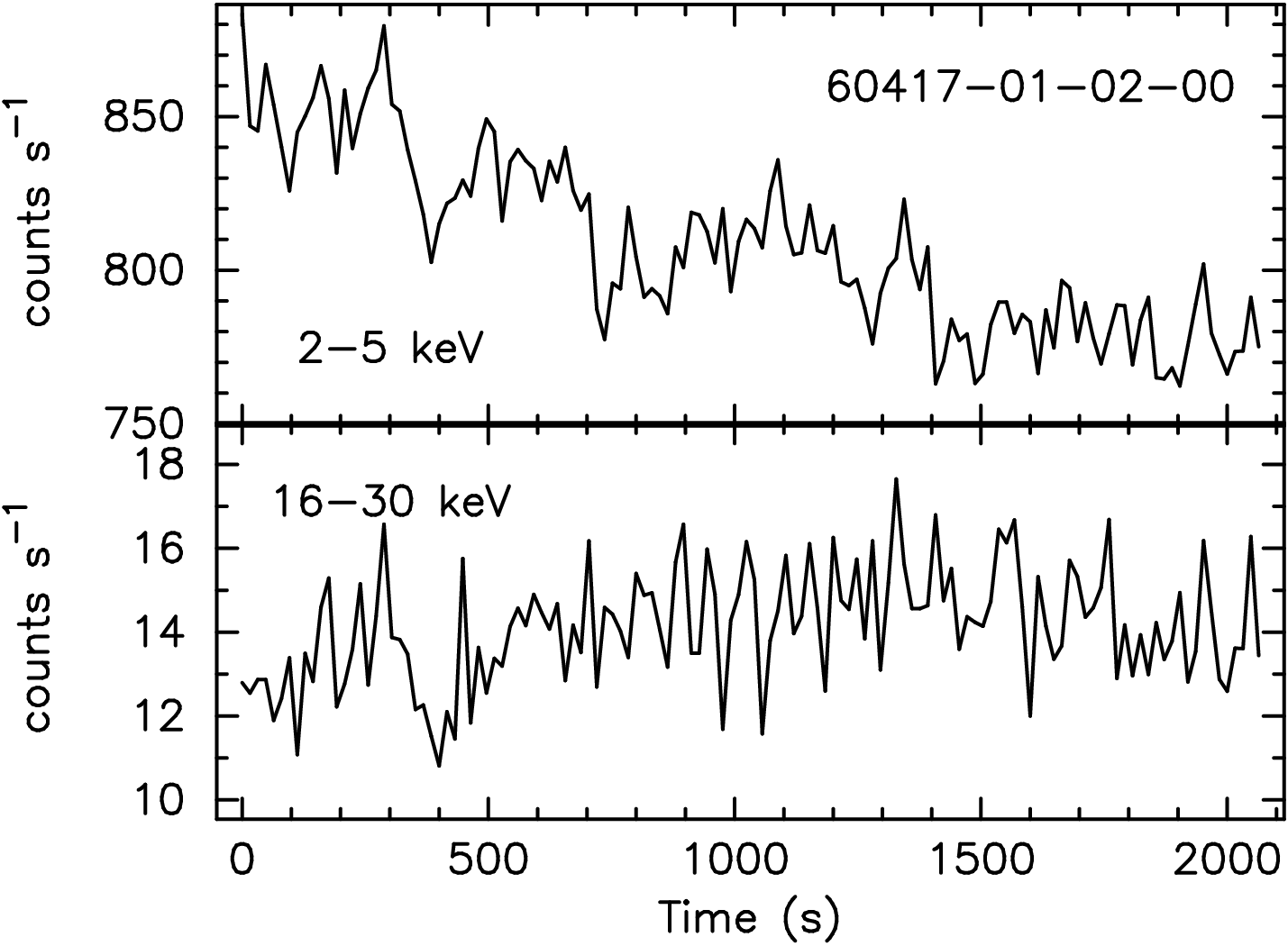}
\includegraphics[width=3.1cm,angle=270,clip]{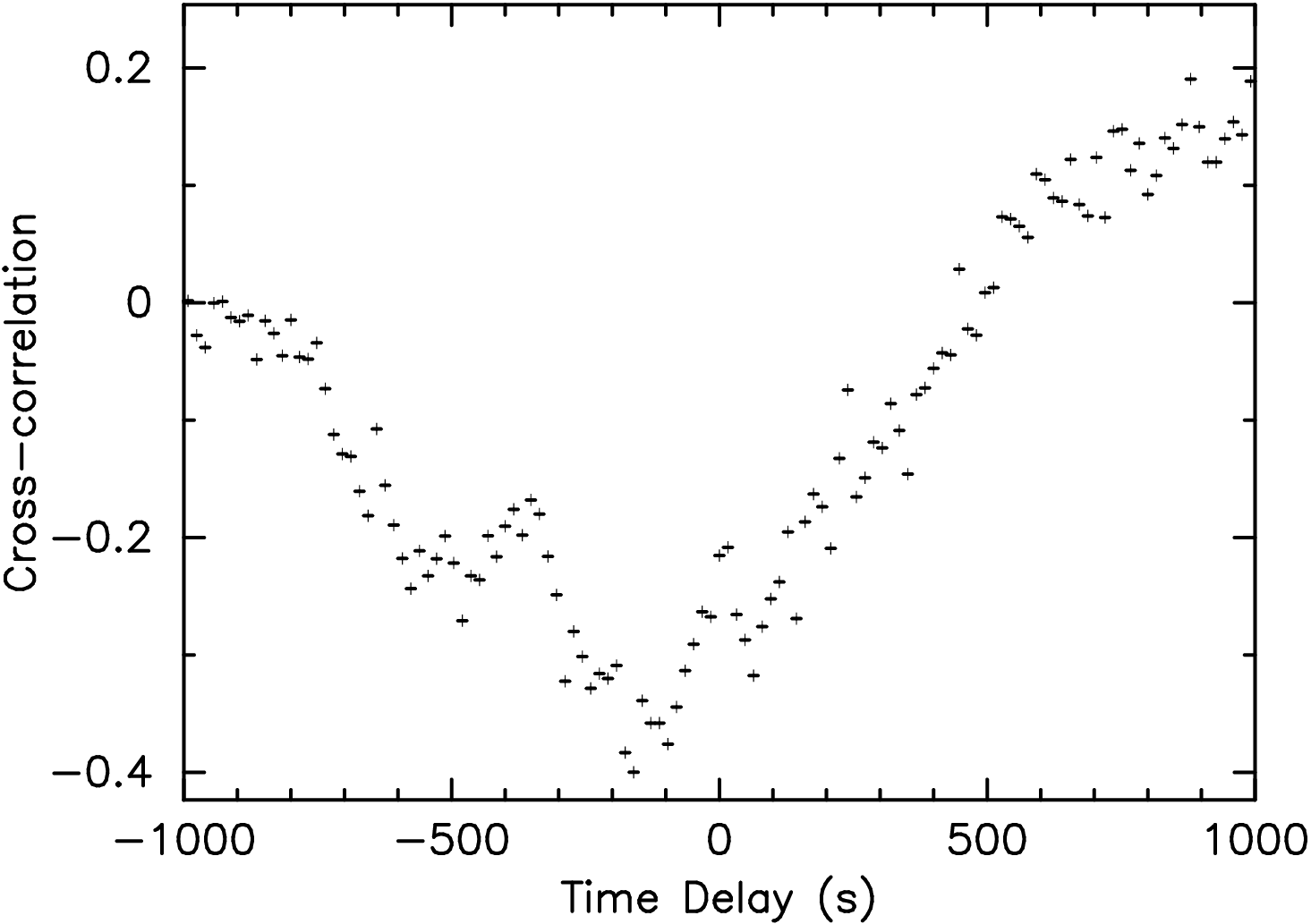}
\includegraphics[width=3.1cm,angle=270,clip]{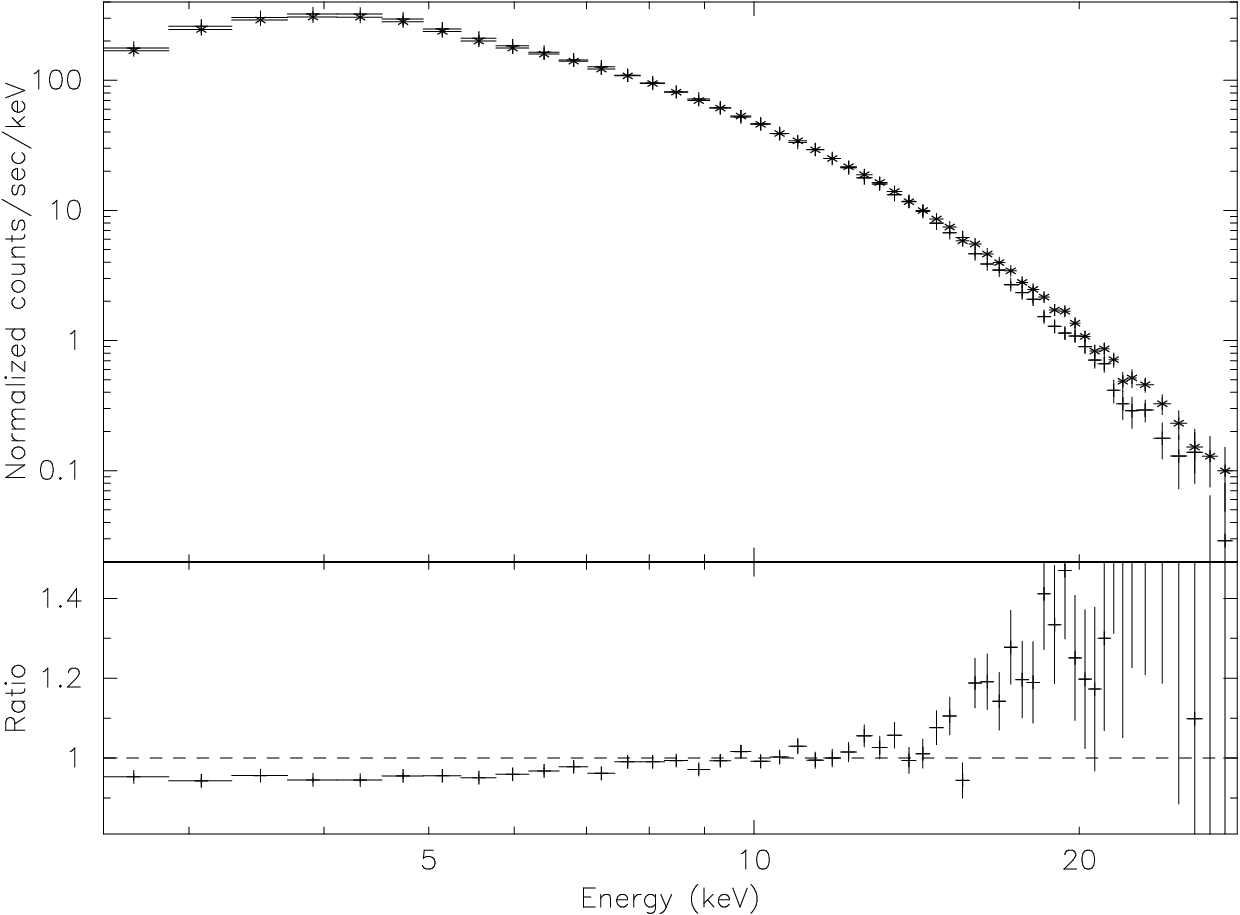}\\
\includegraphics[width=3.1cm,angle=270,clip]{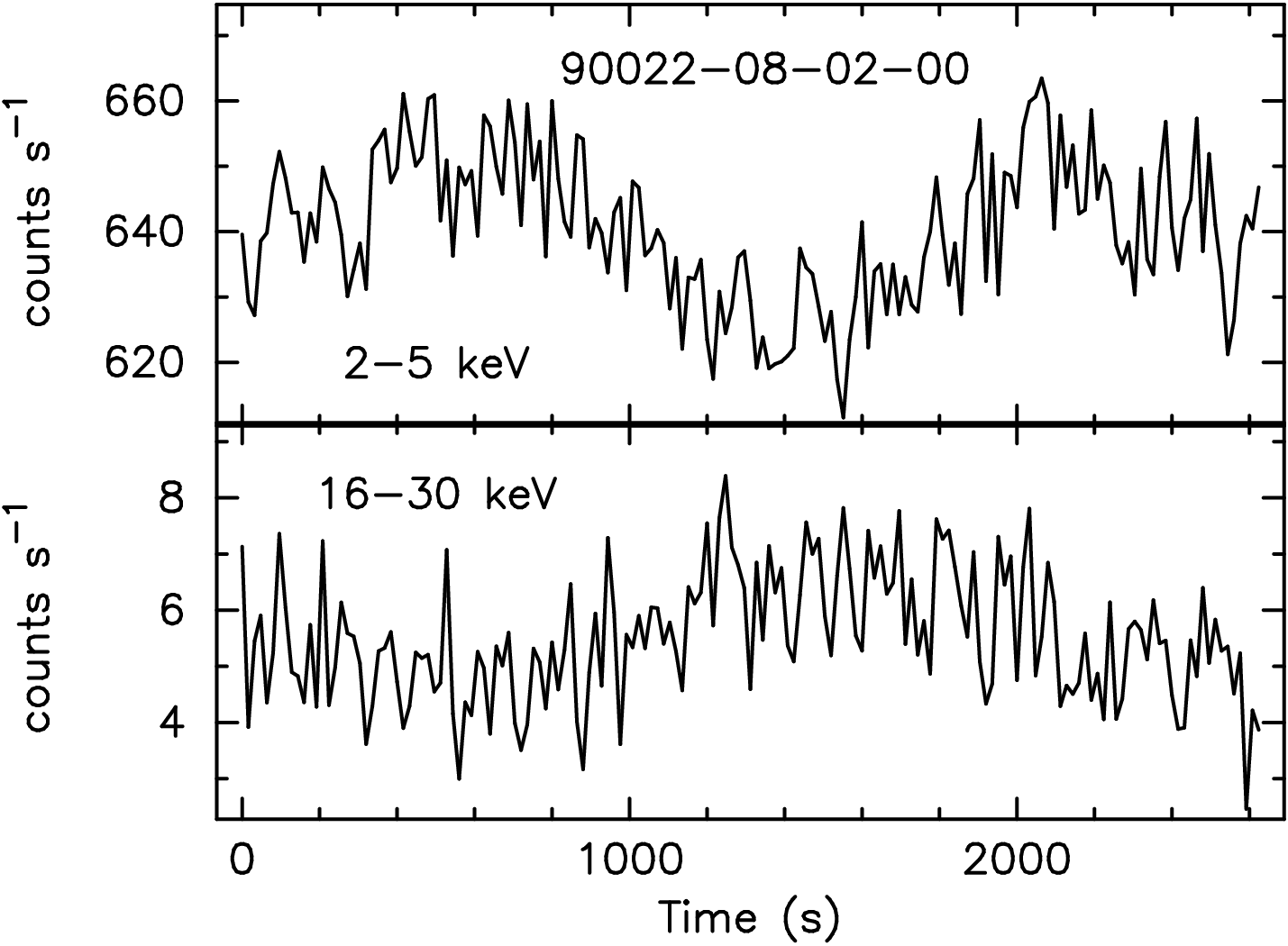}
\includegraphics[width=3.1cm,angle=270,clip]{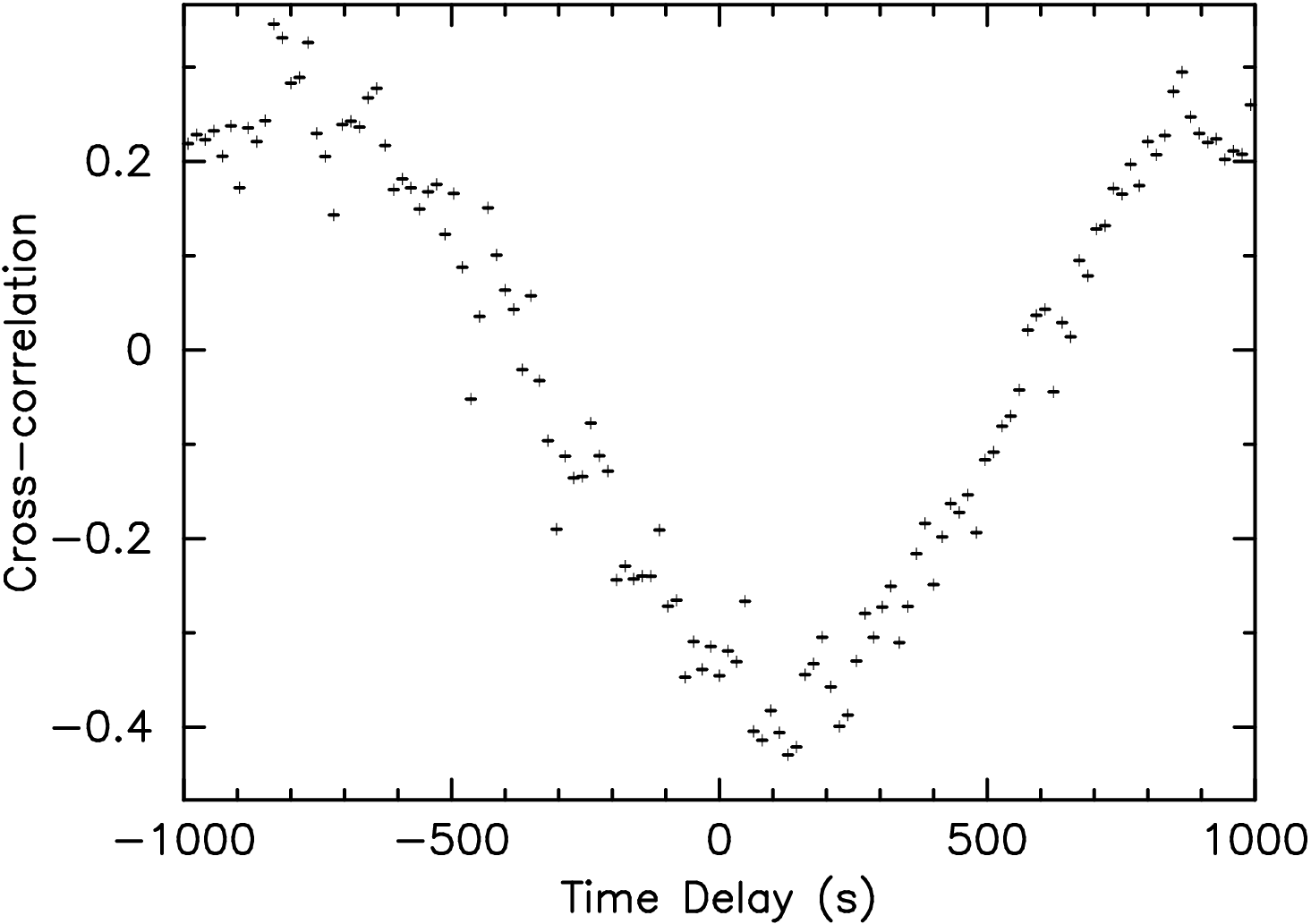}
\includegraphics[width=3.1cm,angle=270,clip]{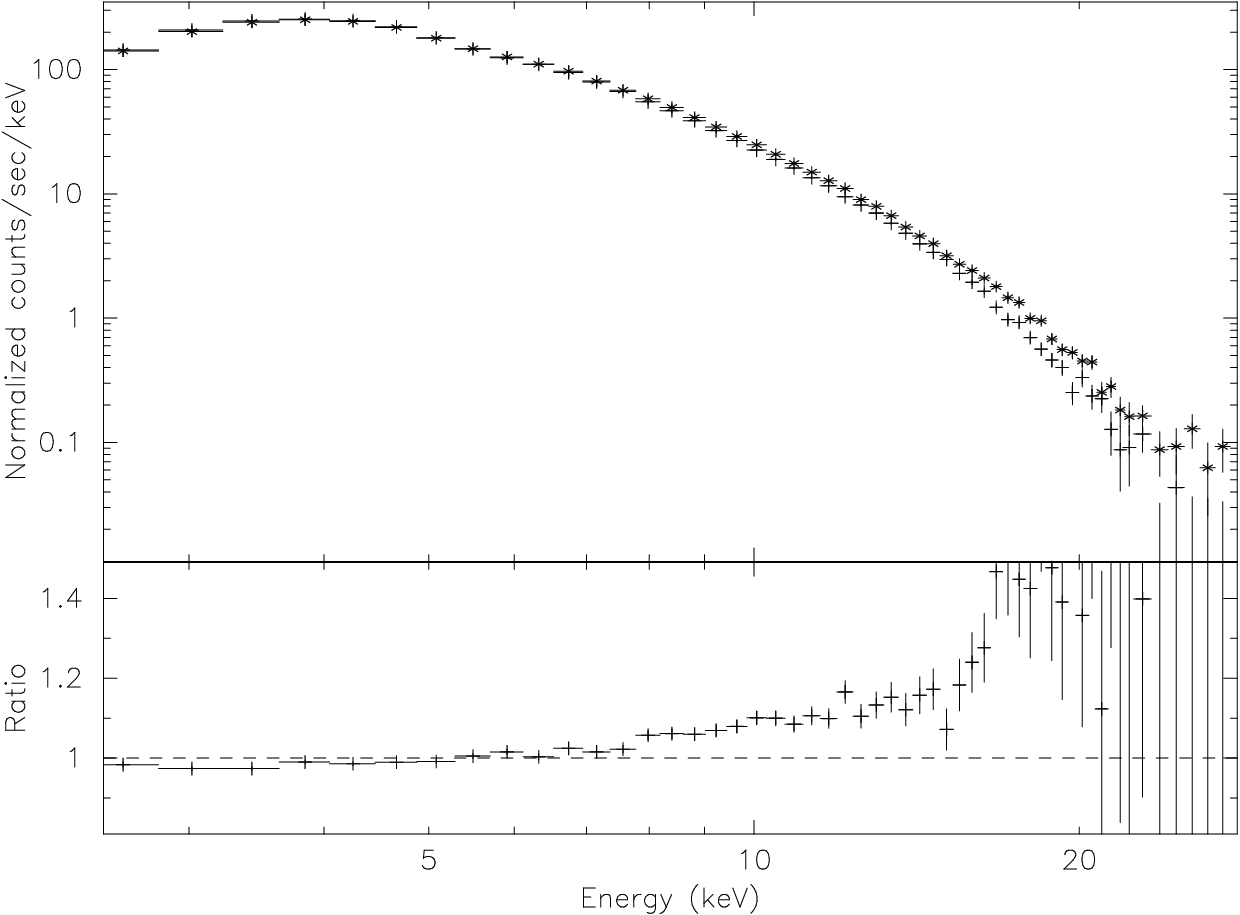}\\
\includegraphics[width=3.1cm,angle=270,clip]{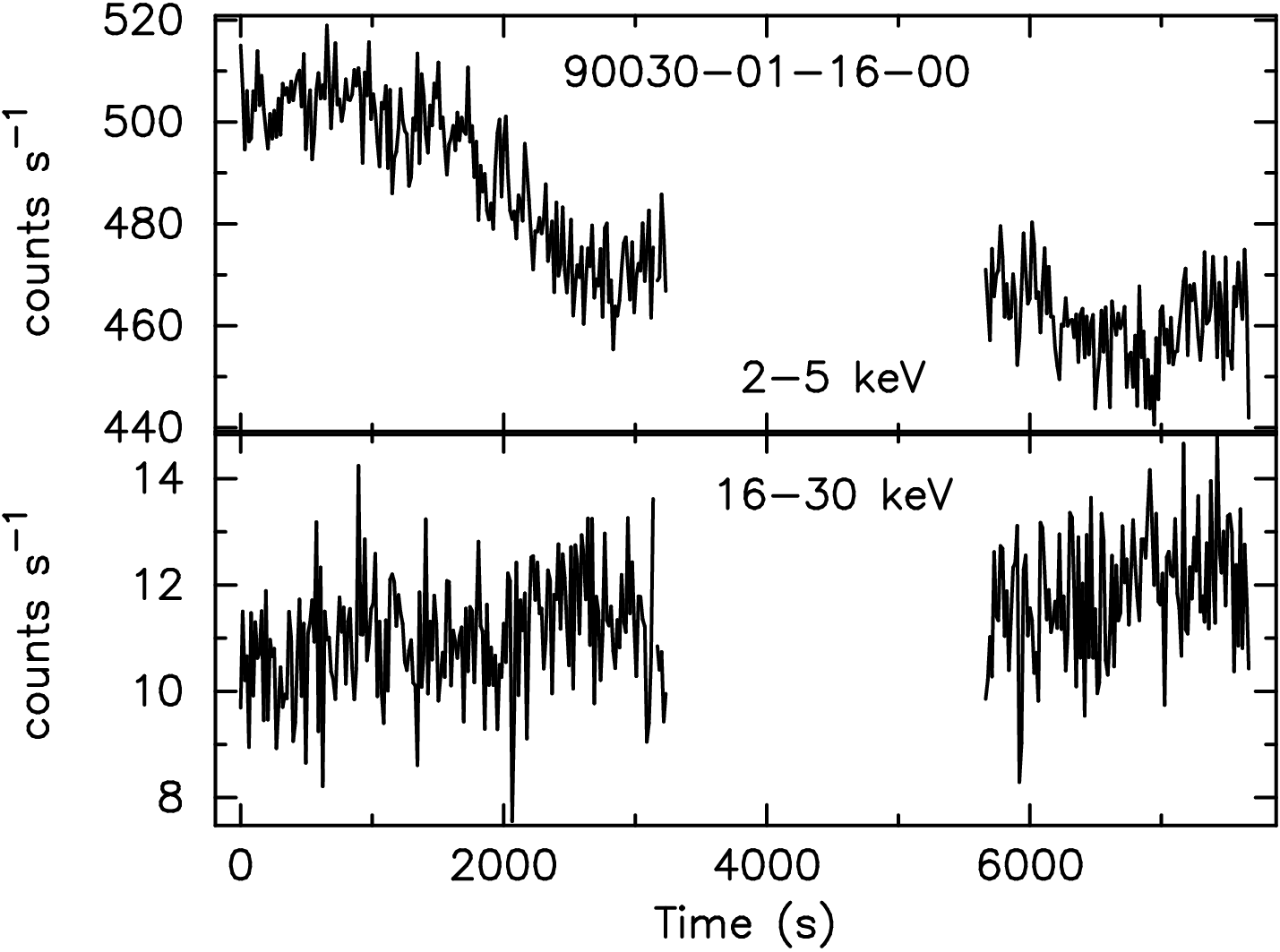}
\includegraphics[width=3.1cm,angle=270,clip]{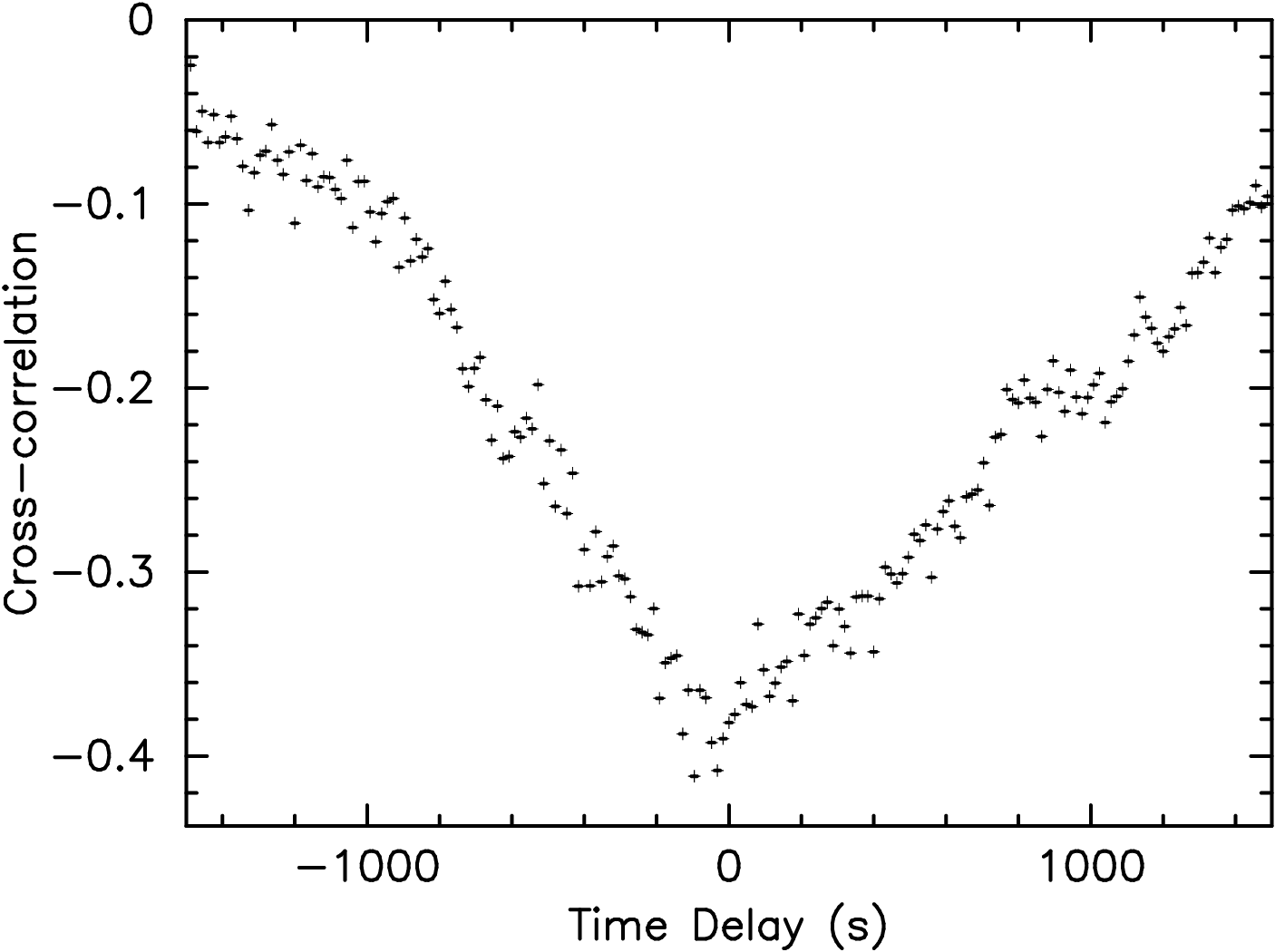}
\includegraphics[width=3.1cm,angle=270,clip]{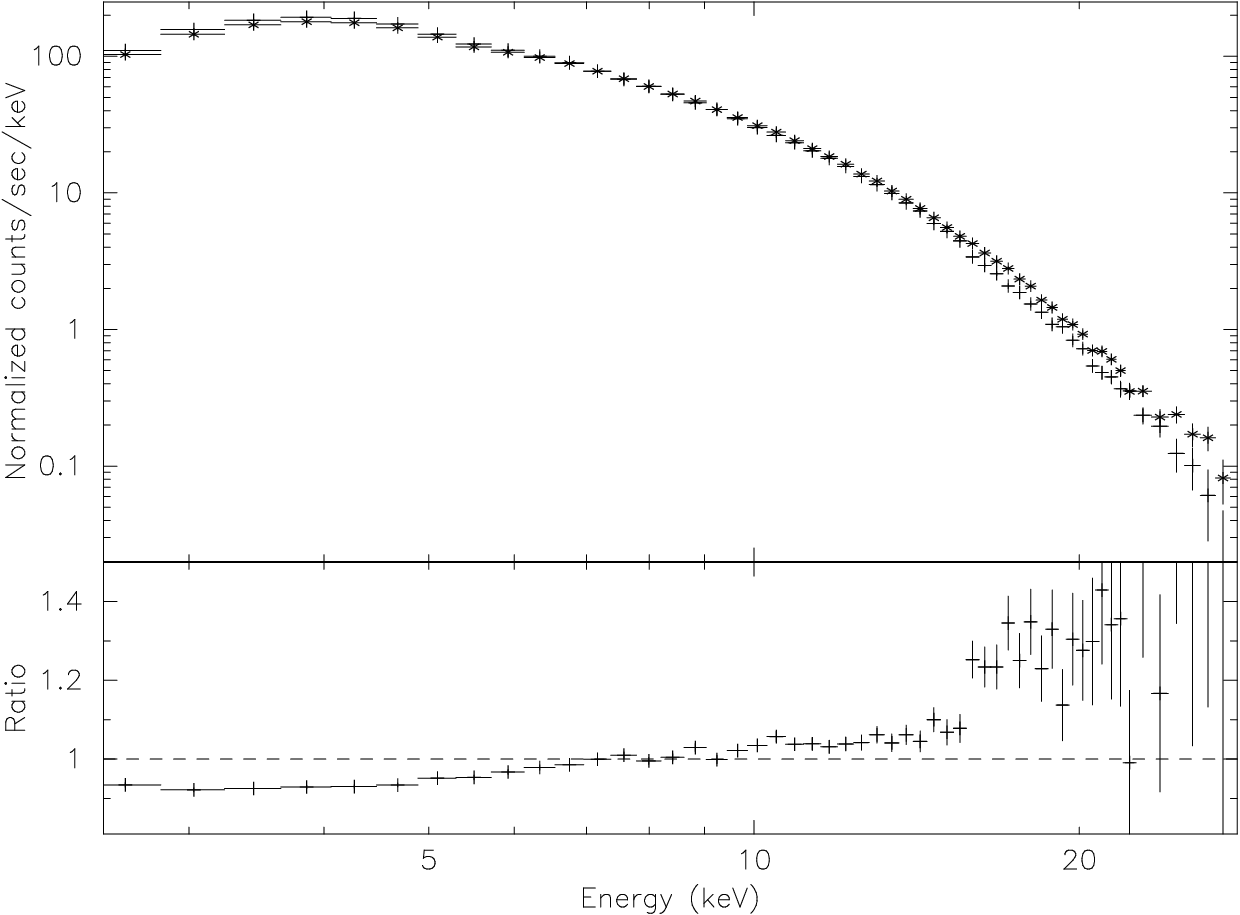}\\
\includegraphics[width=3.1cm,angle=270,clip]{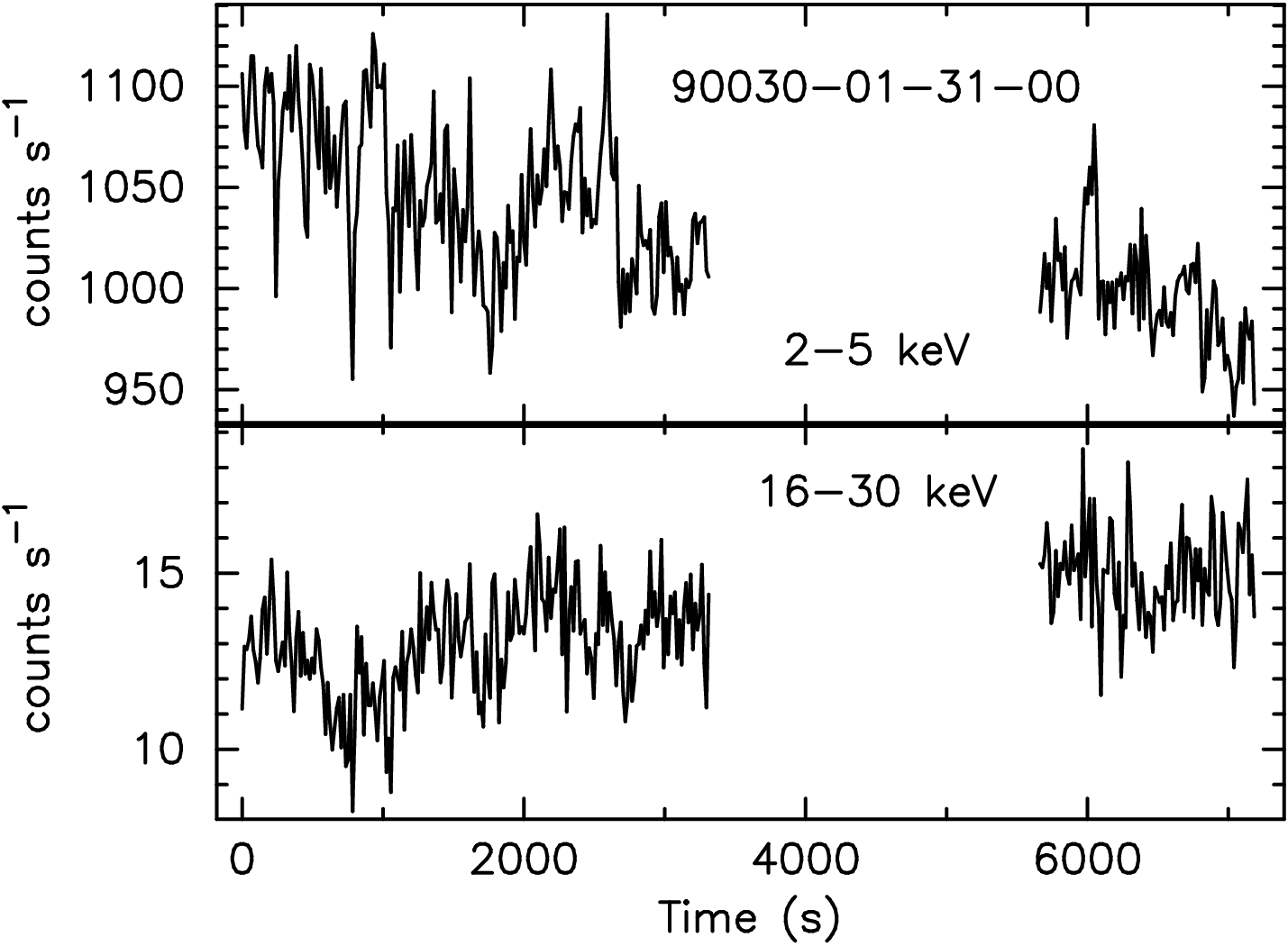}
\includegraphics[width=3.1cm,angle=270,clip]{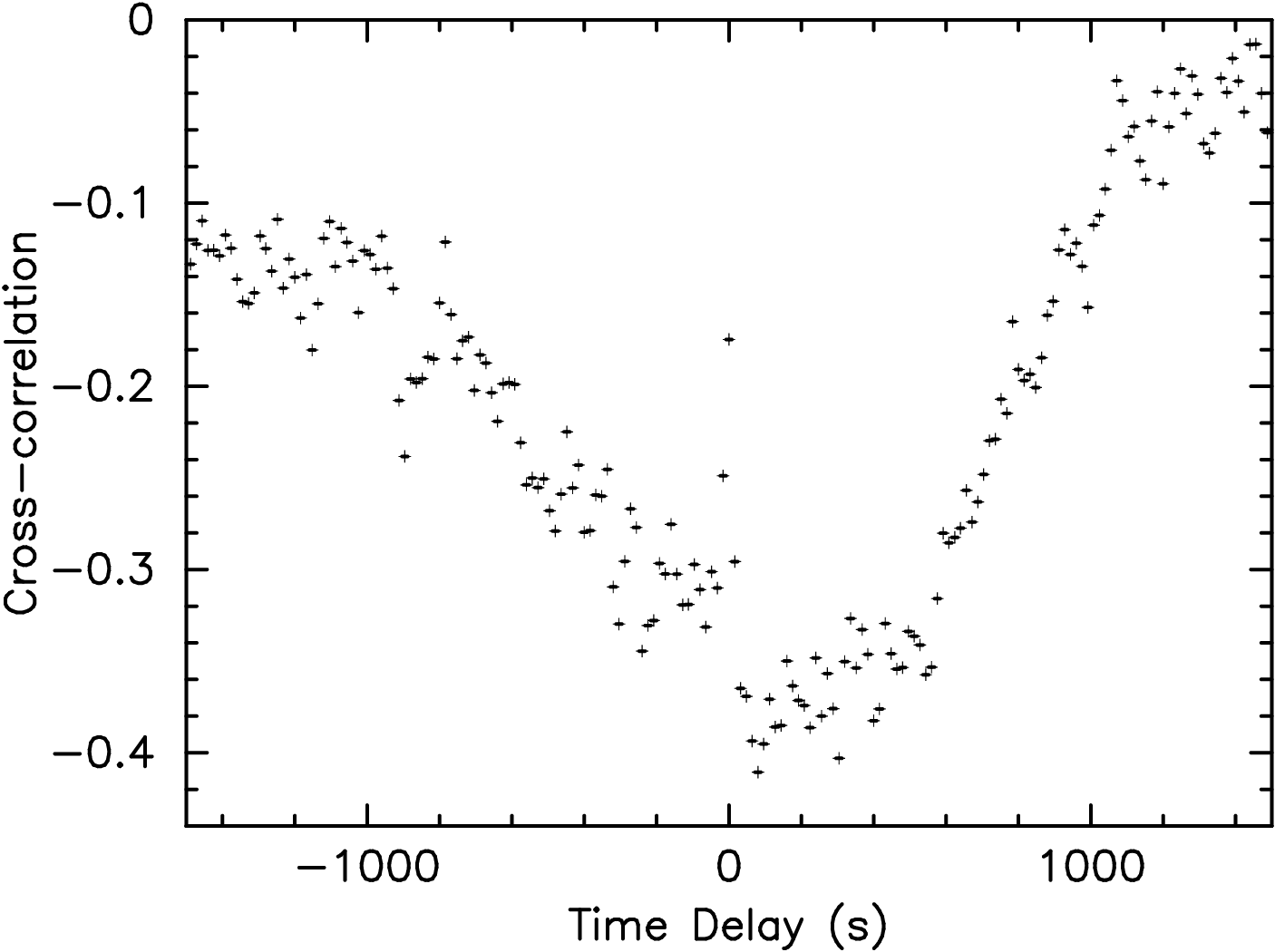}
\includegraphics[width=3.1cm,angle=270,clip]{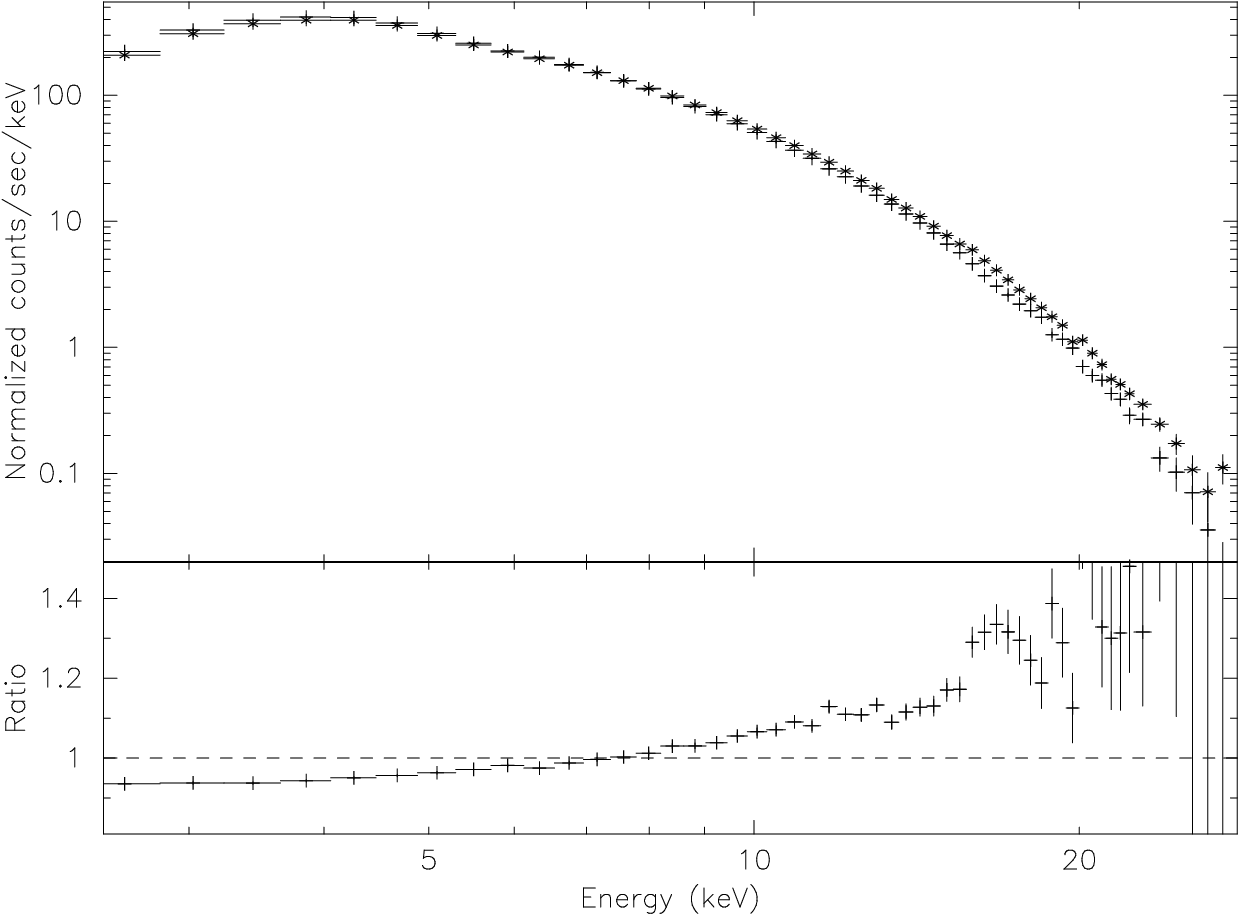}\\[5mm]
{Fig. 4. --- Continued}
\end{center}
\clearpage
\begin{center}
\vspace*{-27mm}
\includegraphics[width=3.1cm,angle=270,clip]{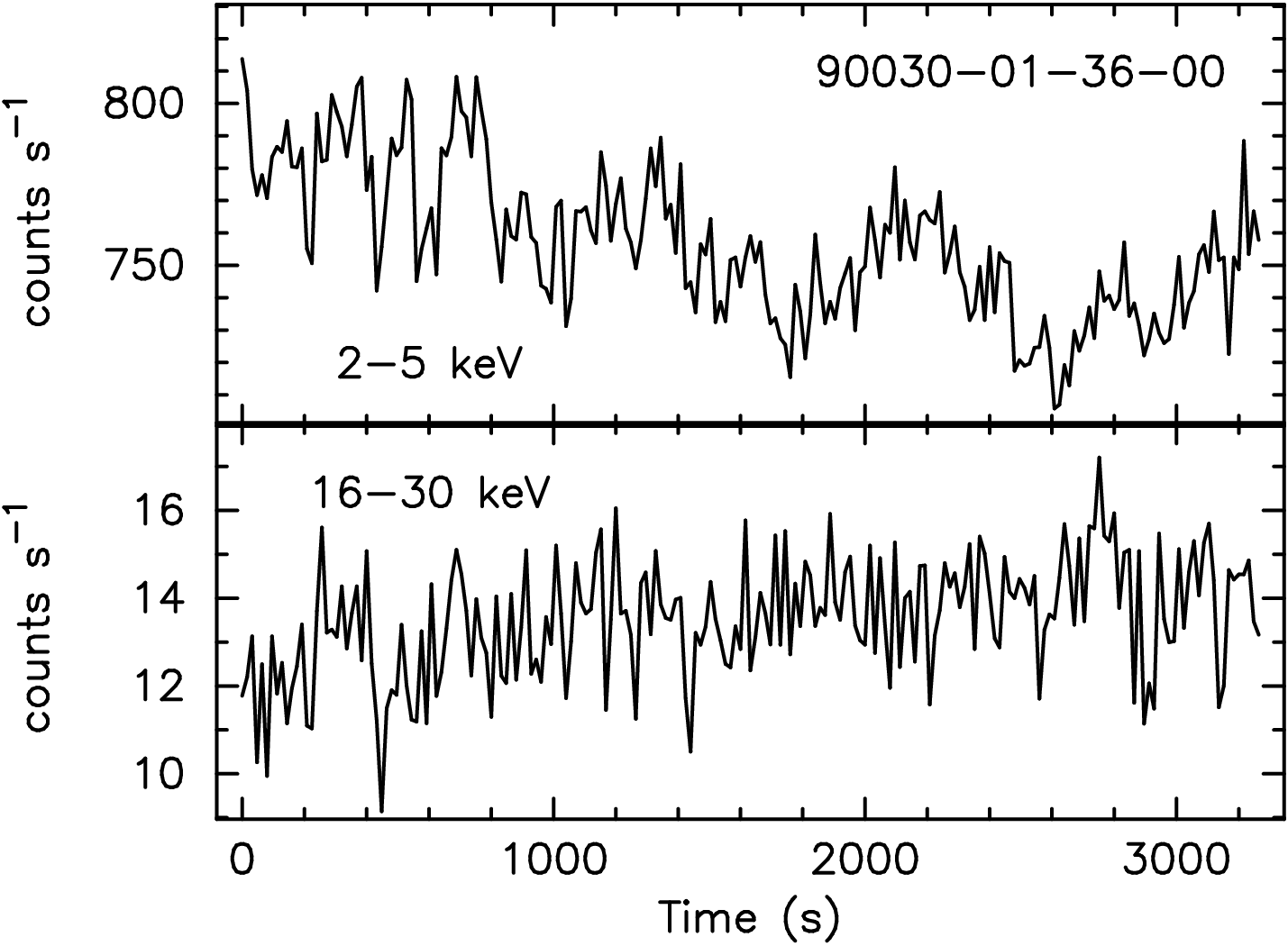}
\includegraphics[width=3.1cm,angle=270,clip]{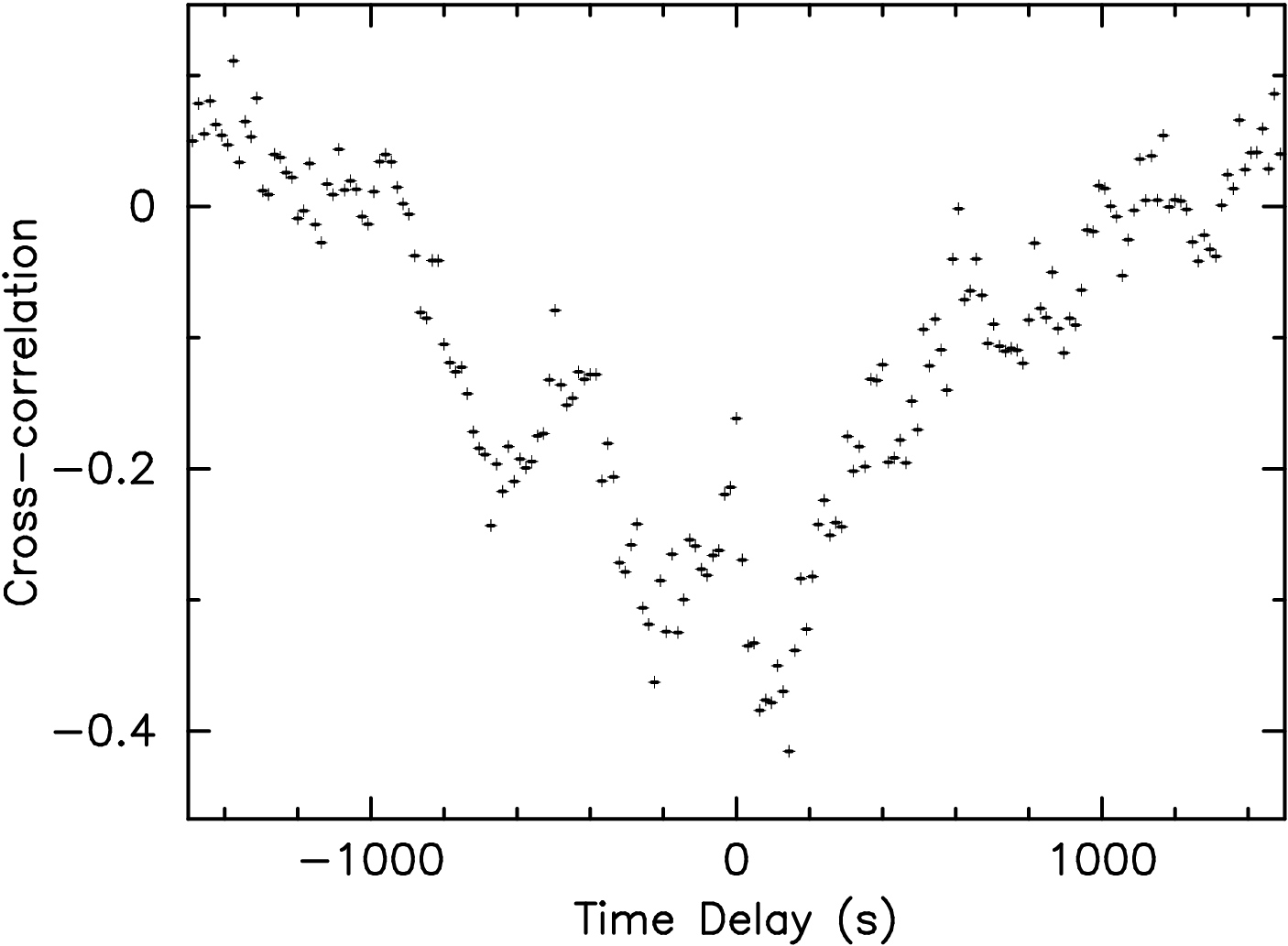}
\includegraphics[width=3.1cm,angle=270,clip]{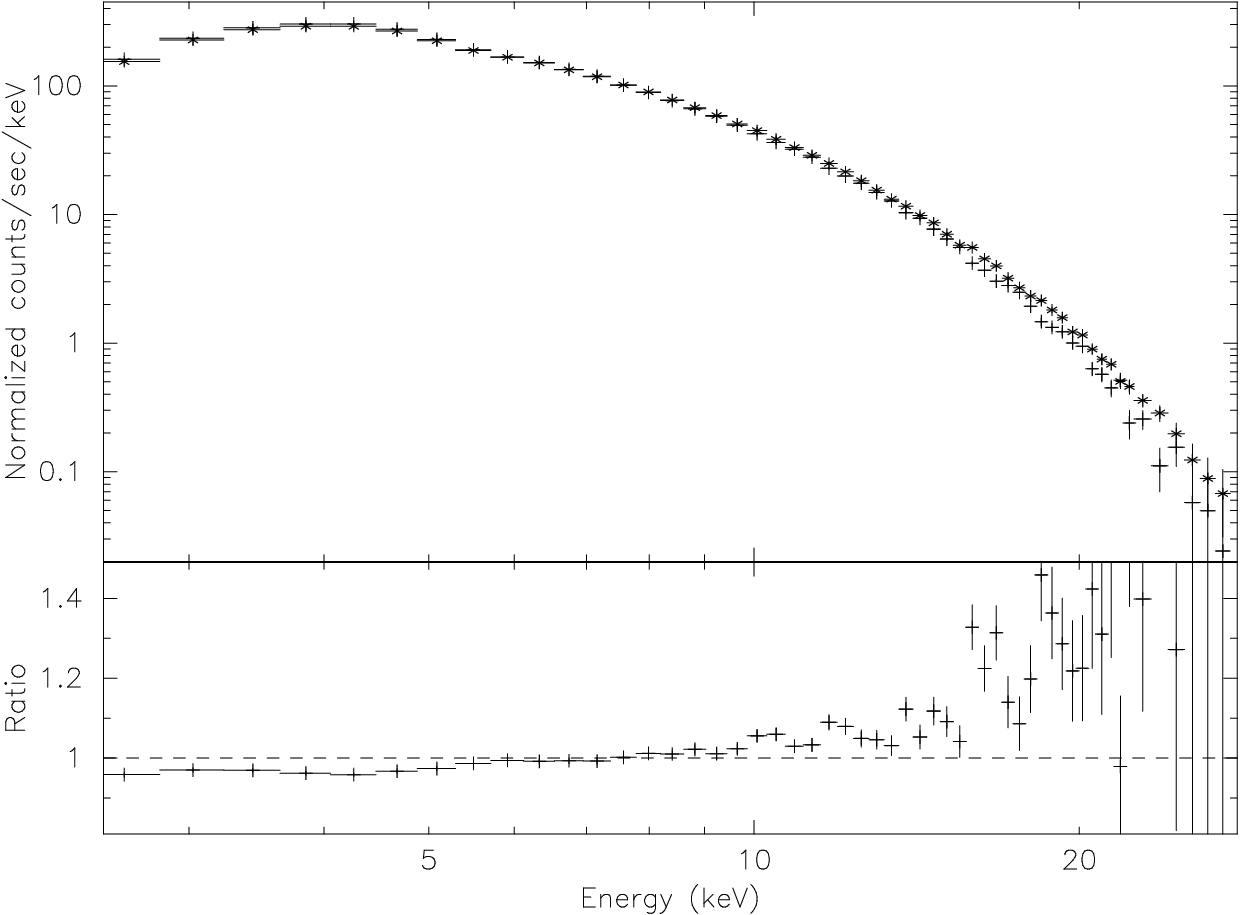}\\
\includegraphics[width=3.1cm,angle=270,clip]{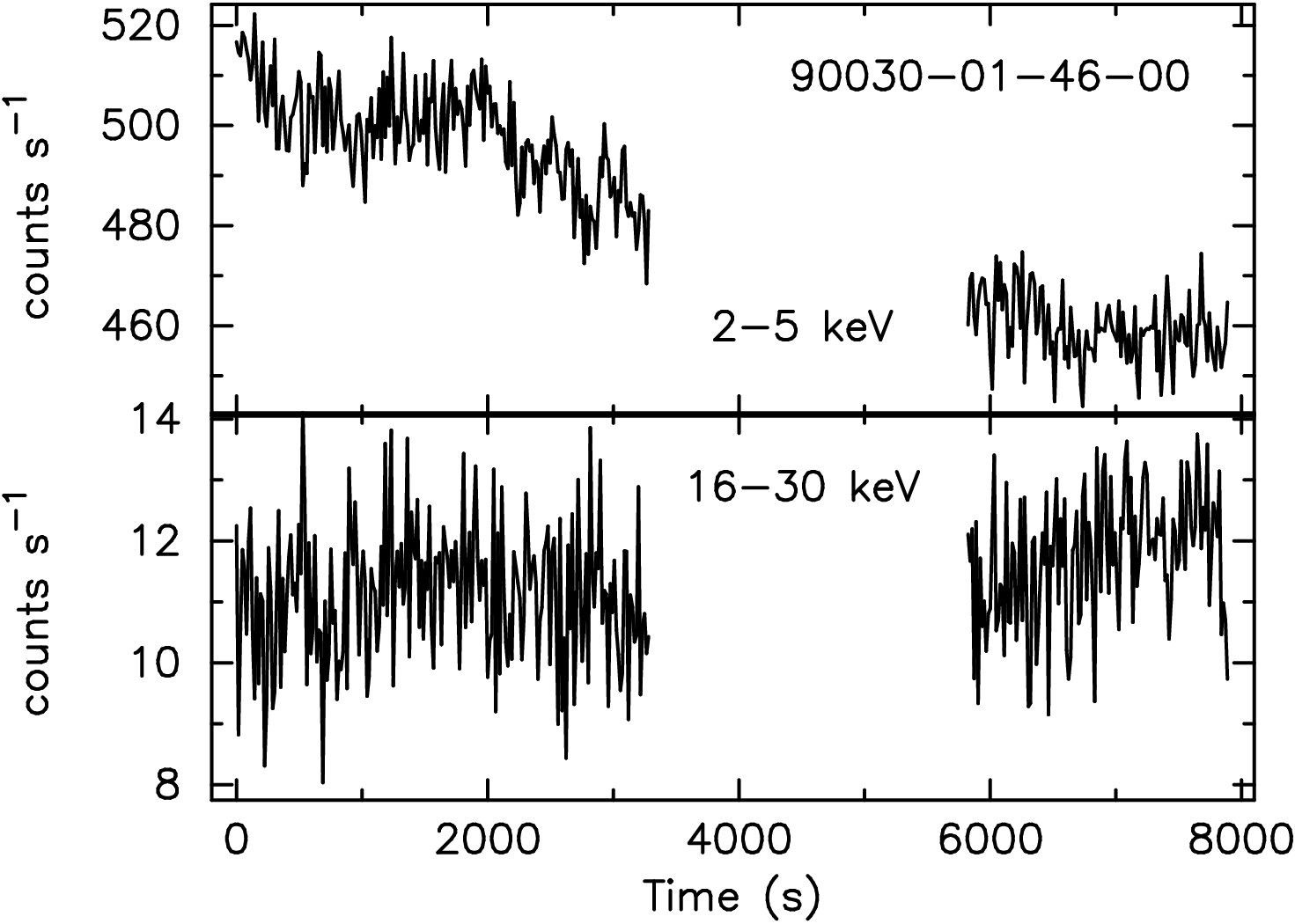}
\includegraphics[width=3.1cm,angle=270,clip]{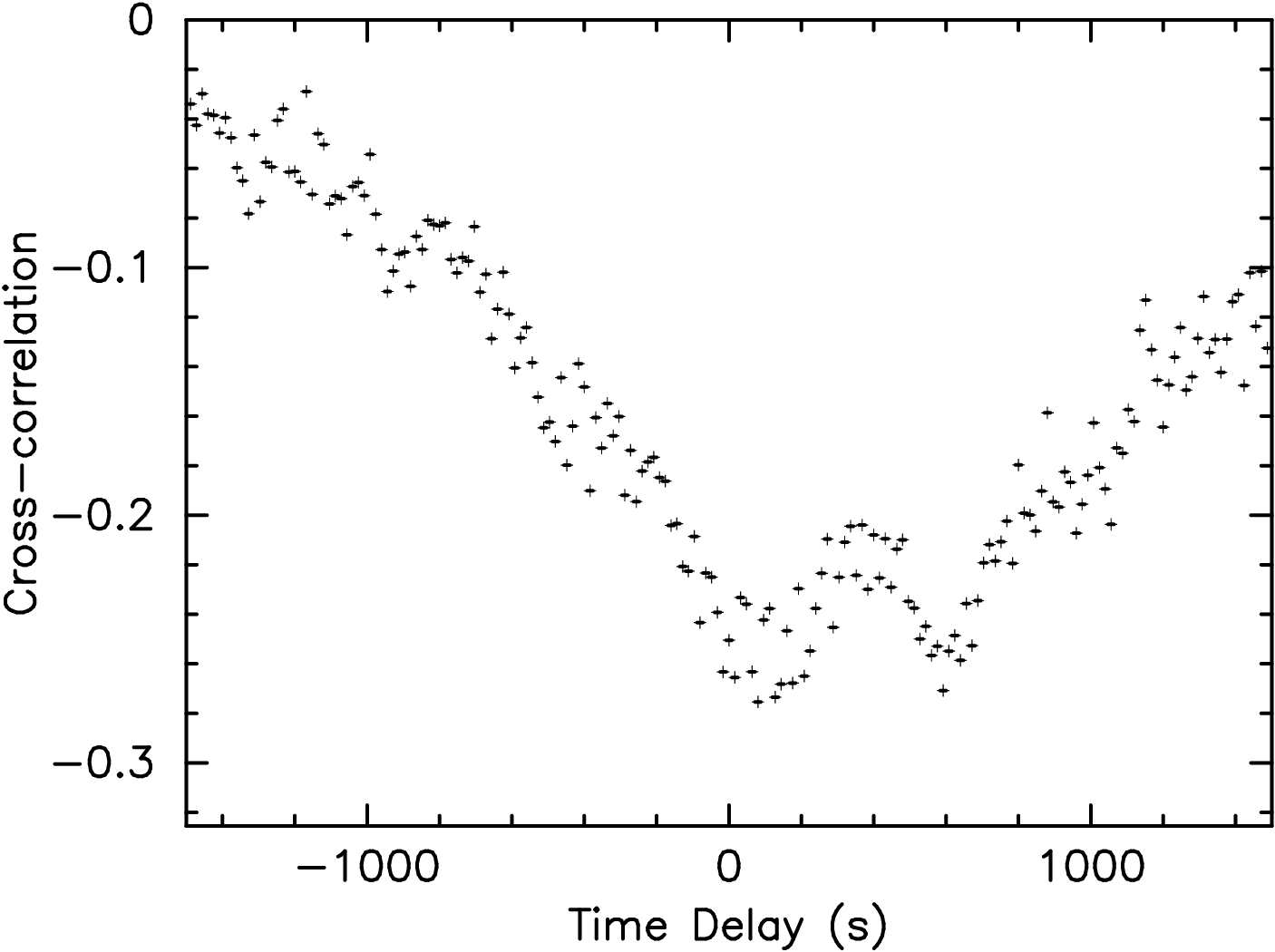}
\includegraphics[width=3.1cm,angle=270,clip]{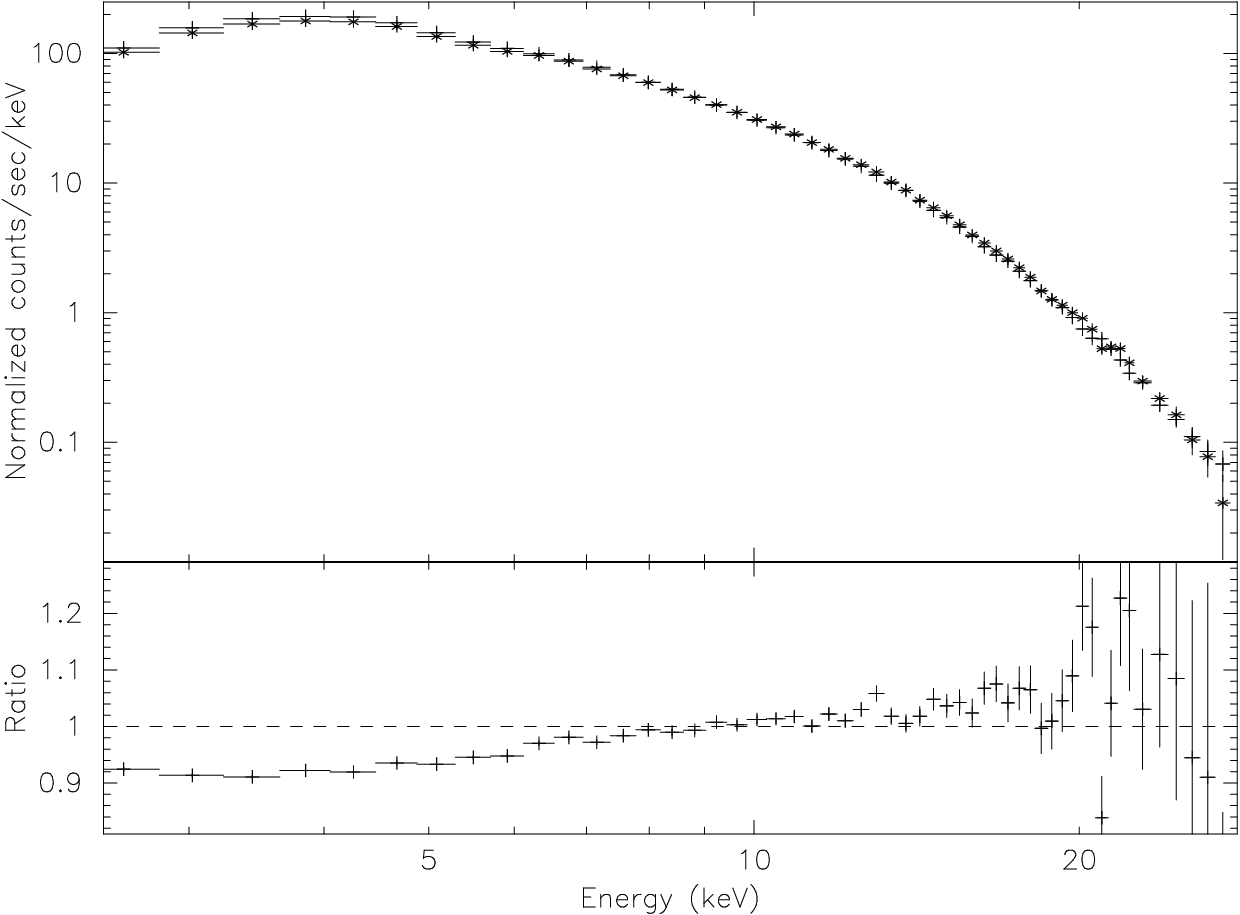}\\
\includegraphics[width=3.1cm,angle=270,clip]{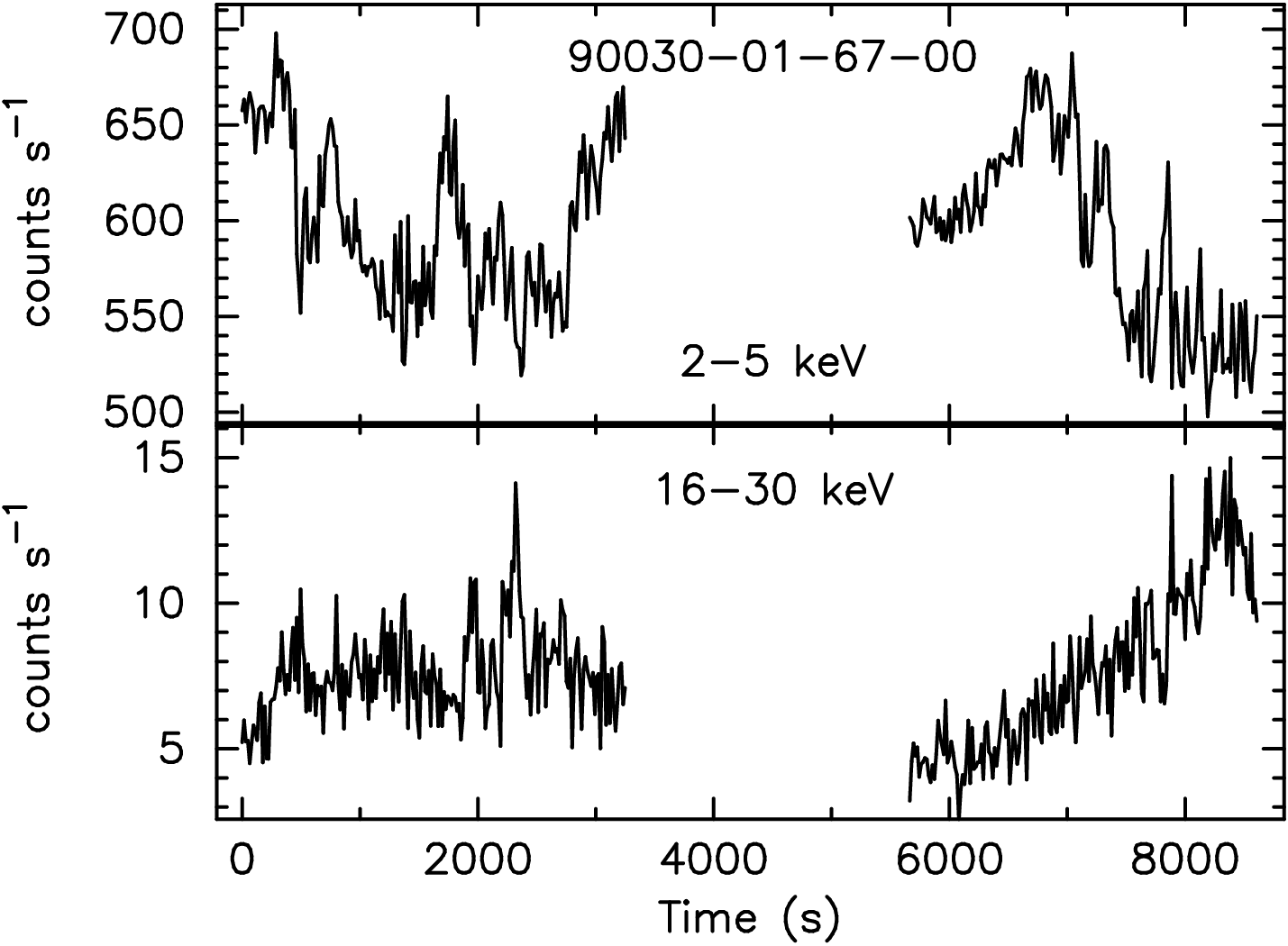}
\includegraphics[width=3.1cm,angle=270,clip]{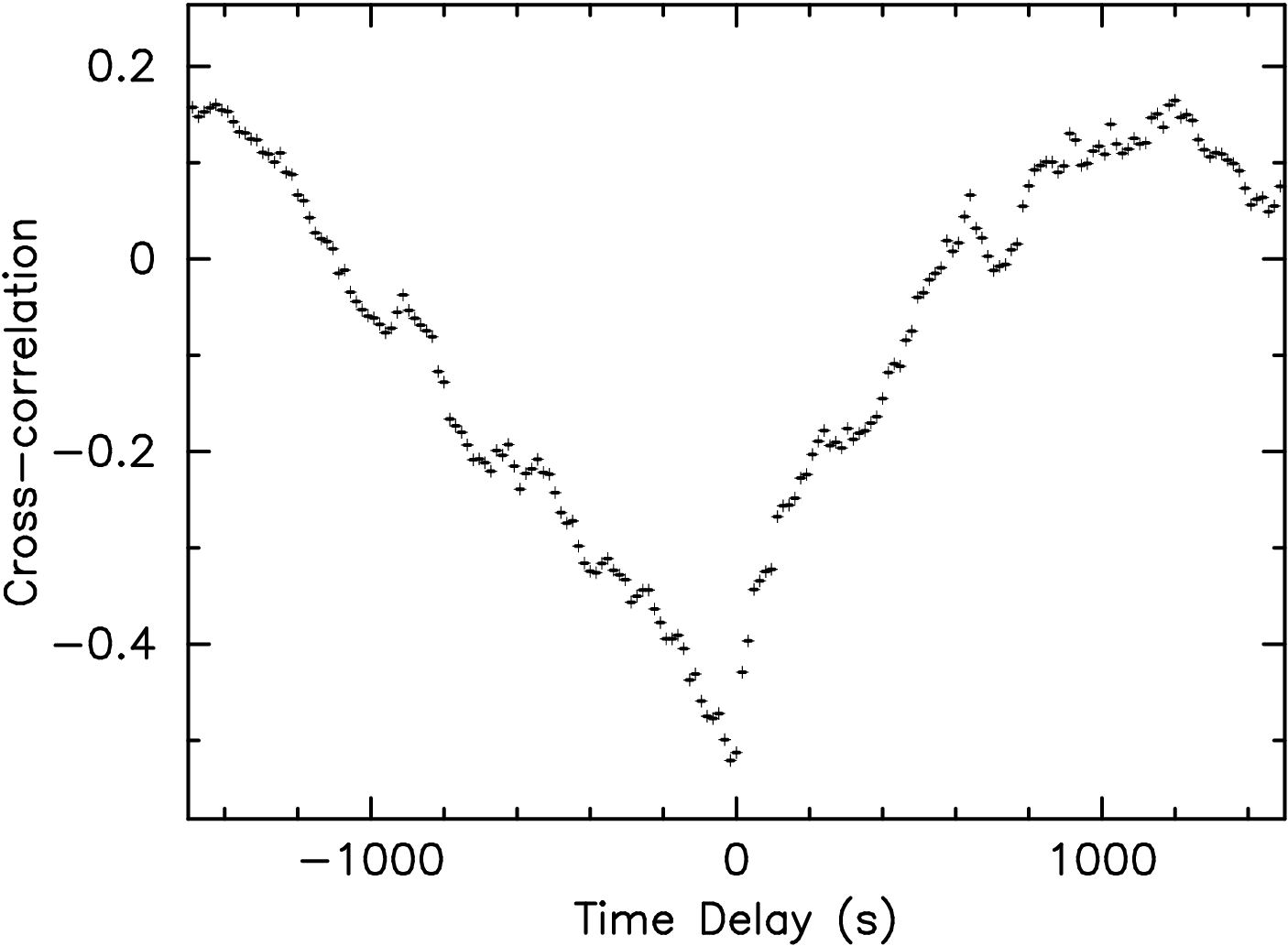}
\includegraphics[width=3.1cm,angle=270,clip]{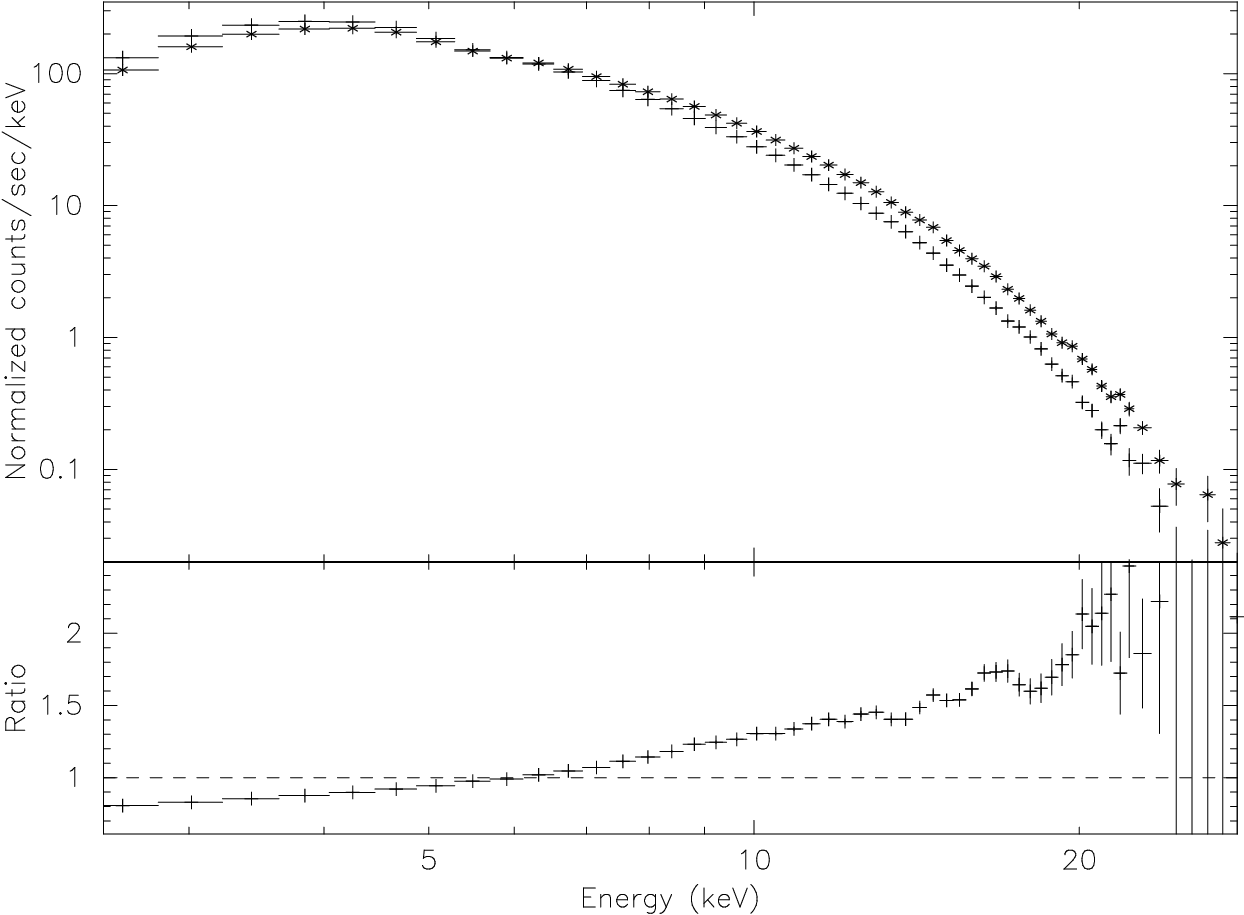}\\
\includegraphics[width=3.1cm,angle=270,clip]{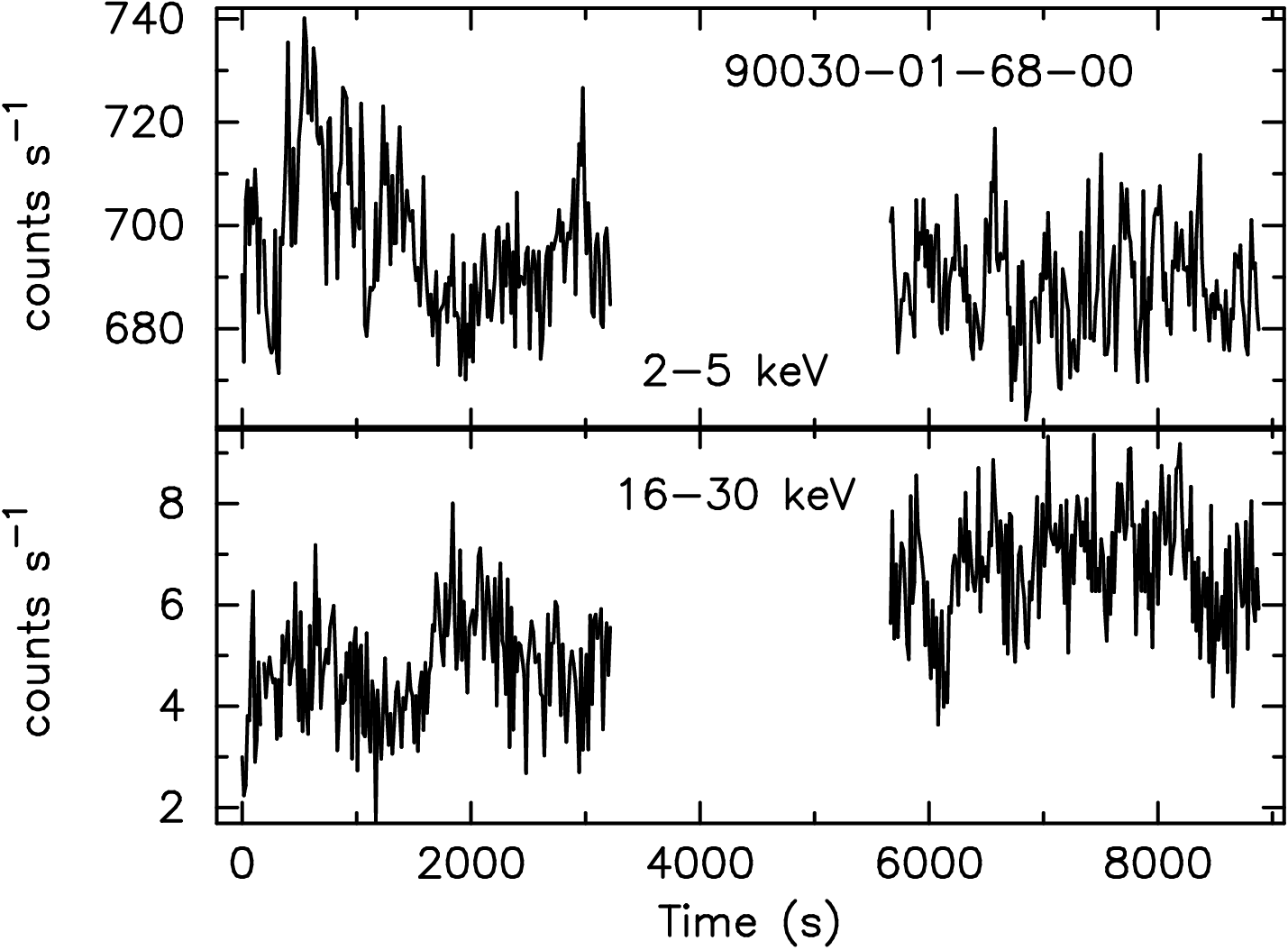}
\includegraphics[width=3.1cm,angle=270,clip]{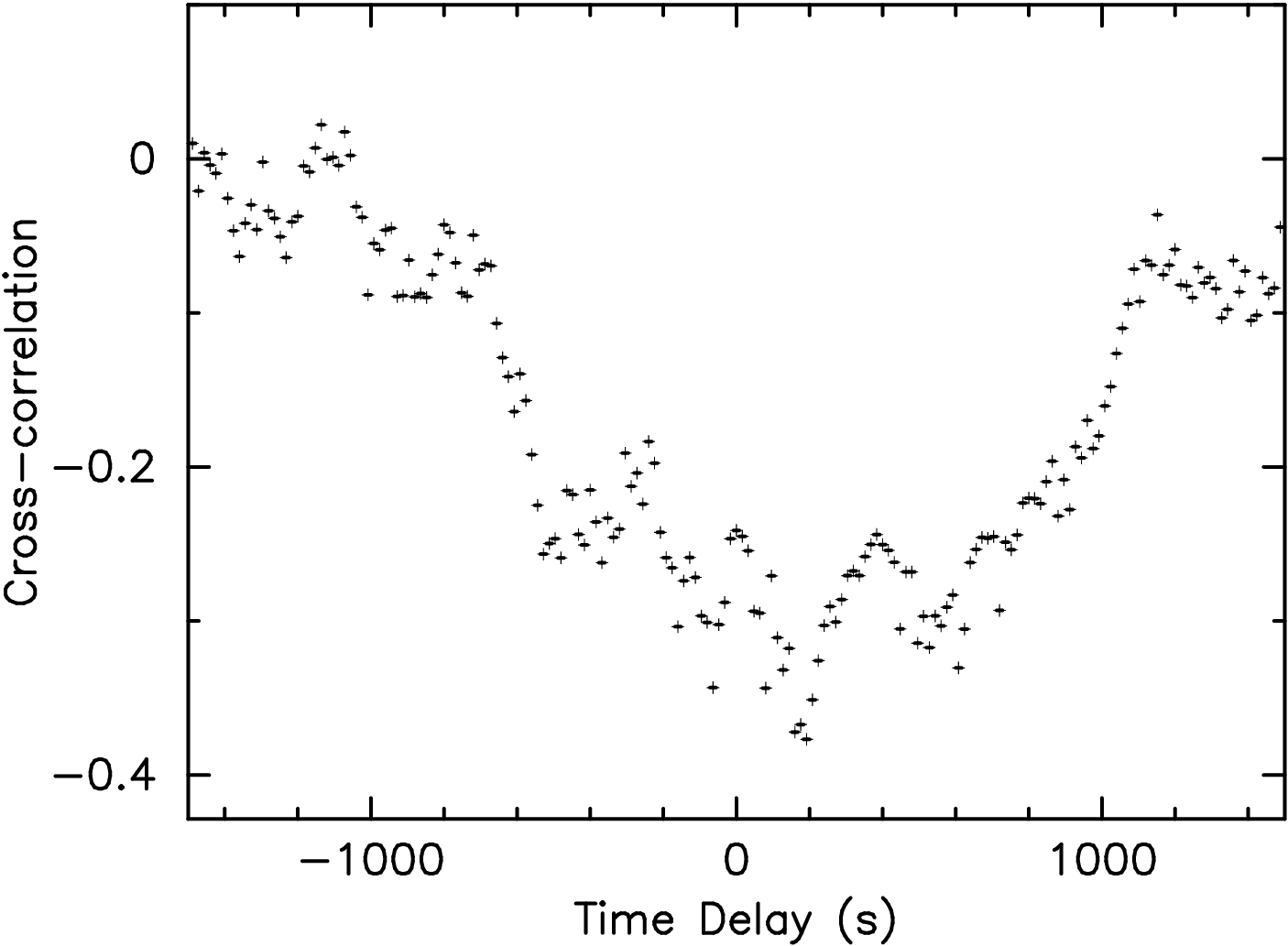}
\includegraphics[width=3.1cm,angle=270,clip]{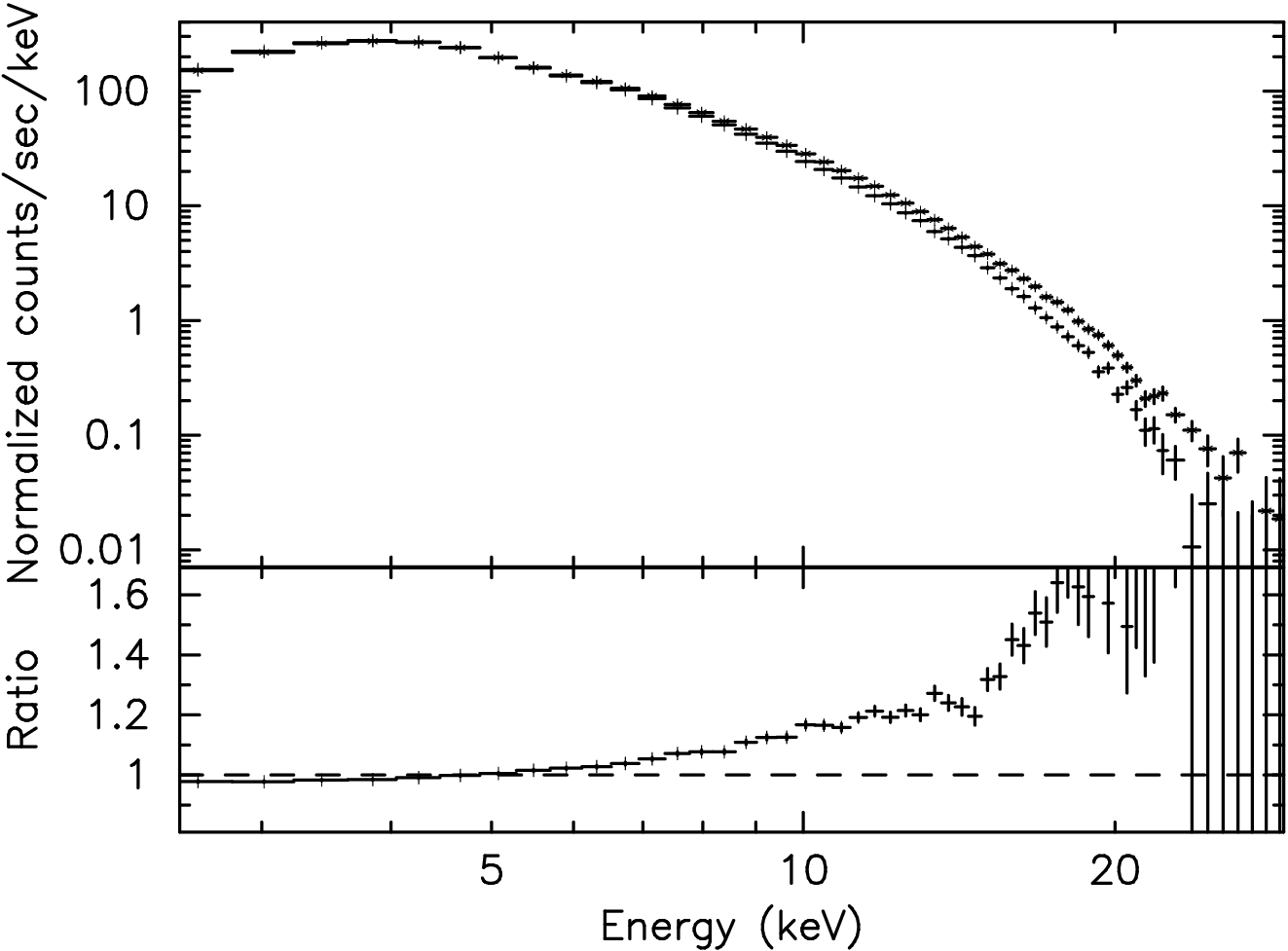}\\
\includegraphics[width=3.1cm,angle=270,clip]{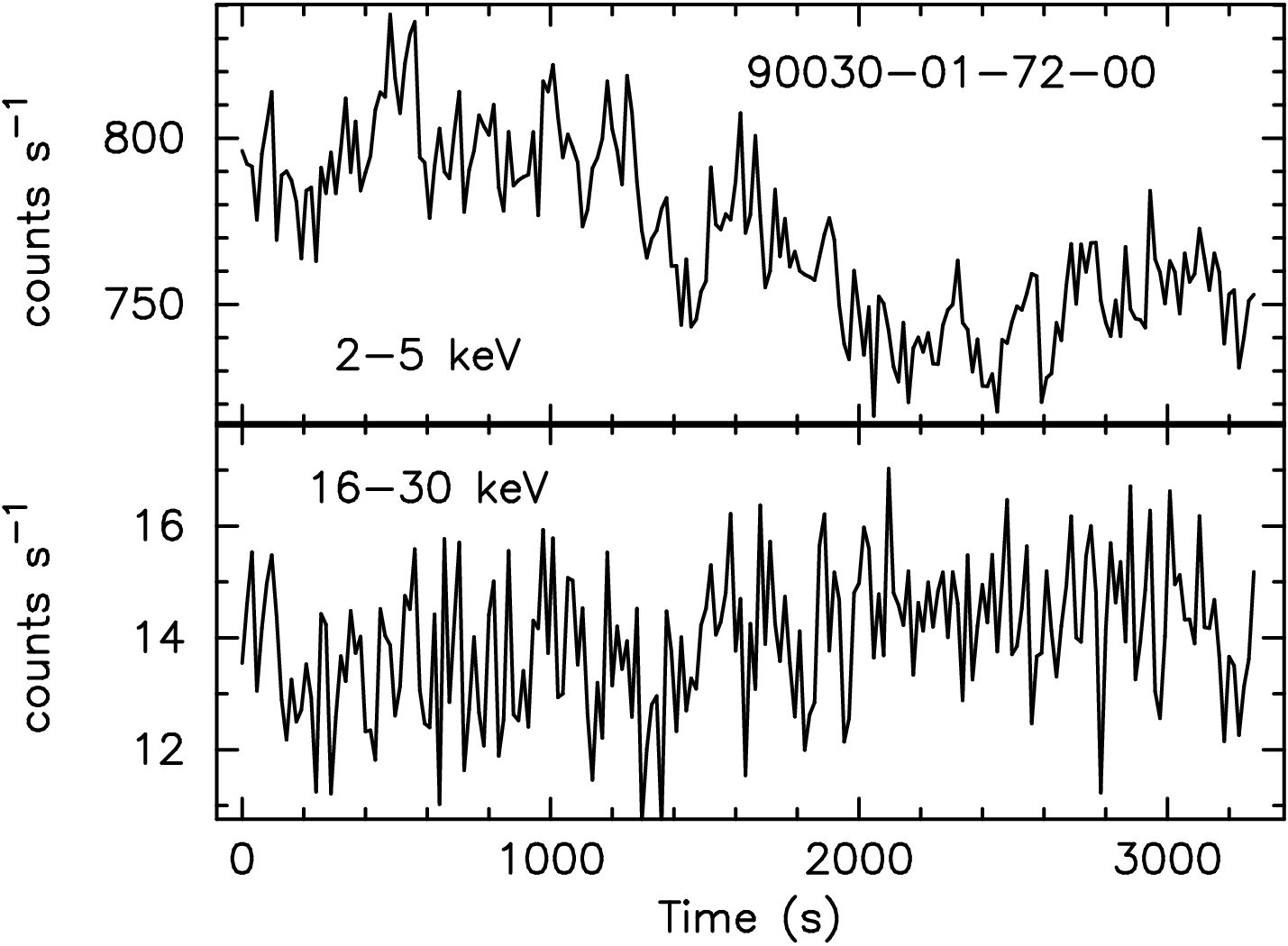}
\includegraphics[width=3.1cm,angle=270,clip]{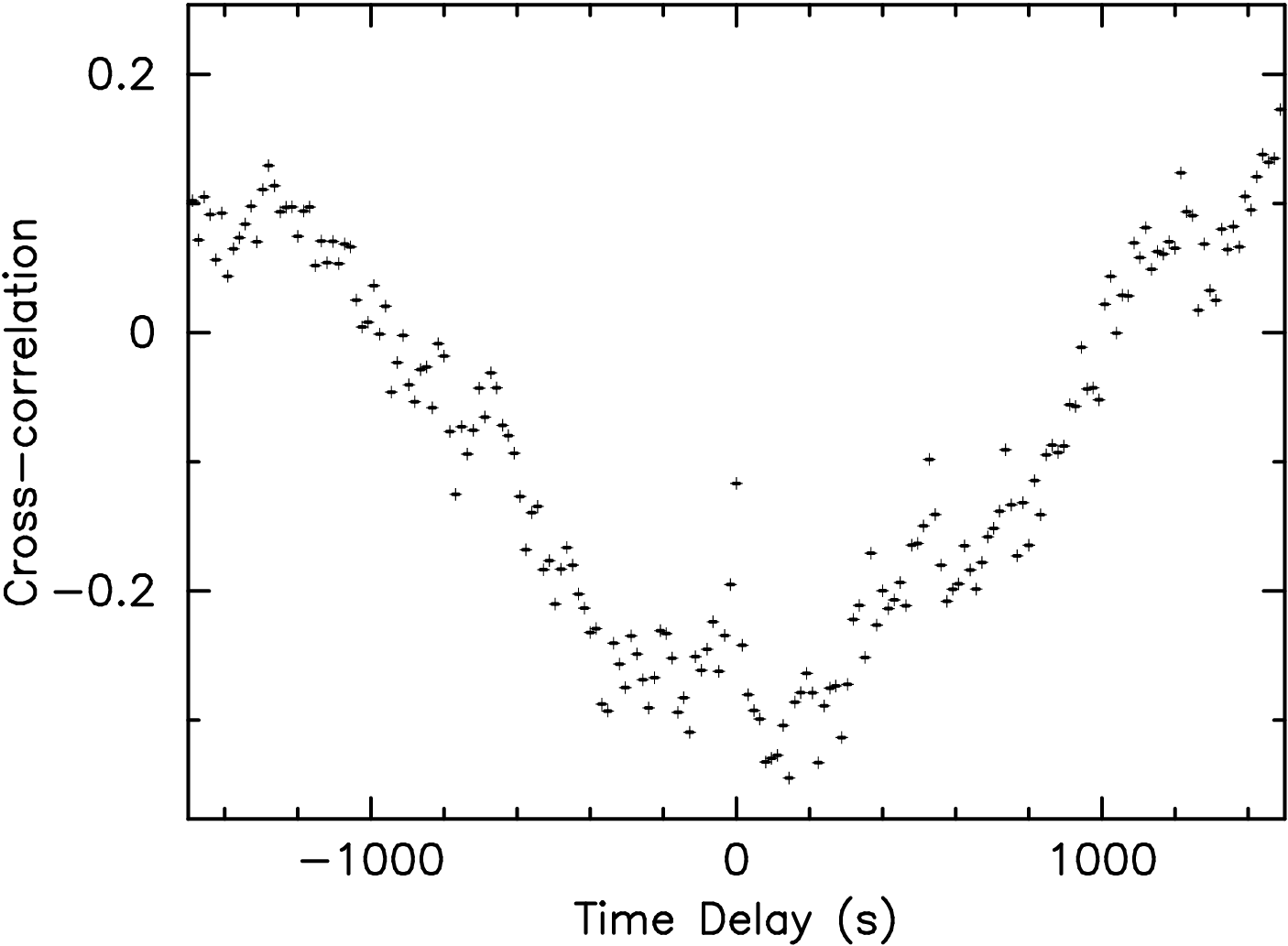}
\includegraphics[width=3.1cm,angle=270,clip]{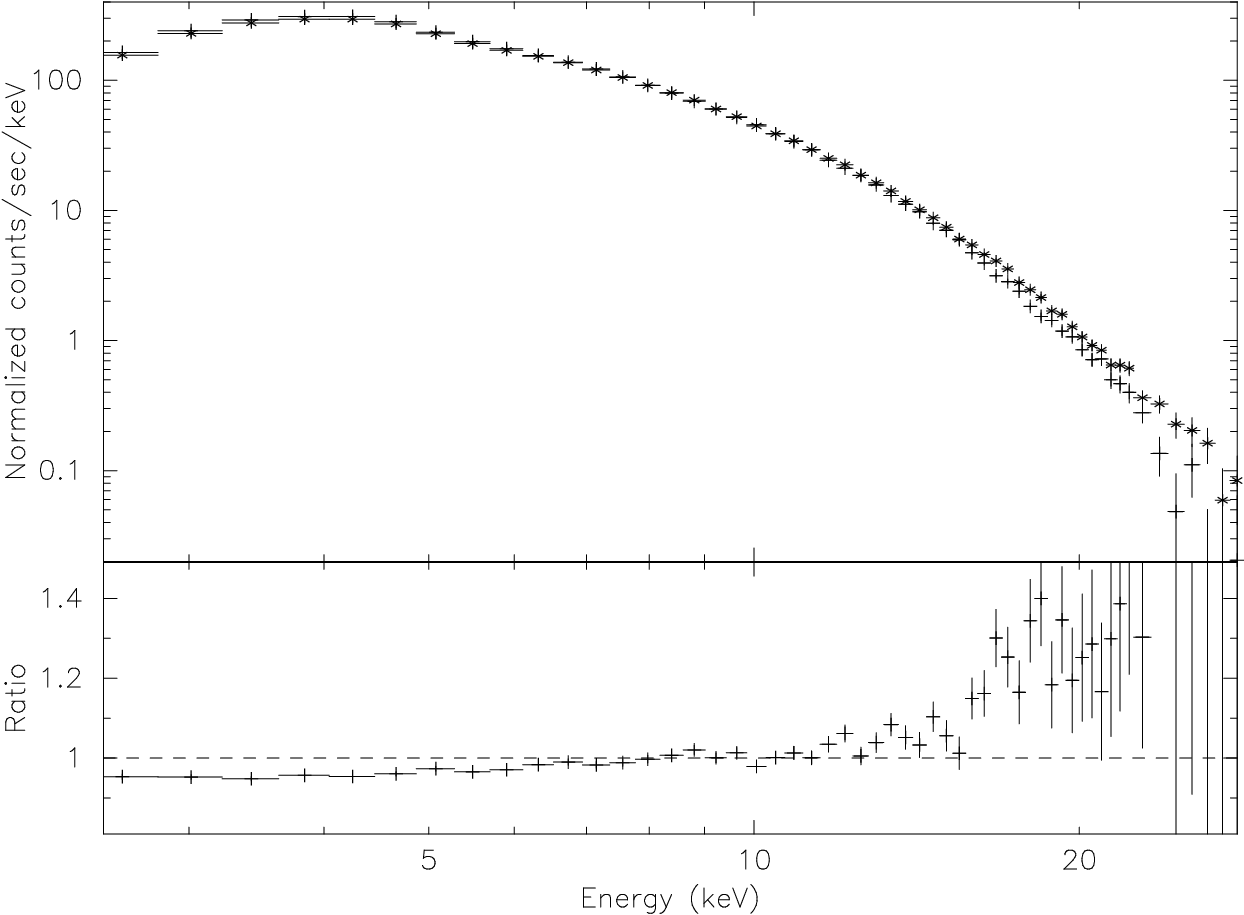}\\
\includegraphics[width=3.1cm,angle=270,clip]{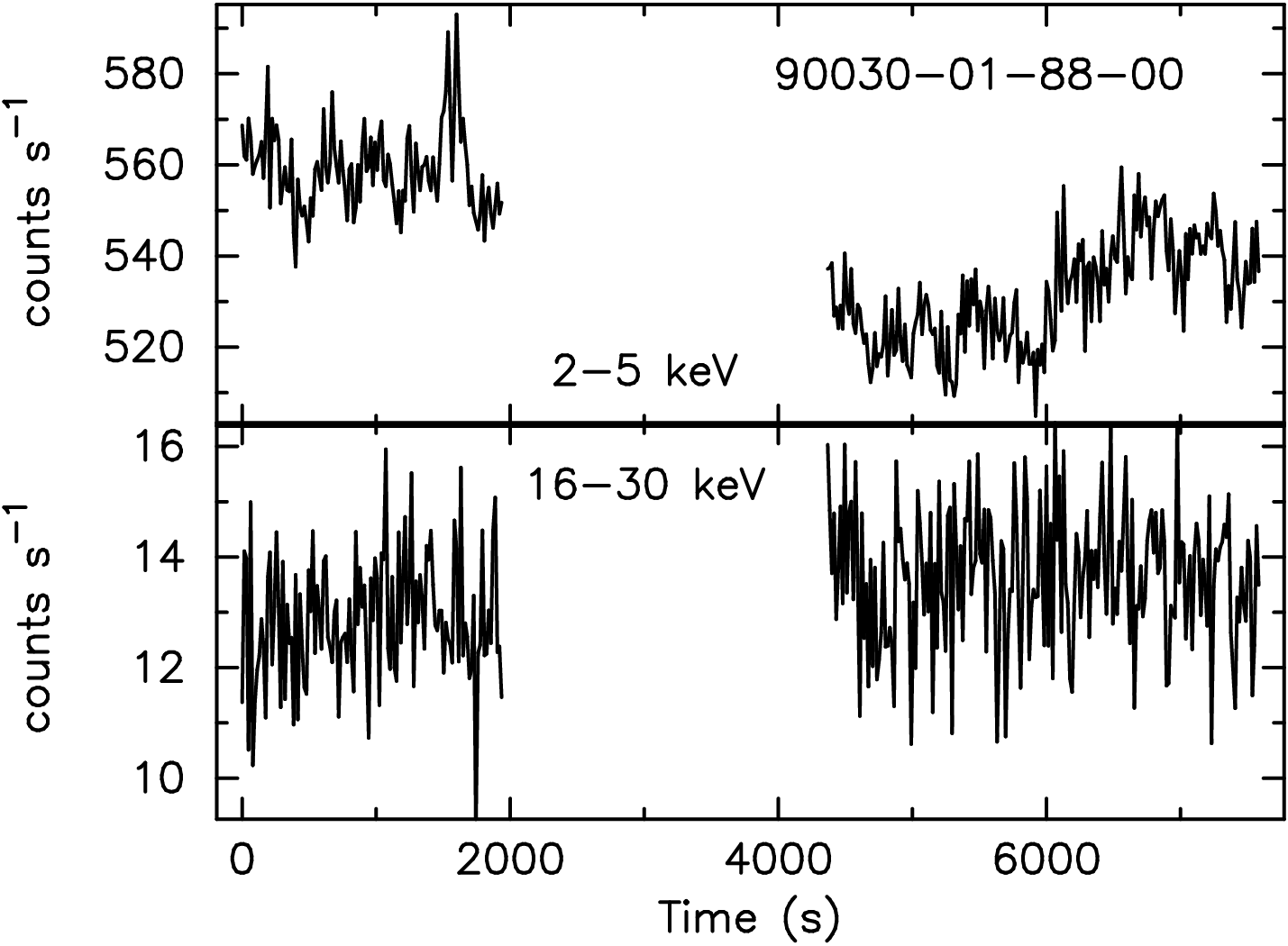}
\includegraphics[width=3.1cm,angle=270,clip]{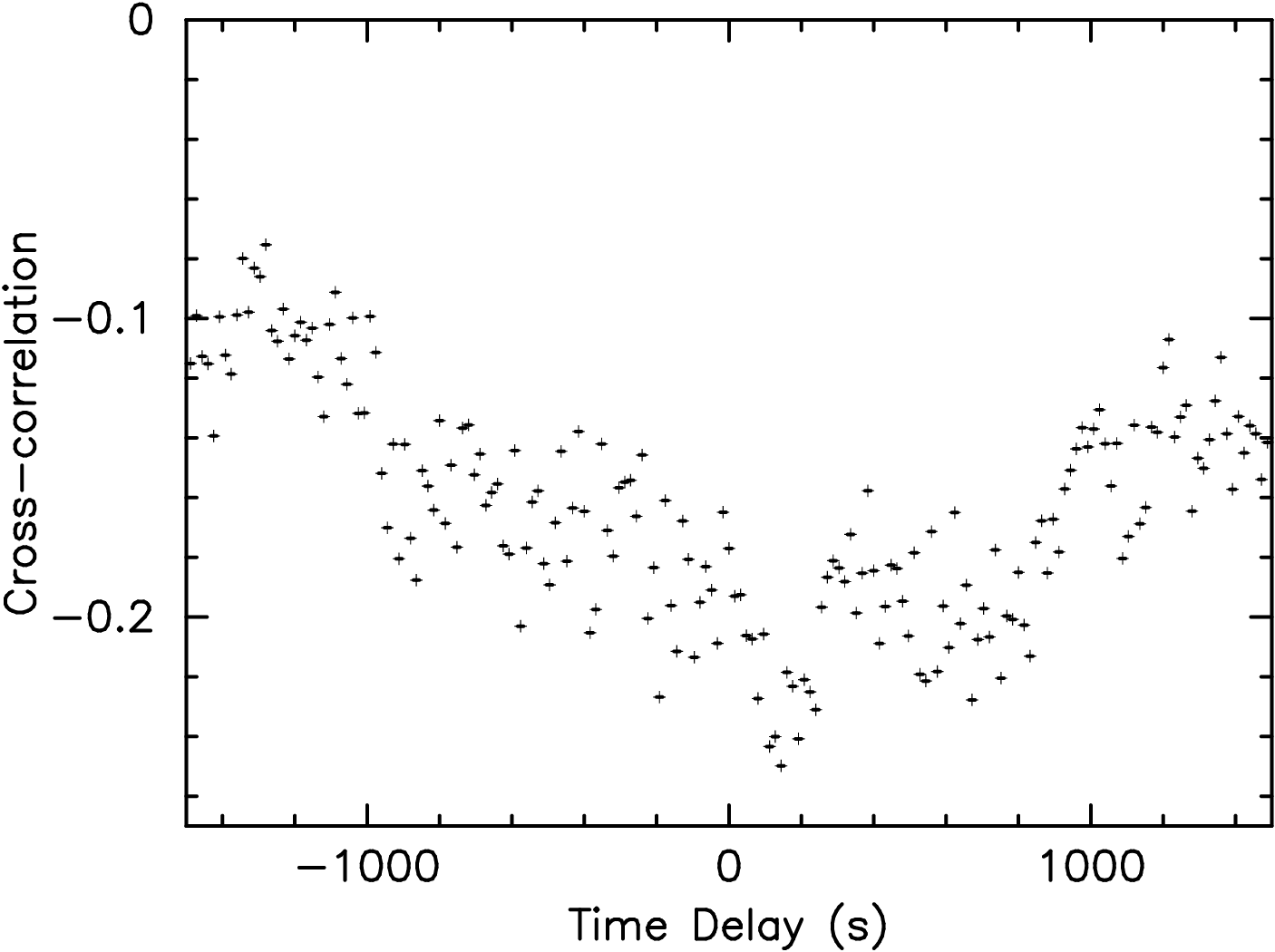}
\includegraphics[width=3.1cm,angle=270,clip]{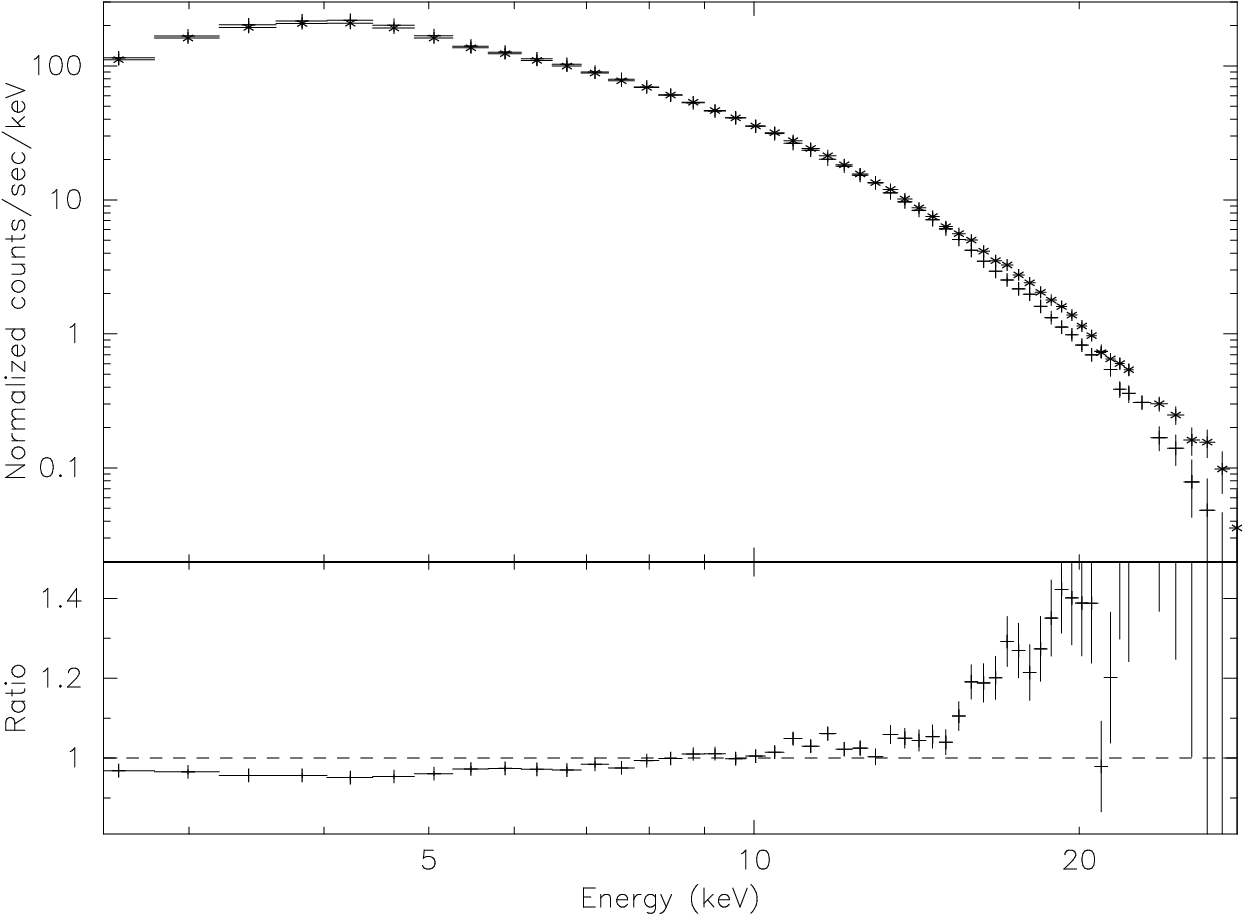}\\
\includegraphics[width=3.1cm,angle=270,clip]{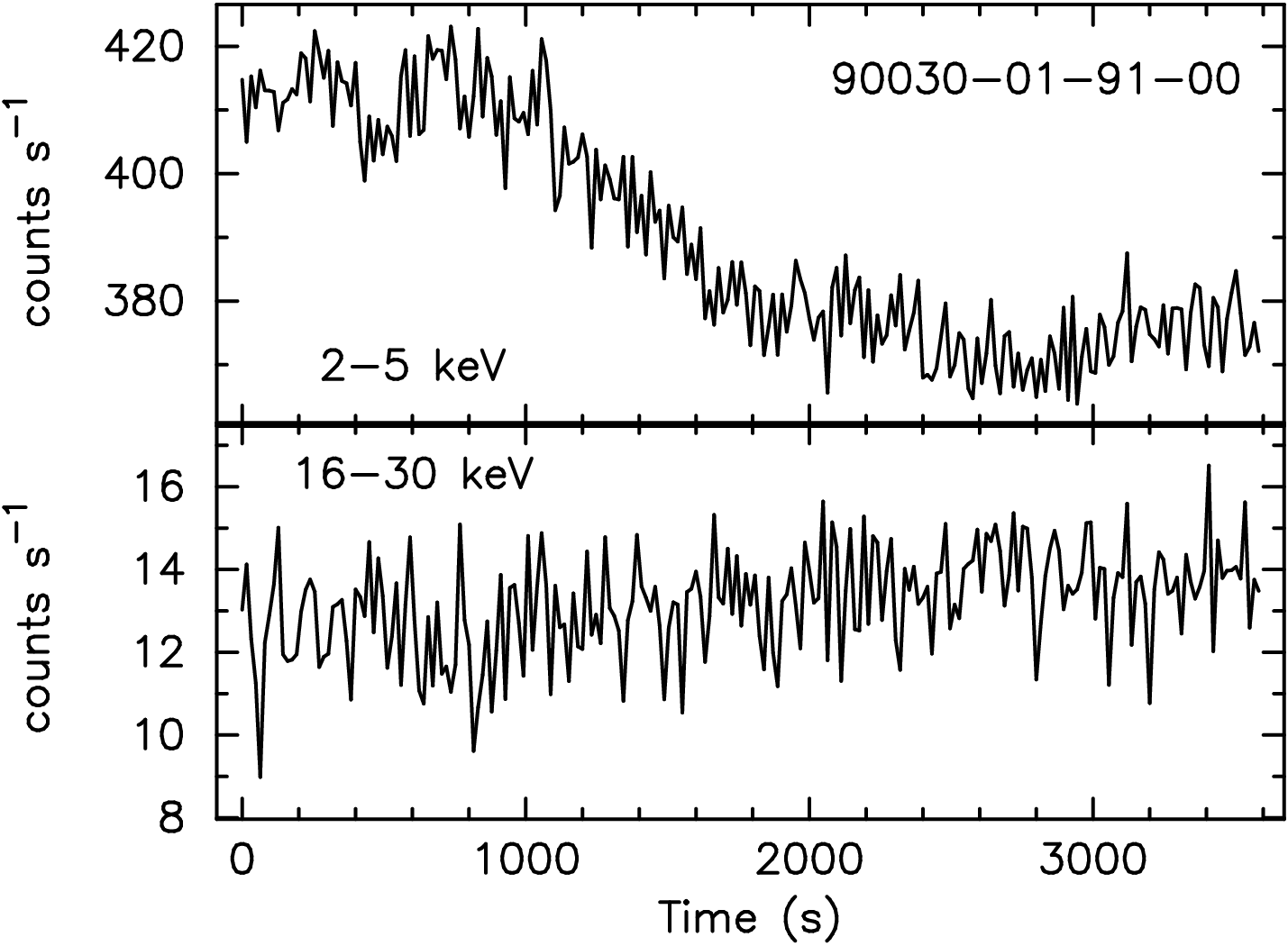}
\includegraphics[width=3.1cm,angle=270,clip]{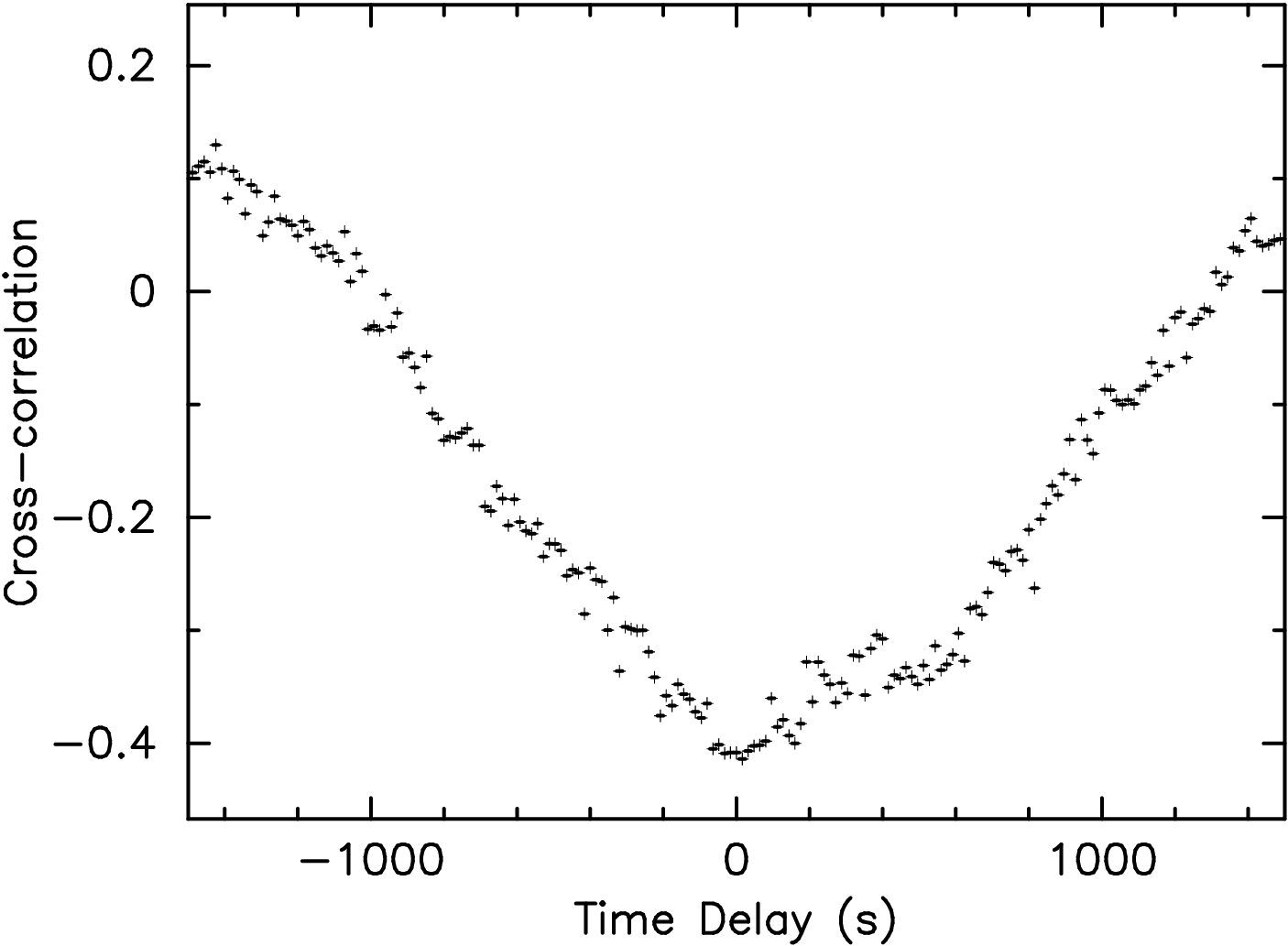}
\includegraphics[width=3.1cm,angle=270,clip]{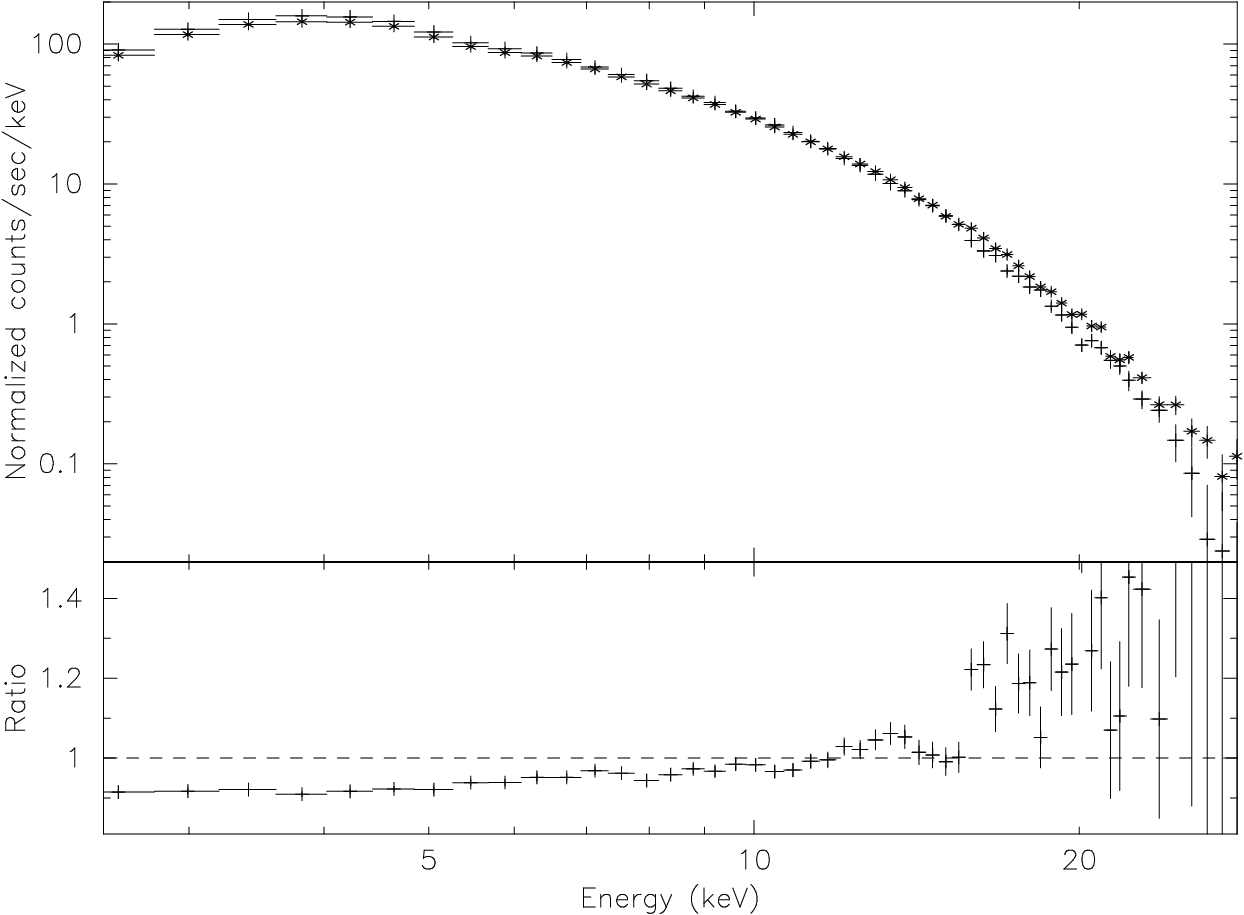}\\[5mm]
\centerline{Fig. 4. --- Continued}
\end{center}
\clearpage
\begin{center}
\includegraphics[width=3.1cm,angle=270,clip]{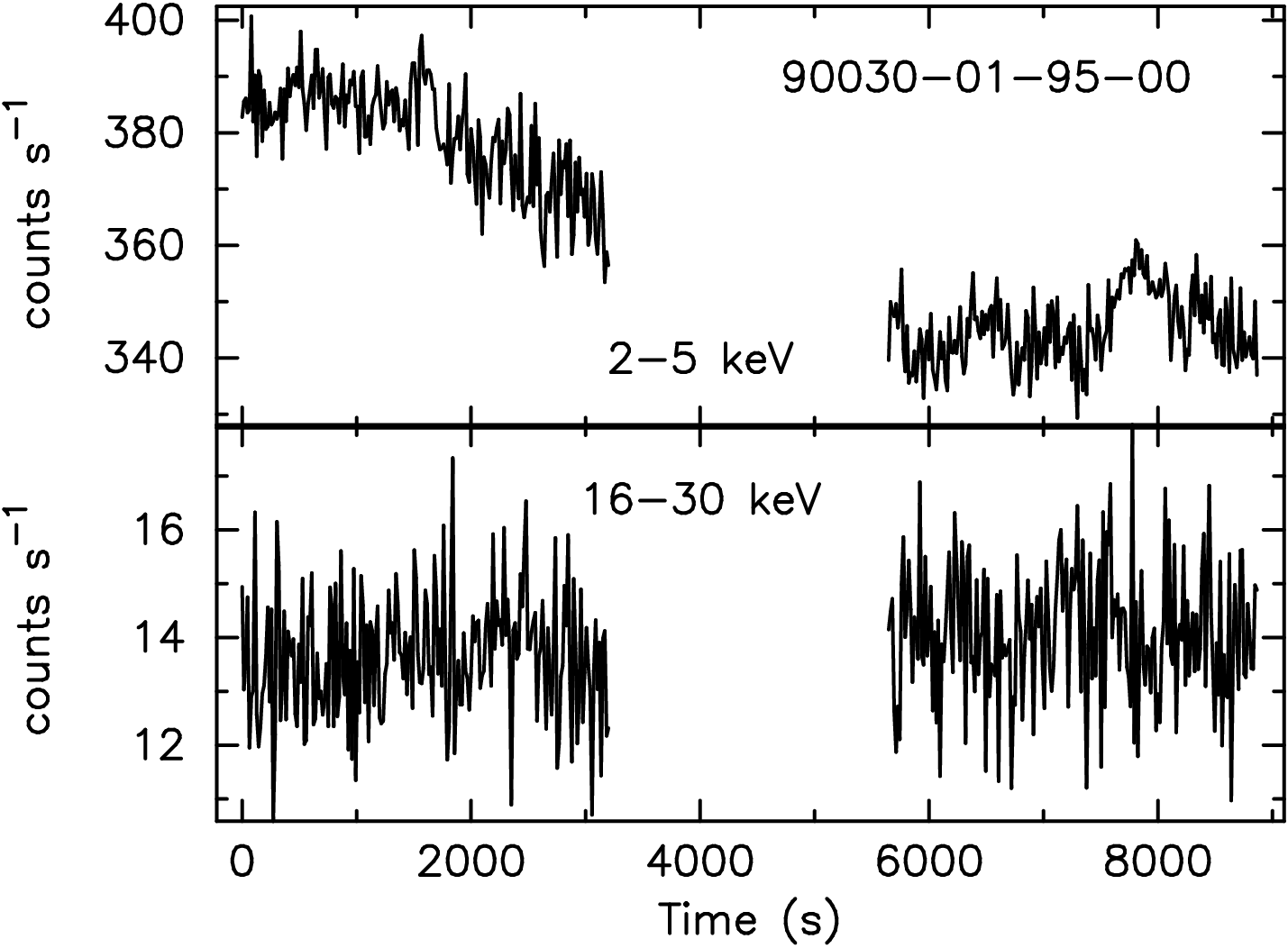}
\includegraphics[width=3.1cm,angle=270,clip]{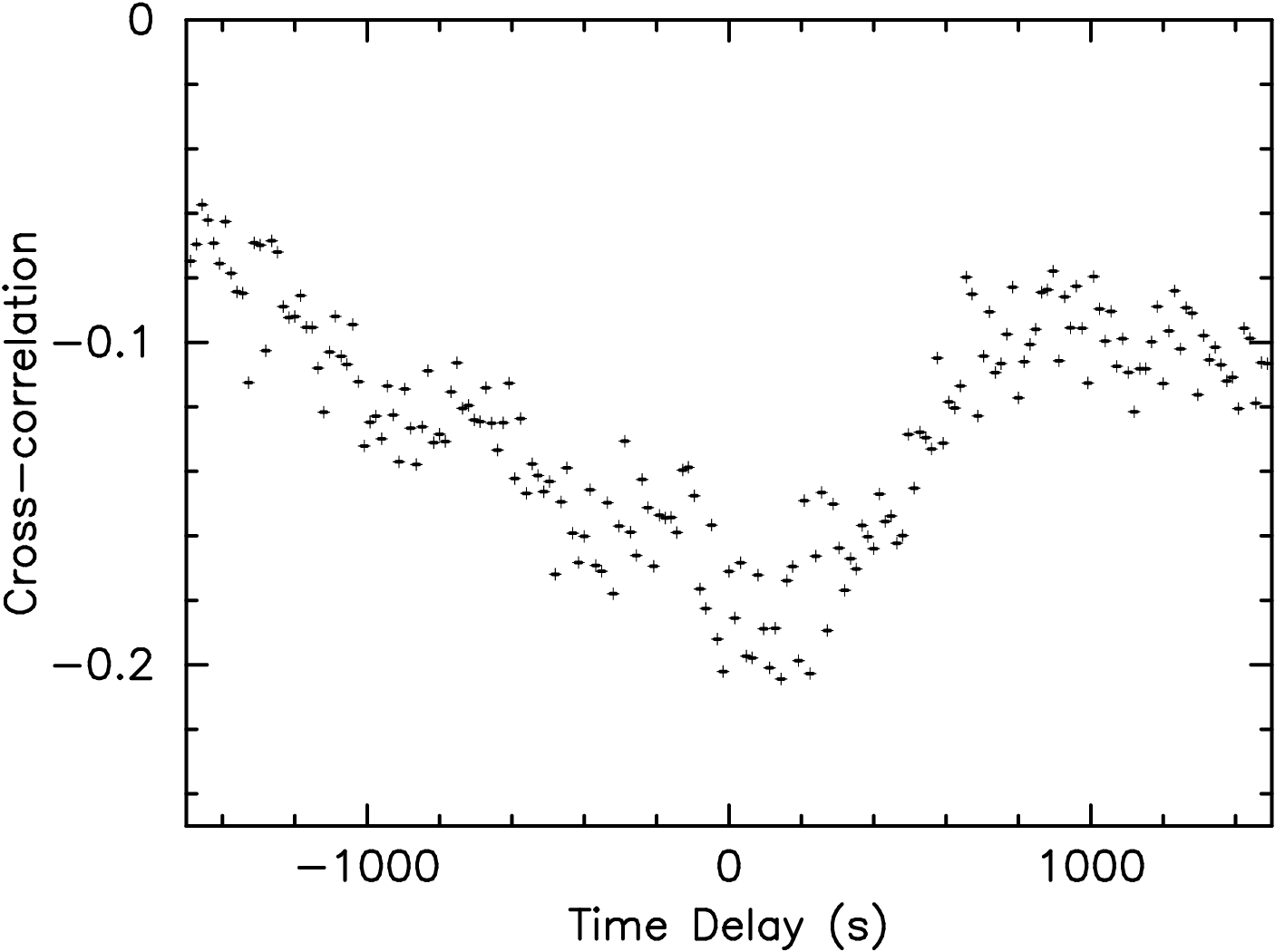}
\includegraphics[width=3.1cm,angle=270,clip]{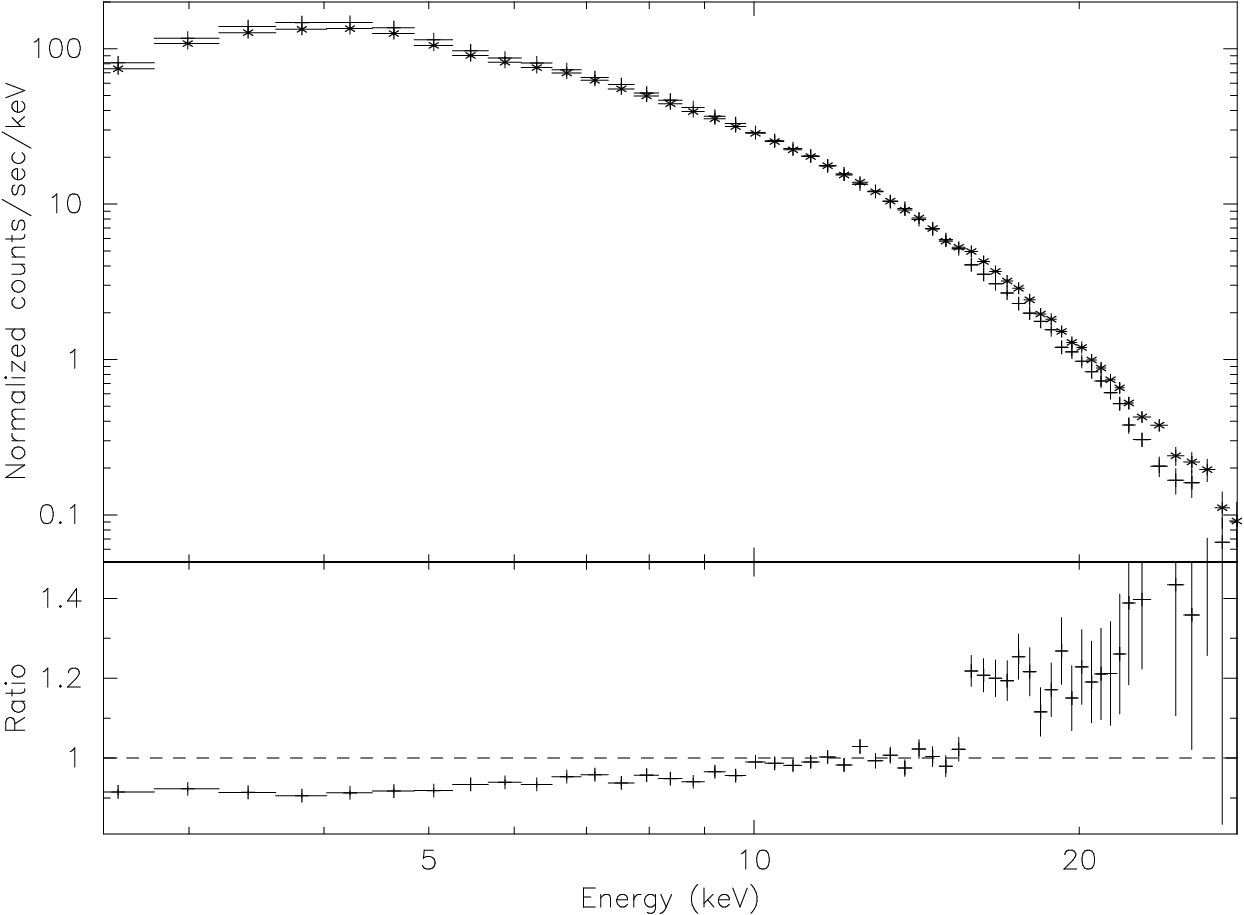}\\
\includegraphics[width=3.1cm,angle=270,clip]{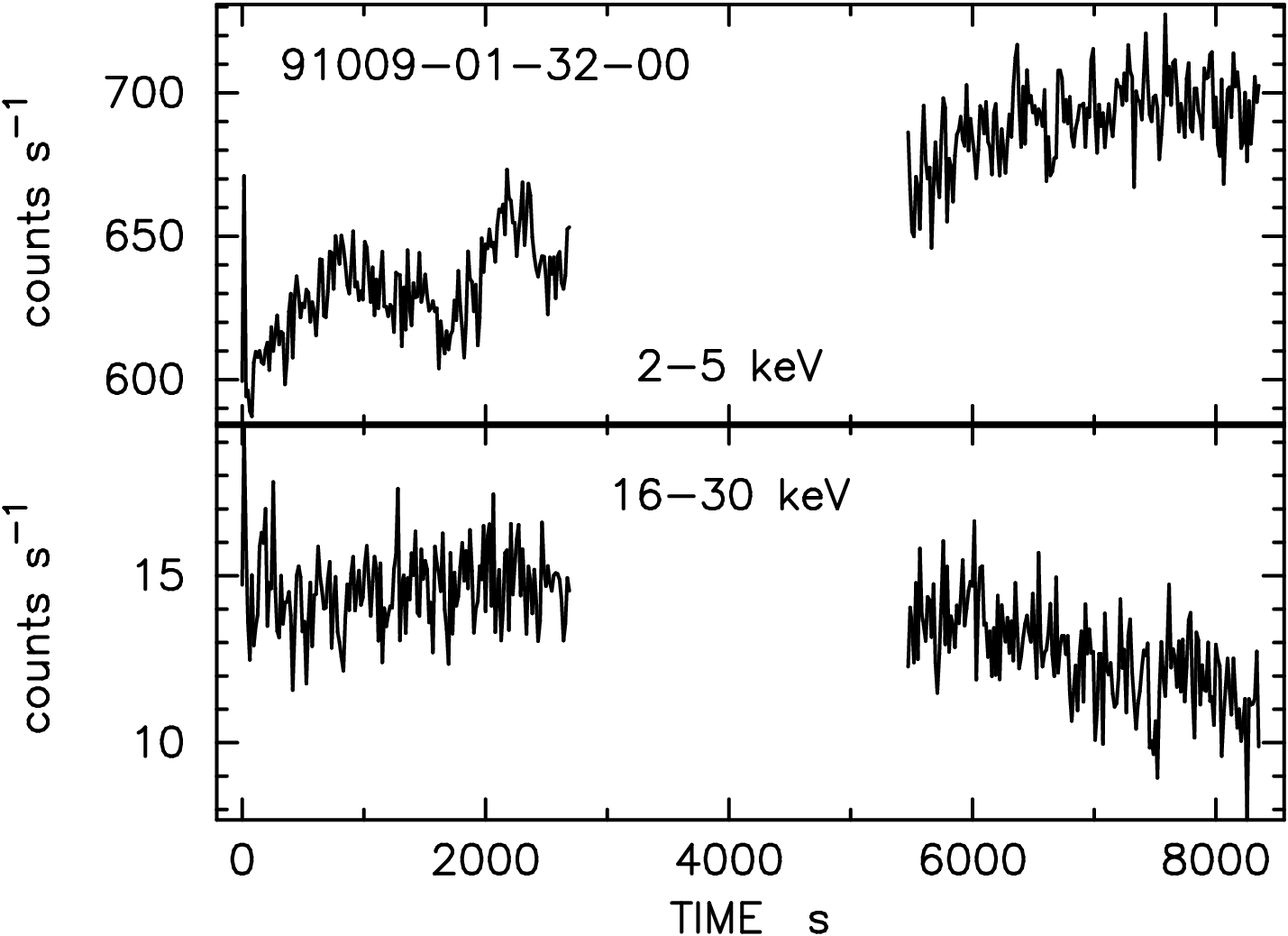}
\includegraphics[width=3.1cm,angle=270,clip]{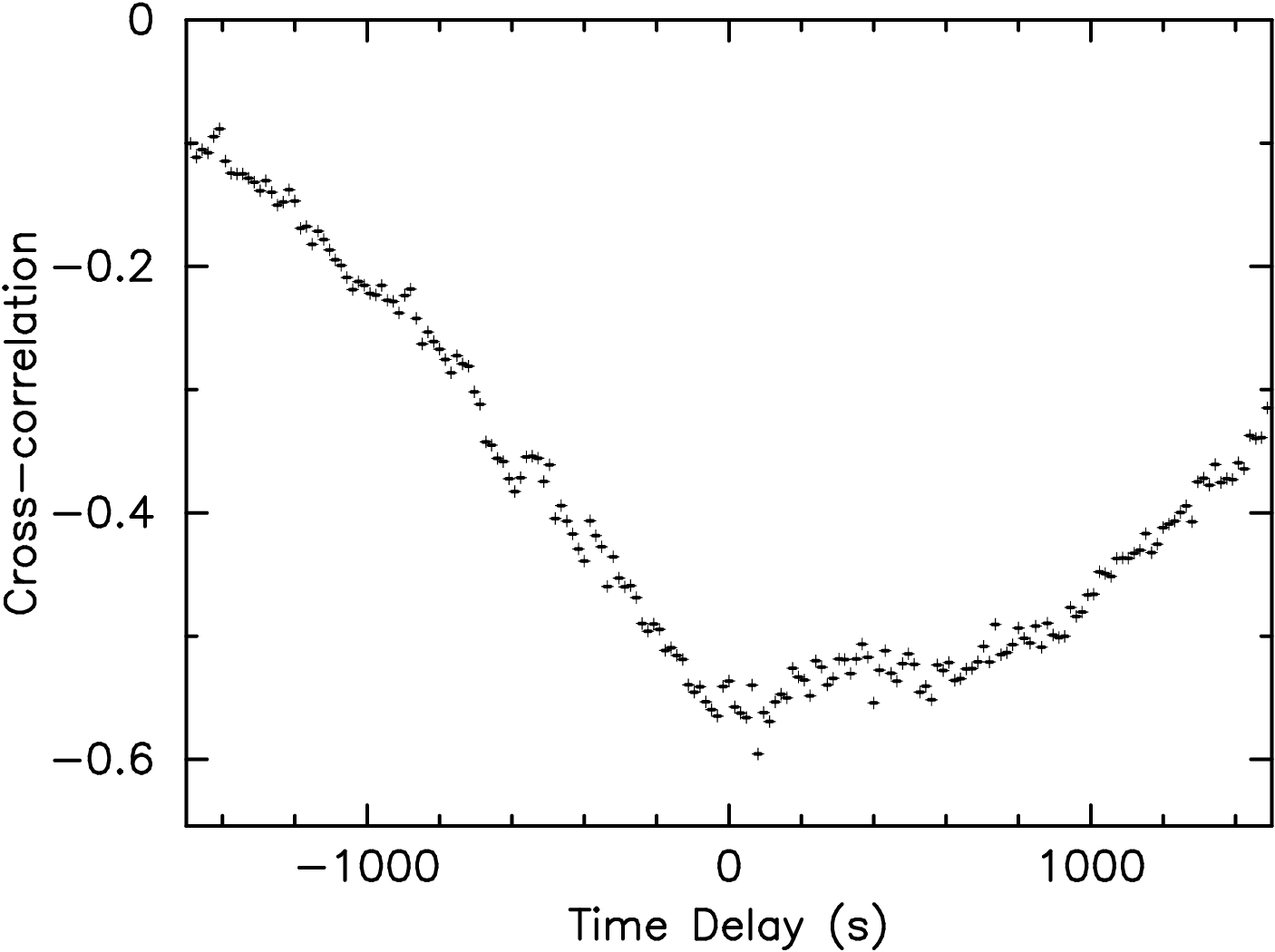}
\includegraphics[width=3.1cm,angle=270,clip]{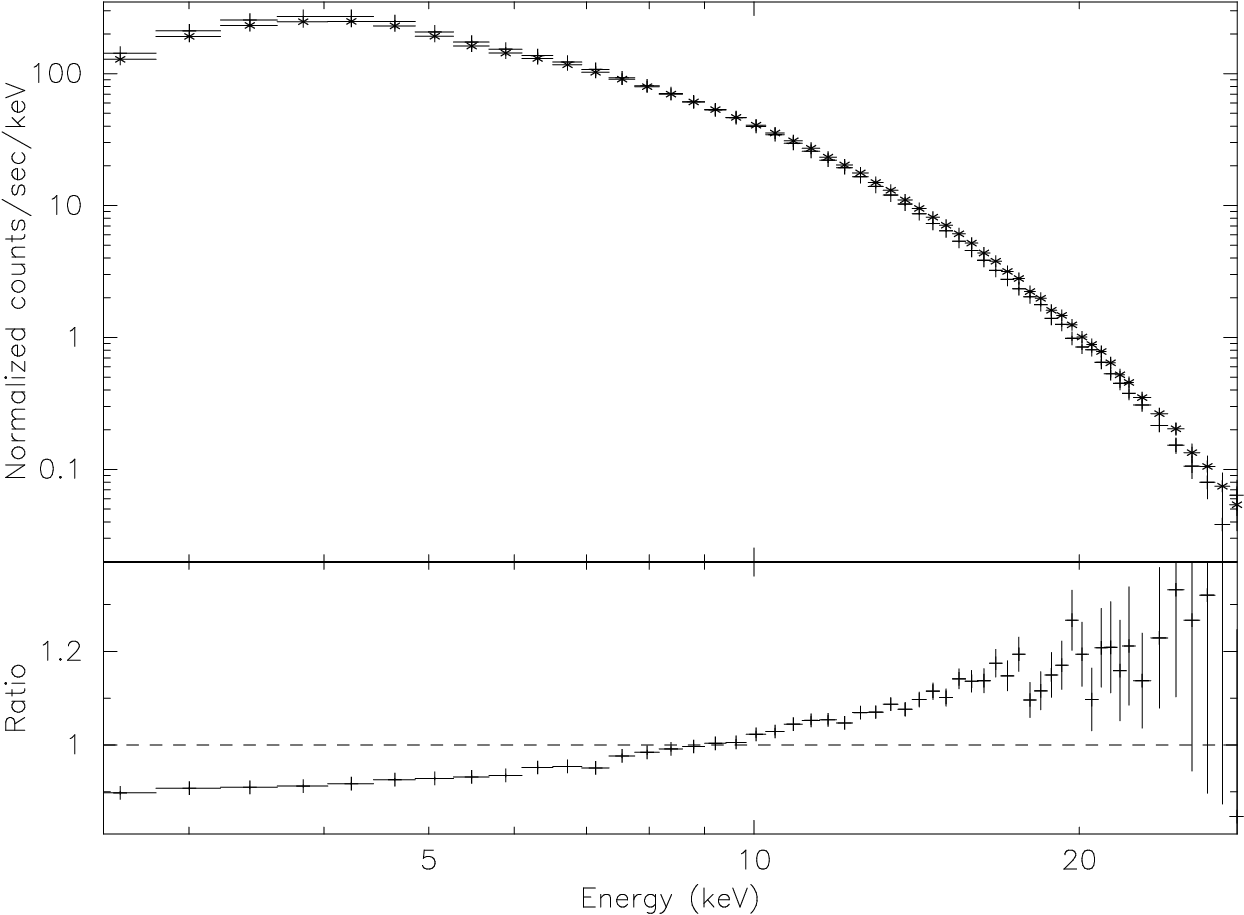}\\[5mm]
\centerline{Fig. 4. --- Continued}
\end{center}
\clearpage
\pagestyle{plaintop}

\begin{figure*}
\includegraphics[width=6cm,angle=270,clip]{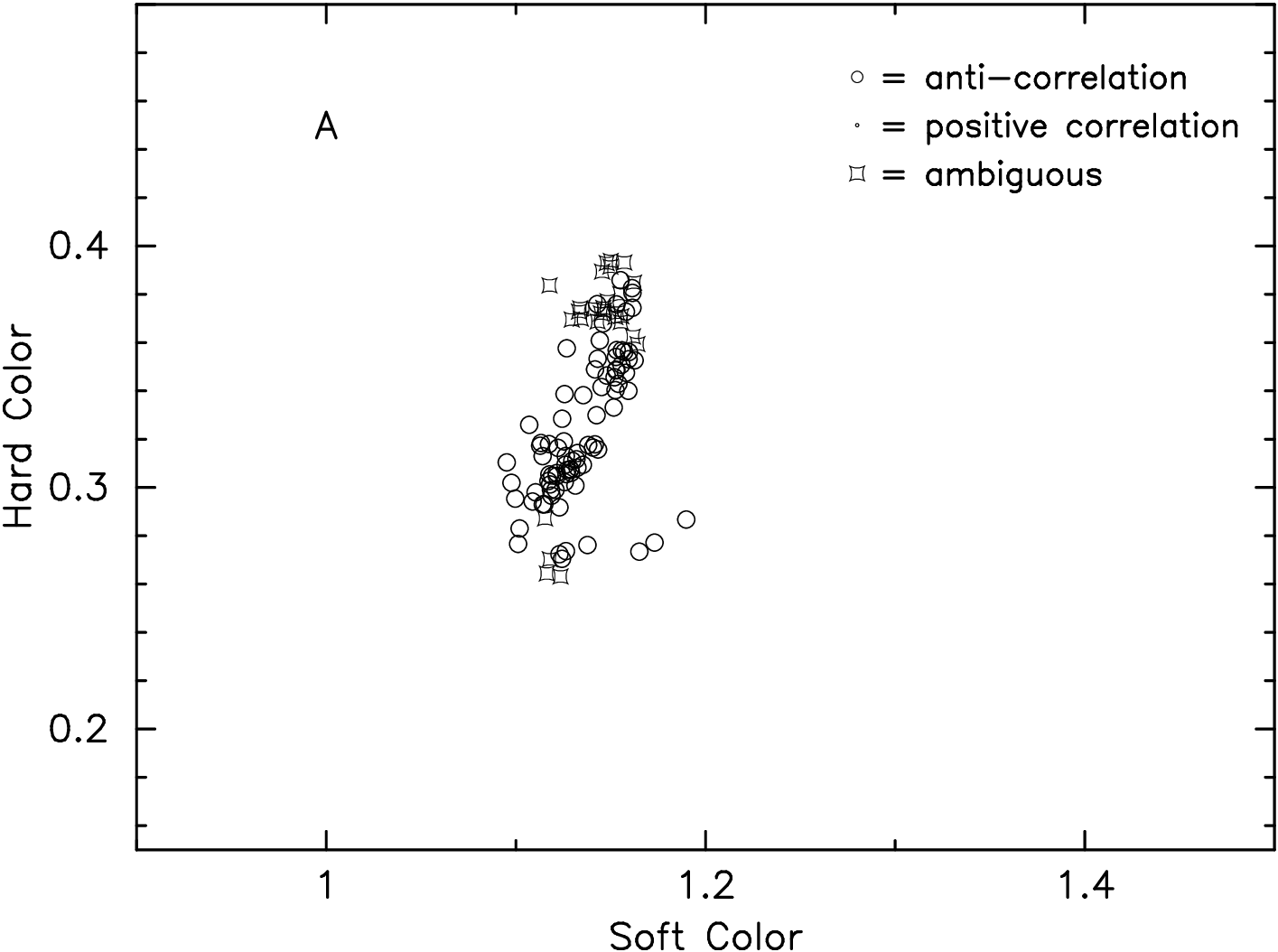}
\includegraphics[width=6cm,angle=270,clip]{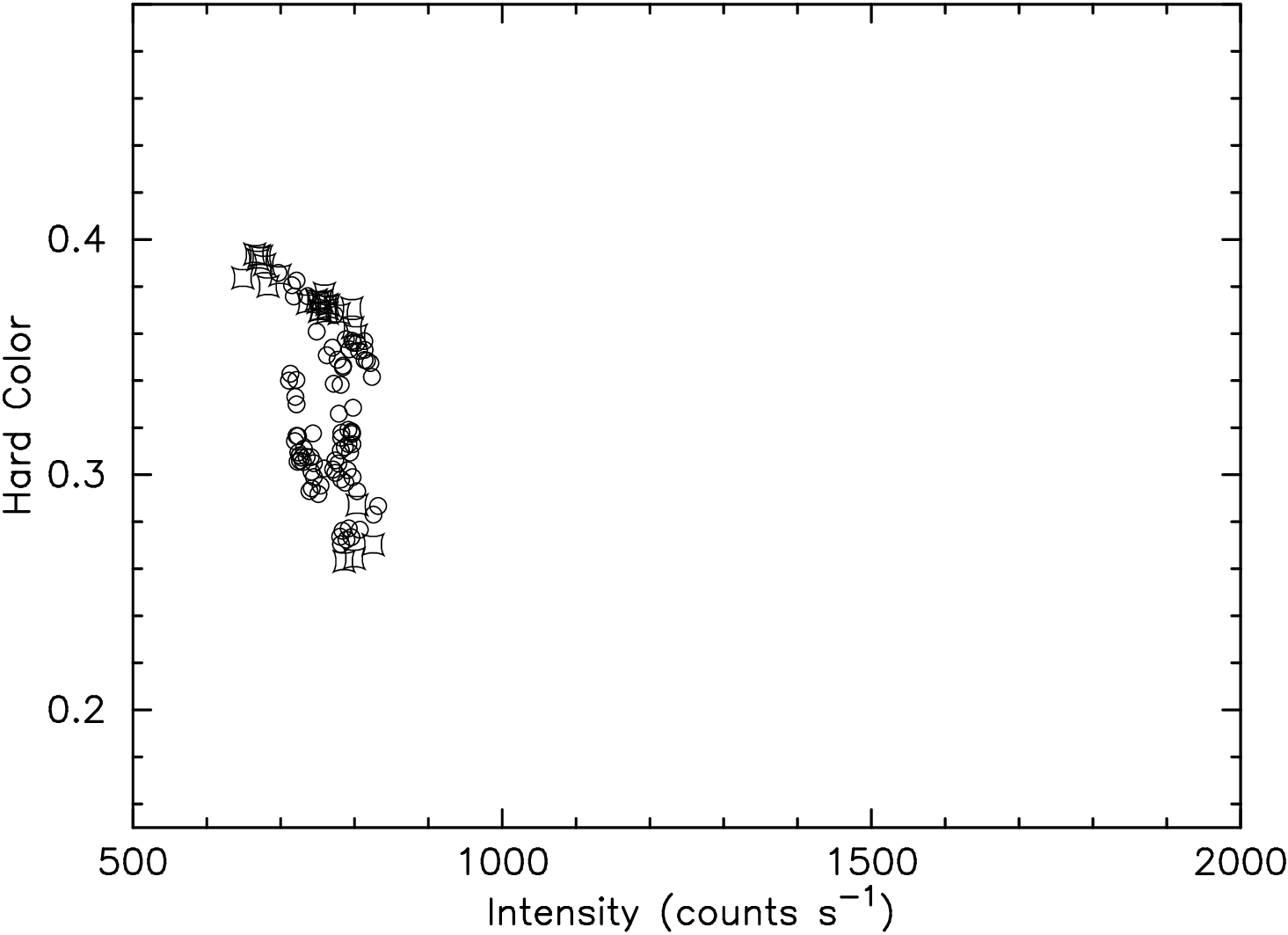}
\includegraphics[width=6cm,angle=270,clip]{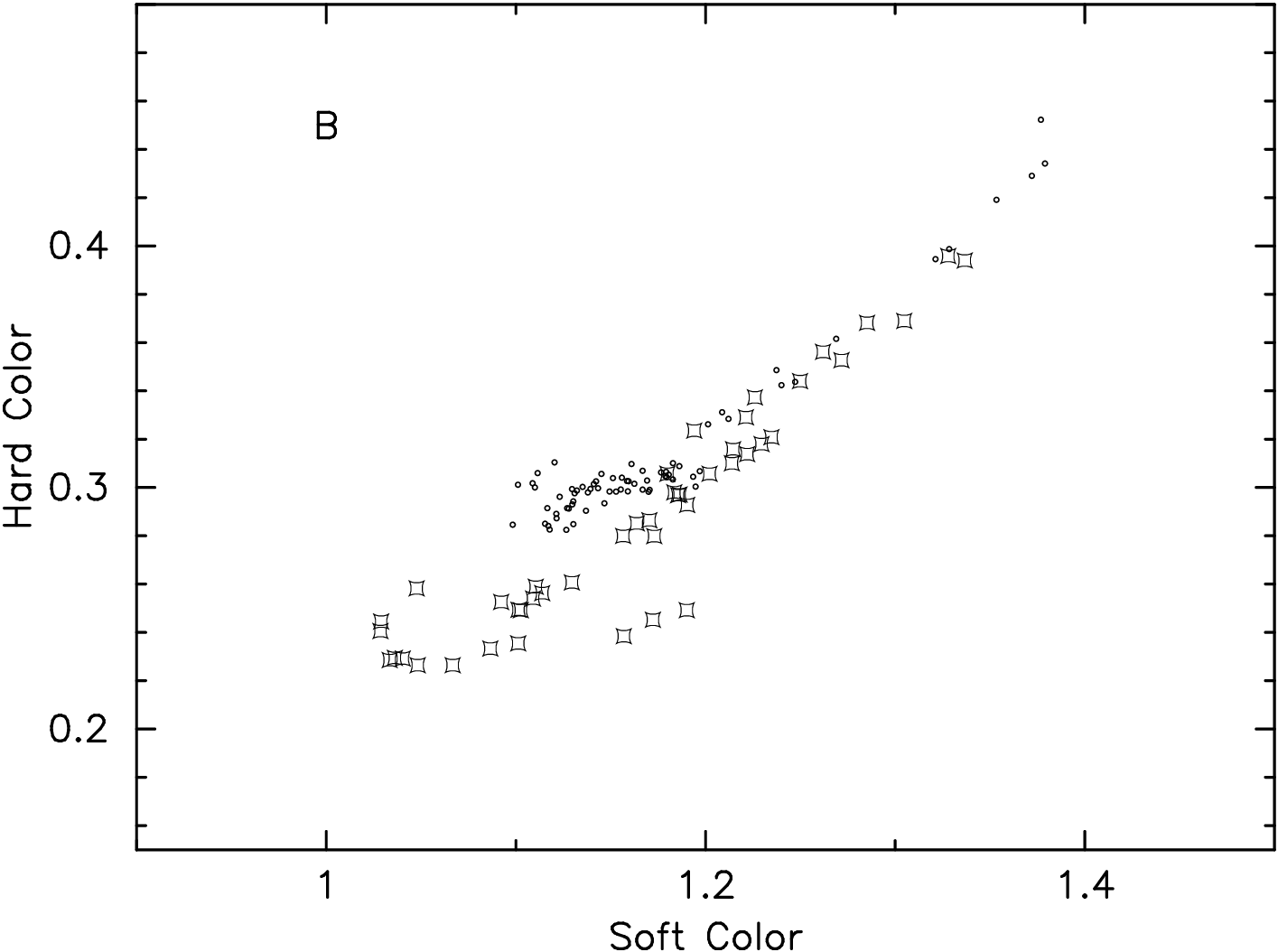}
\includegraphics[width=6cm,angle=270,clip]{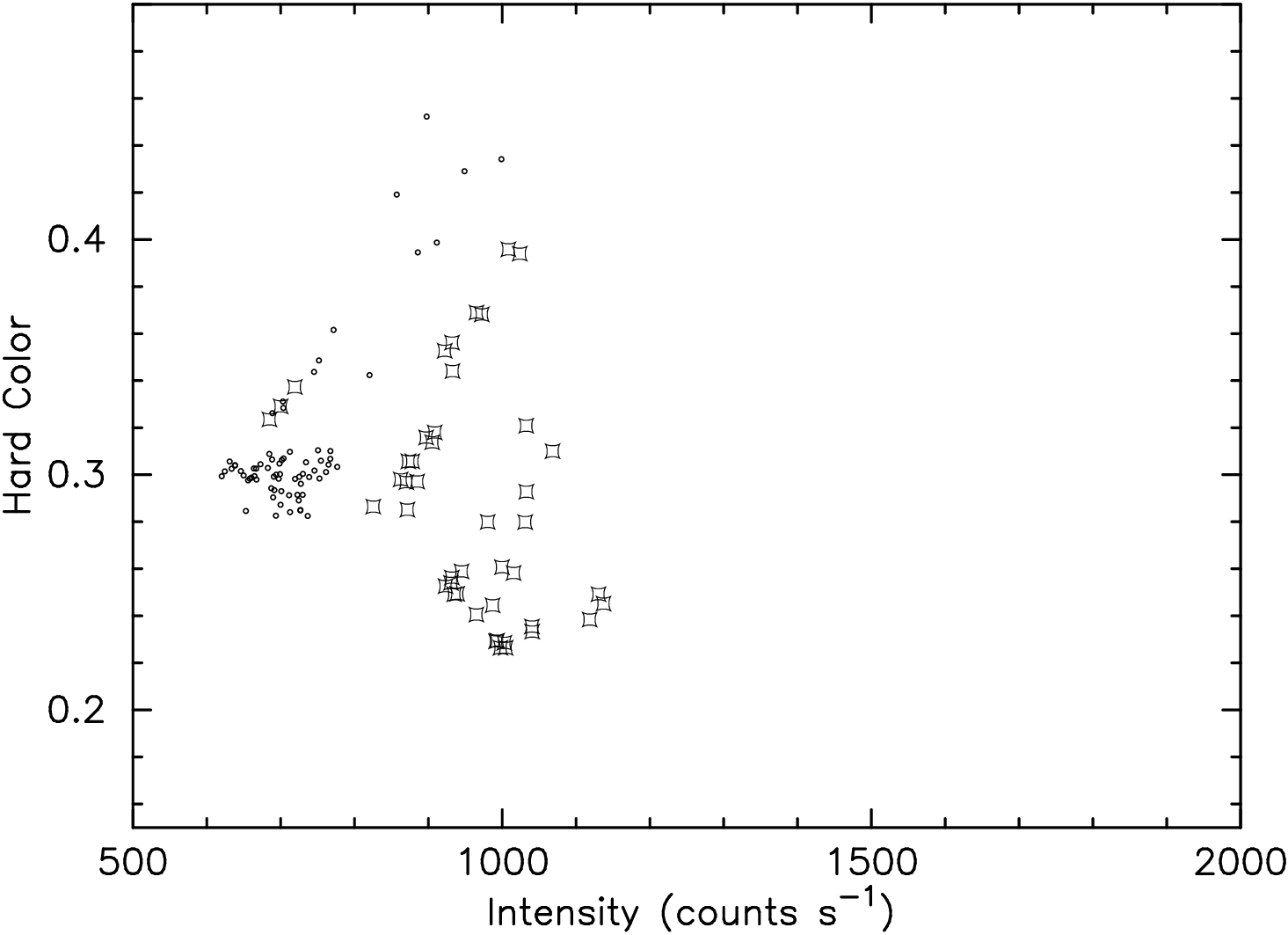}
\caption{The CCDs and HIDs during the low-intensity state from
(A)28/29 September, 1997 and 16/17 December, 1999, (B) 23/24/26 December, 1999. Each point is on average of 1024 s.}
\label{fig5}
\end{figure*}

\begin{figure*}
\begin{center}
\includegraphics[width=10cm,angle=270,clip]{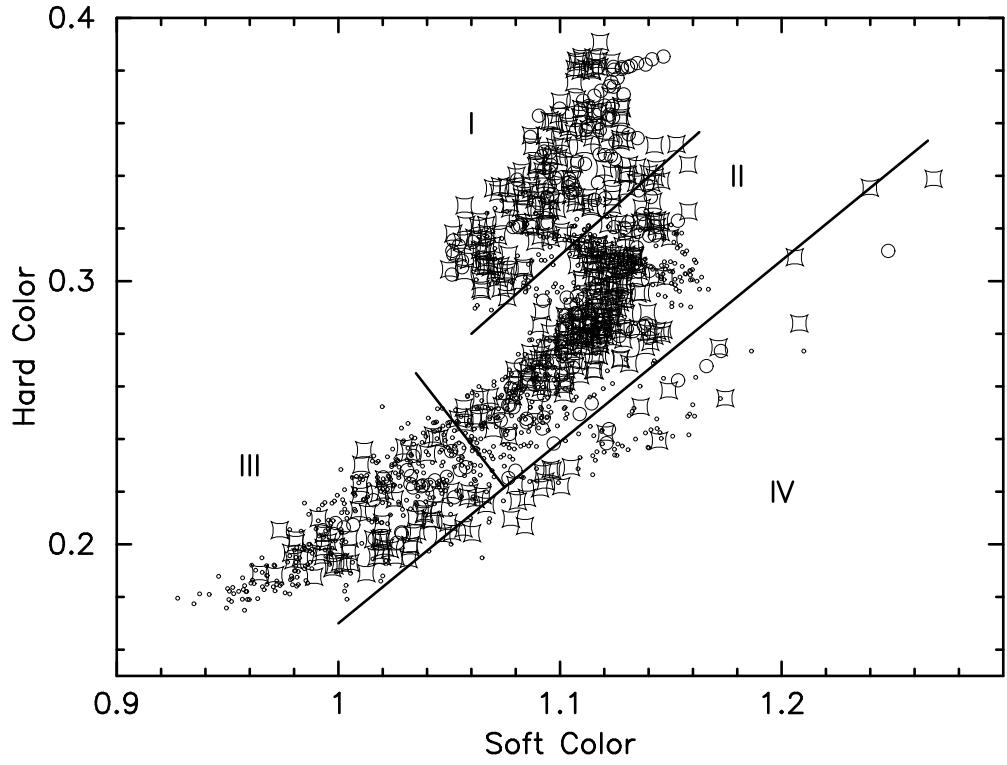}
\caption{The CCD of the observations of the medium-intensity and some low-intensity, 
which is divided into four regions.} \label{fig6}
\end{center}
\end{figure*}

\begin{figure*}
\includegraphics[width=6cm,angle=270,clip]{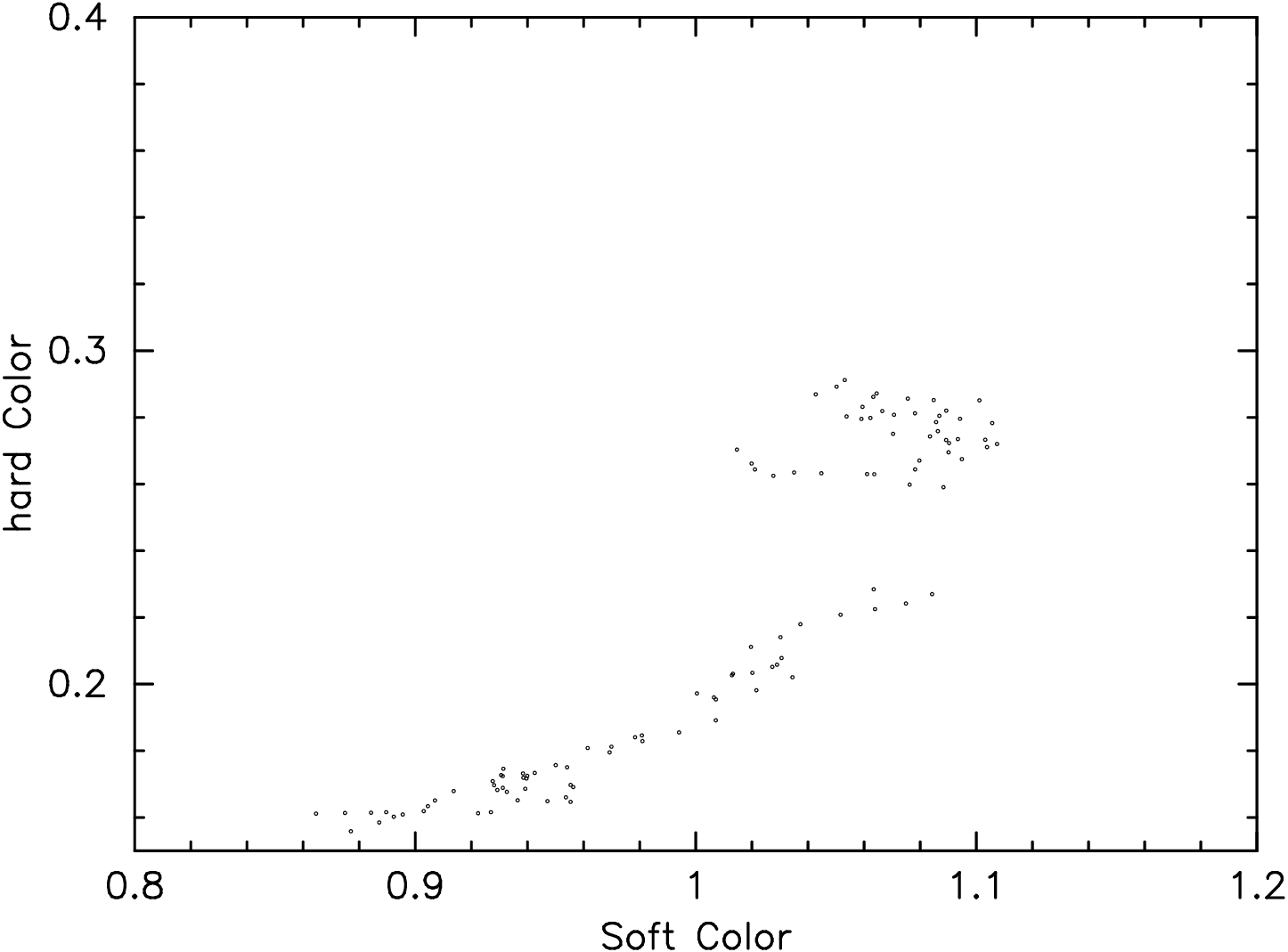}
\includegraphics[width=6cm,angle=270,clip]{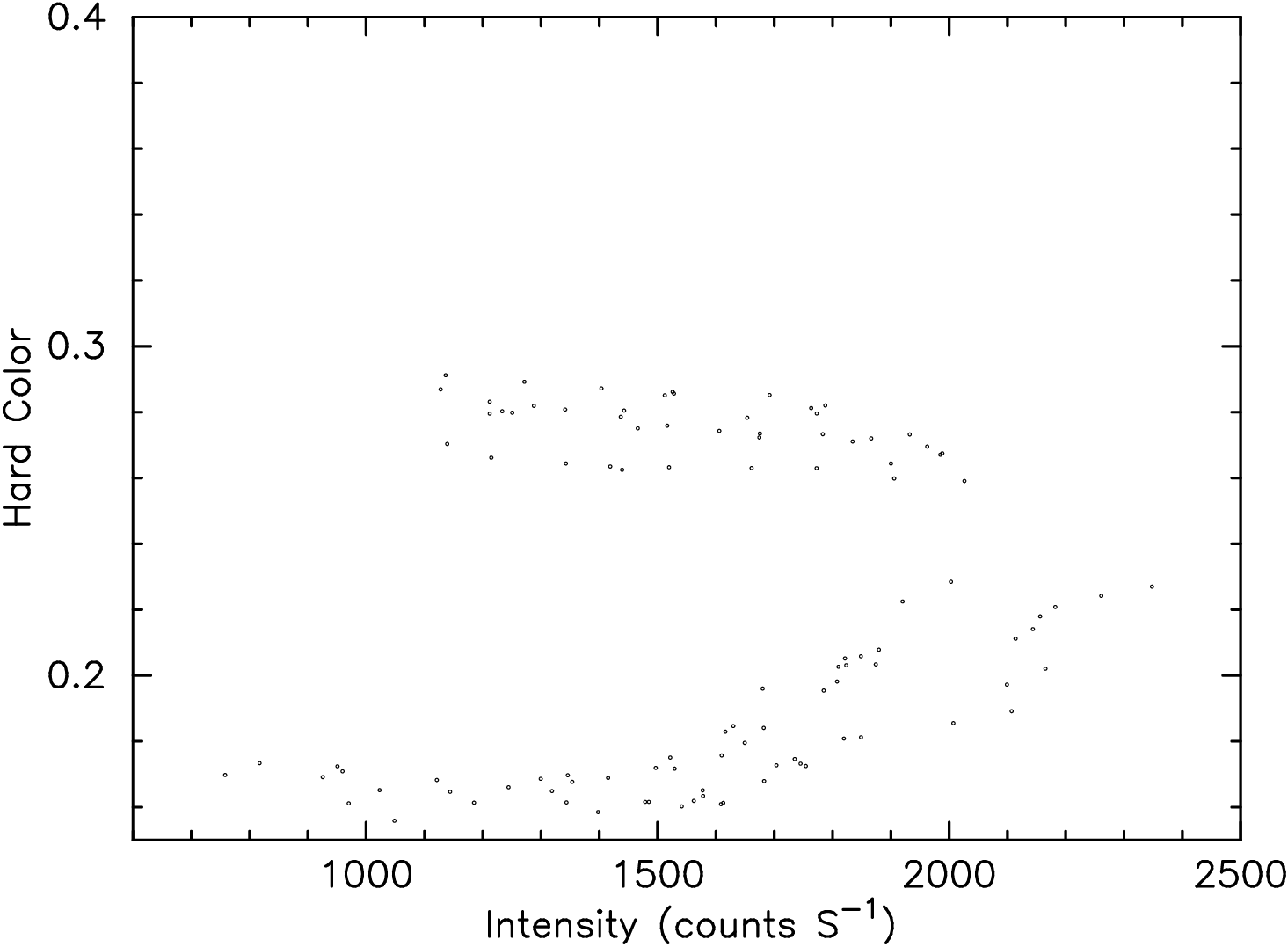}
\caption{The CCD and HID during the high-intensity state of Cyg
X-2.} \label{fig7}
\end{figure*}



\end{document}